\documentclass[aps,prl,onecolumn,amsfonts,superscriptaddress,amssymb,amsmath]{revtex4-2}

\usepackage{graphicx}
\usepackage[separate-uncertainty = false]{siunitx}
\usepackage{color}
\usepackage{lineno}
\usepackage{array}

\newcolumntype{L}[1]{>{\raggedright\let\newline\\\arraybackslash\hspace{0pt}}m{#1}}
\newcolumntype{C}[1]{>{\centering\let\newline\\\arraybackslash\hspace{0pt}}m{#1}}
\newcolumntype{R}[1]{>{\raggedleft\let\newline\\\arraybackslash\hspace{0pt}}m{#1}}

\newcommand\red[1]{{\color{black}#1}}

\newcommand{\calP}{$\mathcal{P}$}
\newcommand{\calT}{$\mathcal{T}$}
\newcommand{\ra}{\rightarrow}
\newcommand{\la}{\leftarrow}

\renewcommand{\baselinestretch}{1.8} 

\newcommand\authors[1]{#1}

\setcitestyle{super}

\begin{document}
\title{Nonlinear optical diode effect in a magnetic Weyl semimetal}%

\authors{\author{Christian Tzschaschel}\email[]{}
\affiliation{Department of Chemistry and Chemical Biology, Harvard University, Massachusetts 02138, USA}

\author{Jian-Xiang Qiu}\affiliation{Department of Chemistry and Chemical Biology, Harvard University, Massachusetts 02138, USA}

\author{Xue-Jian Gao}\affiliation{Department of Physics, Hong Kong University of Science and Technology, Clear Water Bay, Hong Kong, China}

\author{Hou-Chen Li}\affiliation{Department of Chemistry and Chemical Biology, Harvard University, Massachusetts 02138, USA}

\author{Chunyu Guo}
\affiliation{Laboratory of Quantum Materials (QMAT), Institute of Materials (IMX), École Polytechnique Fédérale de Lausanne (EPFL), CH-1015, Lausanne, Switzerland}
\affiliation{Max Planck Institute for the Structure and Dynamics of Matter, Hamburg, Germany}

\author{Hung-Yu Yang}\affiliation{Department of Physics, Boston College, Chestnut Hill, MA, USA}

\author{Cheng-Ping Zhang}\affiliation{Department of Physics, Hong Kong University of Science and Technology, Clear Water Bay, Hong Kong, China}
\author{Ying-Ming Xie}\affiliation{Department of Physics, Hong Kong University of Science and Technology, Clear Water Bay, Hong Kong, China}
\author{Yu-Fei Liu}\affiliation{Department of Chemistry and Chemical Biology, Harvard University, Massachusetts 02138, USA}
\author{Anyuan Gao}\affiliation{Department of Chemistry and Chemical Biology, Harvard University, Massachusetts 02138, USA}
\author{Damien B\'erub\'e}\affiliation{Department of Chemistry and Chemical Biology, Harvard University, Massachusetts 02138, USA}
\author{Thao Dinh}\affiliation{Department of Chemistry and Chemical Biology, Harvard University, Massachusetts 02138, USA}
\author{Sheng-Chin Ho}\affiliation{Department of Chemistry and Chemical Biology, Harvard University, Massachusetts 02138, USA}

\author{Yuqiang Fang}\affiliation{State Key Laboratory of High Performance Ceramics and Superfine Microstructure, Shanghai Institute of Ceramics, Chinese Academy of Science, Shanghai, China}\affiliation{State Key Laboratory of Rare Earth Materials Chemistry and Applications, College of Chemistry and Molecular Engineering Peking University, Beijing, China}

\author{Fuqiang Huang}\affiliation{State Key Laboratory of High Performance Ceramics and Superfine Microstructure, Shanghai Institute of Ceramics, Chinese Academy of Science, Shanghai, China}\affiliation{State Key Laboratory of Rare Earth Materials Chemistry and Applications, College of Chemistry and Molecular Engineering Peking University, Beijing, China}
\author{Johanna Nordlander}\affiliation{Department of Physics, Harvard University, Massachusetts 02138, USA}

\author{Qiong Ma}\affiliation{Department of Physics, Boston College, Chestnut Hill, MA, USA}

\author{Fazel Tafti}\affiliation{Department of Physics, Boston College, Chestnut Hill, MA, USA}

\author{Philip J.W. Moll}
\affiliation{Laboratory of Quantum Materials (QMAT), Institute of Materials (IMX), École Polytechnique Fédérale de Lausanne (EPFL), CH-1015, Lausanne, Switzerland}
\affiliation{Max Planck Institute for the Structure and Dynamics of Matter, Hamburg, Germany}

\author{Kam Tuen Law}\affiliation{Department of Physics, Hong Kong University of Science and Technology, Clear Water Bay, Hong Kong, China}
\author{Su-Yang Xu}
\email[]{ }\affiliation{Department of Chemistry and Chemical Biology, Harvard University, Massachusetts 02138, USA}}


\maketitle

\begin{center}
	$^\ast$ current address: Max-Born Institute for Nonlinear Optics and Short Pulse Spectroscopy, Berlin, Germany. Email: christian.tzschaschel@mbi-berlin.de\\
	$^\dagger$ Email: suyangxu@fas.harvard.edu
\end{center}

\section{Abstract}

\textbf{Diode effects are of great interest for both fundamental physics and modern technologies. Electrical diode effects (nonreciprocal transport) have been observed in Weyl systems. Optical diode effects arising from the Weyl fermions have been theoretically considered but not probed experimentally. Here, we report the observation of a nonlinear optical diode effect (NODE) in the magnetic Weyl semimetal CeAlSi, where the magnetization introduces a pronounced directionality in the nonlinear optical second-harmonic generation (SHG). We demonstrate a six-fold change of the measured SHG intensity between opposite propagation directions over a bandwidth exceeding \SI{250}{meV}. Supported by density-functional theory, we establish the linearly dispersive bands emerging from Weyl nodes as the origin of this broadband effect. We further demonstrate current-induced magnetization switching and thus electrical control of the NODE. Our results advance ongoing research to identify novel nonlinear optical/transport phenomena in magnetic topological materials and further opens new pathways for the unidirectional manipulation of light.}

\section{Introduction}

Diode effects, i.e., phenomena that exhibit a preferred direction, are at the heart of new fundamental physics and modern technologies. A primary example is an electrical diode, which readily conducts current in one direction, but has a high resistance in the opposite direction. In analogy to electrical diodes, optical diodes are characterized by a directionally asymmetric propagation of light. Recent theoretical studies have shown that such an optical diode effect can reveal fundamentally new physics, such as the quantum metric of Bloch wavefunctions \red{\cite{Lapa2019,Gao2019, Zhang2023, Nagaosa2023}}. Moreover, the optical diode would be a crucial component of various optical technologies including photonics, where light instead of electrical current is envisioned as information carrier. For instance, in the microwave regime, the optical diode effect has been proposed for telecommunications \cite{Sabharwal2014,Reiskarimian2016}. Such prospects have inspired a flurry of experimental studies of optical diode effects (such as the nonreciprocal directional dichroism) especially in magnetoelectric wide-bandgap insulators \cite{Fiebig2005Revival, Spaldin2019}. However, the observed optical diode effect is typically very weak with the difference between forward and backward propagation being less than 1\% unless the photon energy is tuned to specific narrow electronic or magnetic resonances with bandwidths often below \SI{5}{meV}.

In addition to its propagation direction, light has another important degree of freedom, the color (frequency). Therefore, it is possible to conceptualize the nonlinear optical diode effect \red{(NODE)}, where the photon frequency changes during the diode process. A second-order diode effect is a process where the optical second-harmonic generation (SHG) of a material along the forward and backward propagation directions differ. In terms of fundamental physics, the NODE may reveal novel quantum geometrical phenomena distinct from the linear optical diode \cite{Nagaosa2020,Orenstein2021,Ma2021,Nagaosa2022, Nagaosa2023}. Also, nonlinear optical processes (e.g. SHG) can lead to a much stronger diode effect (i.e., a larger contrast between the forward and backward direction) as SHG supports much stronger crystallographic selection rules \cite{Toyoda2021,Mund2021}. In terms of technology, the NODE may also enable novel optical applications such as unidirectional and mode-locked lasers \cite{Fan2012} and novel designs for optical isolators for optical communication \cite{Feng2011}.  

\red{We show in Fig.~\ref{Fig1}a a schematic representation of the NODE. From the point of view of symmetry, we note that the two opposite propagation directions are related by a mirror operation (mirror perpendicular to $\hat{\mathbf{x}}$ in Fig.~\ref{Fig1}a). The observation of a directionally dependent SHG intensity, i.e., a NODE, thus requires the presence of different contributions to the SHG intensity that are even and odd with respect to the mirror operation. This distinguishes the NODE from other SHG interference phenomena; most notably magnetic SHG contrast \cite{Fiebig05} and nonreciprocal SHG \cite{Toyoda2021, Mund2021}. Magnetic contrast arises due to  interference between i-type (time reversal \calT\ even) and c-type (\calT\ odd) contributions to the SHG response and thus strictly requires \calT\ symmetry to be broken. Nonreciprocal SHG arises from a $\mathbf{k}$-dependent nonlinear optical susceptibility ($\mathbf{k}$ being the photon momentum). Therefore, nonreciprocal SHG contains $\mathbf{k}$-even contributions, such as electric-dipole SHG (which is \calP\ odd) and $\mathbf{k}$-odd contributions, such as as magnetic-dipole or electric-quadrupole SHG (which are \calP\ even) \cite{Toyoda2021, Mund2021}. In contrast, the NODE is observable in the absence of \calT\ or \calP\ symmetry breaking or can be realized purely based on electric-dipole SHG. The NODE generalizes the ground-breaking concept of nonreciprocal SHG and may occur concomitantly with a magnetic contrast (see Supplementary Section 1D for an extended discussion of SHG interference phenomena and a minimal example to realize a NODE).}

Recently, Weyl semimetals have emerged as an attractive system to explore various novel electrical and optical phenomena, due to its combination of nontrivial quantum geometry and strong symmetry breaking (to generate Weyl fermions, the system has to break space-inversion, or time-reversal, or both). In the electrical regime, a giant electrical diode effect (the nonreciprocal transport) has been observed in Weyl systems \cite{Tokura2018}. In the optical regime, novel optical diode effect has been theoretically considered \cite{Lapa2019, Gao2019, Nandy2022} but never probed experimentally. In contrast to conventional magnetoelectric insulators \cite{Toyoda2021, Mund2021} (Fig.~\ref{Fig1}a), the Weyl semimetals host linearly dispersive Weyl cones, which can allow for a wide range of photon energies to excite interband optical transitions in the material (Fig.~\ref{Fig1}b).

In this paper, we report the observation and manipulation of a giant, broadband NODE in the magnetic Weyl semimetal CeAlSi. We demonstrate that at least a sixfold change in SHG intensity is achievable between forward and backward propagating light over a broad spectral range covering \SI{250}{meV}. Supported by first-principles calculations, we establish a link between the broadband characteristics and the linearly dispersive Weyl fermions in CeAlSi. We show that the directionality of the NODE is directly related to the magnetic order in CeAlSi. We can switch the magnetic order by passing a current and thus demonstrate electrical control of the NODE. We reveal that this basic opto-spintronic functionality persists in technologically relevant, micromachined device structures with the additional advantage of drastically reduced switching currents.

\section{Results}

\subsection{Observation of the nonlinear optical diode effect in CeAlSi}

CeAlSi belongs to the family of Weyl semimetals with chemical formula $R$Al$X$ ($R=\textrm{La-Nd}$ and $X$=\textrm{Si or Ge}) that recently gained considerable attention as magnetic Weyl semimetals\cite{Chang2018Magnetic, Hodovanets2018single, Suzuki2019singular, Puphal2020Topological, Yang2021, Sun2021Mapping, Su2021Multiple, Gaudet2021Weyl, Yao2023, Alam2023, Zhang2023a, Hodovanets2022}. This family is closely related to TaAs \cite{Xu2015Discovery, Lv2015Experimental}, as they have the same noncentrosymmetric, tetragonal crystal structure (point group $4mm$) and the same valence condition ($R^{3+}[\textrm{Al}X]^{3-}$ vs. $\textrm{Ta}^{3+}\textrm{As}^{3-}$). The introduction of magnetic rare earth elements (Ce, Pr, or Nd) in $R$Al$X$ compounds leads to long-range magnetic order. Depending on the rare earth element, a variety of magnetic structures ranging from simple collinear ferromagnetism to noncollinear magnetism and complex multi-\textbf{k} helical spin structures has been found \cite{Hodovanets2018single, Suzuki2019singular, Puphal2020Topological, Yang2021, Sun2021Mapping, Su2021Multiple, Gaudet2021Weyl}. The interplay between Weyl fermions, inversion symmetry-breaking, and magnetism provides a fertile ground to discover novel emergent quantum electromagnetism in this class of Weyl semimetals.

\begin{figure}[h]
	\includegraphics[width = 110 mm]{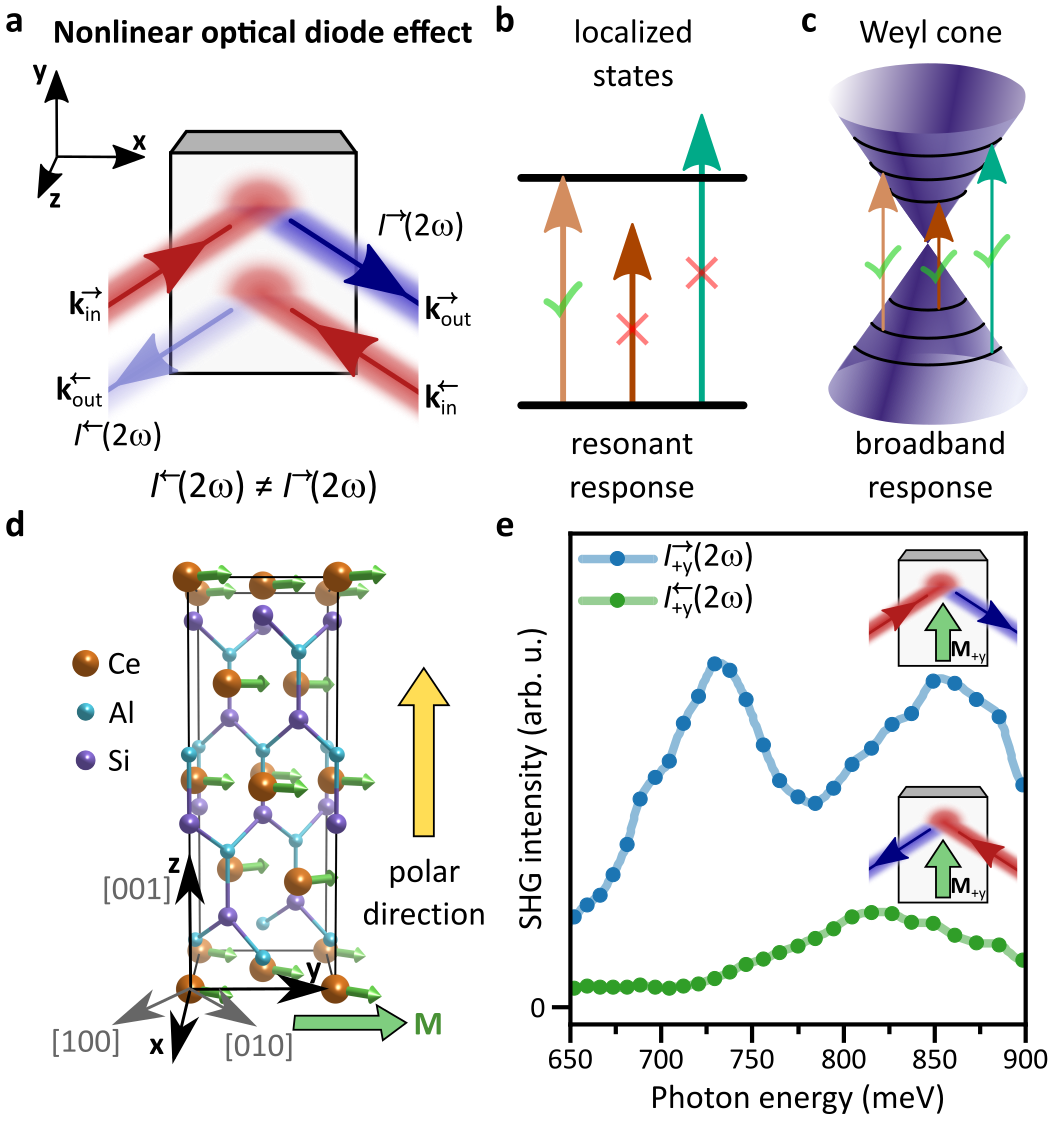}
	\caption{\textbf{Observation of a broadband NODE in the magnetic Weyl semimetal CeAlSi.} \textbf{a} Definition of the nonlinear optical diode effect (NODE) in reflection. Reversing the light path by swapping light source and detector corresponds to changing the light $\mathbf{k}$ vectors $\mathbf{k}^{\ra}_\mathrm{in}$ and $\mathbf{k}^{\ra}_\mathrm{out}$ to $\mathbf{k}^{\la}_\mathrm{in}$ and $\mathbf{k}^{\la}_\mathrm{out}$. A NODE is evidenced by a difference between intensities $I^{\ra}(2\omega)$ in forward direction and $I^{\la}(2\omega)$ in reversed direction. \textbf{b} For a \red{transition between localized electronic states}, such as in wide-gap insulators, only specific photon energies can drive the transition resulting in a resonant response. \textbf{c} For dispersive bands such as Weyl cones, different photon energies can drive transitions leading to a broadband response. \textbf{d} Crystal structure of the magnetic Weyl semimetal CeAlSi. We define $+\hat{\mathbf{x}} = [110]$, $+\hat{\mathbf{y}} = [\bar{1}10]$, and $+\hat{\mathbf{z}} = [001]$ relative to the tetragonal unit cell of (paramagnetic) CeAlSi. \textbf{e} Observation of a broadband NODE in the $\mathbf{M}_\mathrm{+y}$ state of CeAlSi. $I^{\ra}_\mathrm{+y}(2\omega)~\gg~I^{\la}_\mathrm{+y}(2\omega)$ over a broad range $>$\SI{250}{meV}.}
	\label{Fig1}
\end{figure}

CeAlSi exhibits a canted ferromagnetic order below $T_{\textrm{C}}\sim \SI{8.4}{K}$ (Fig.~\ref{Fig1}c). The net magnetization points along one of the four symmetry-equivalent in-plane crystallographic directions, $\pm[110]$ and $\pm[\bar{1}10]$ (We define $[110]=+\hat{\mathbf{x}}$, $[\bar{1}10]=+\hat{\mathbf{y}}$, and $[001]=+\hat{\mathbf{z}}$, Fig.~\ref{Fig1}c). Hence, there are four distinct ferromagnetic states, denoted as $\mathbf{M}_\mathrm{+x}$, $\mathbf{M}_\mathrm{-x}$, $\mathbf{M}_\mathrm{+y}$, and $\mathbf{M}_\mathrm{-y}$. Such a \calT-breaking magnetic ordering on a \calP-breaking crystal structure makes CeAlSi a promising candidate for the observation of the NODE. 

We consider the NODE explicitly by experimentally reversing the light path. As CeAlSi is semimetallic and therefore highly reflective, we consider the NODE here in a reflection geometry (Fig.~\ref{Fig1}d). We study in particular SHG from the $(001)$ facet of a CeAlSi single crystal. Therefore, all four magnetic states of CeAlSi exhibit an in-plane net magnetization. We orient the crystal such that the s-polarization of light is parallel to the net magnetization of two out of the four possible magnetic states. We denote those states as $\mathbf{M}_\mathrm{\pm y}$ states and the states with net magnetization perpendicular to the s-polarization $\mathbf{M}_\mathrm{\pm x}$ states (see coordinate system in Fig.~\ref{Fig1}d).

In Fig. \ref{Fig1}e, we directly compare the spectra for the recorded SHG intensity for the forward propagation direction ($I^{\ra}(2\omega)$) and the backward propagation direction ($I^{\la}(2\omega)$) for the magnetic $\mathbf{M}_\mathrm{+y}$ state. The SHG intensity $I^{\ra}(2\omega)$ in forward direction is more than twice as large as the intensity $I^{\la}(2\omega)$ in backward direction over the entire considered spectral range of the incident light from \SI{650}{meV} to \SI{900}{meV}. Reversing the magnetization reverses the situation (Supplementary Section 4A). We thus find that the SHG response exhibits a NODE over a wide range of fundamental photon energies spanning at least \SI{250}{meV}. 

\begin{figure}[h]
	\includegraphics[width = 110 mm]{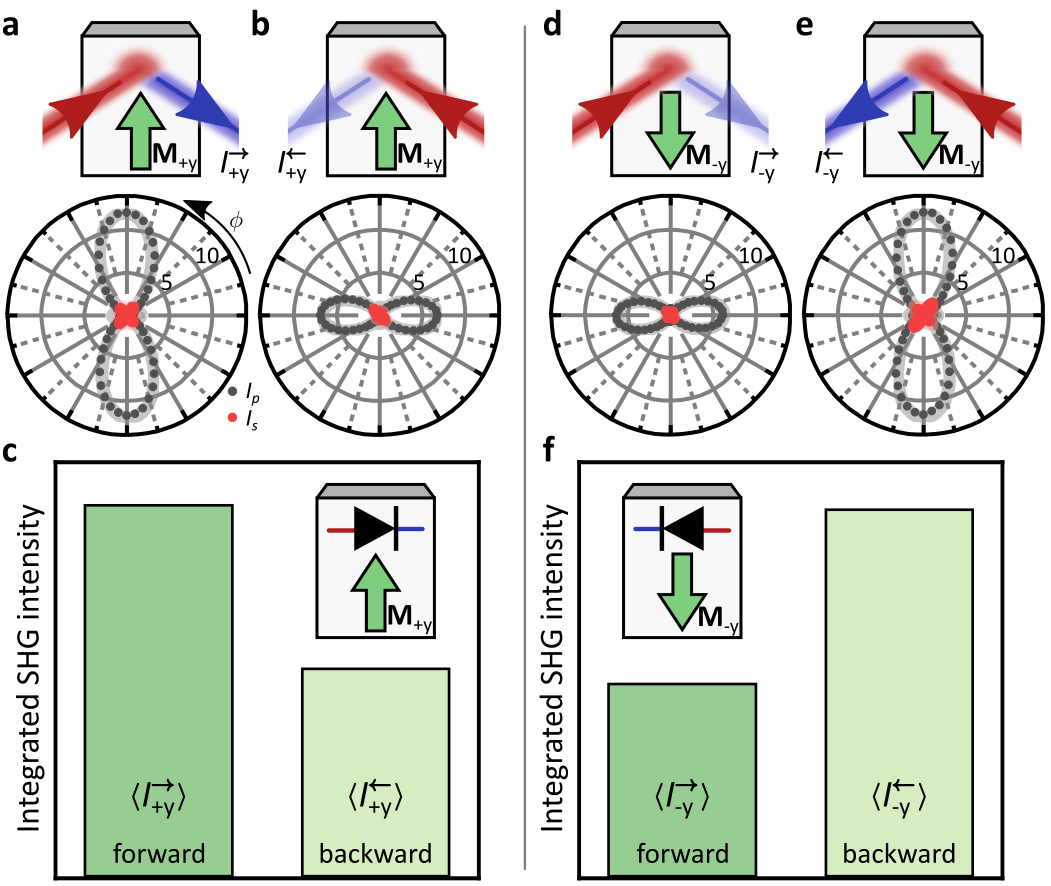}
	\caption{\textbf{Magnetic control of the NODE.} \textbf{a,b} Polarization-dependent SHG for both propagation directions of the $\mathbf{M}_\mathrm{+y}$ state (fundamental photon energy \SI{715}{meV}). \textbf{c}  Integrated SHG intensity of the $\mathbf{M}_\mathrm{+y}$ state for both propagation directions. The integrated SHG intensity is defined as the intensity measured without polarization analysis, thus simultaneously detecting s- and p-polarized SHG light, and averaging over all polarizations of the incident fundamental light (see Supplementary Section 8 for details). \textbf{d-f} Analogous to panels a-c but for the $\mathbf{M}_\mathrm{-y}$ state. Solid lines are fits with one consistent set of ED-SHG $\mathcal{X}$ tensor components (see \red{Supplementary Sections  1D and 2 for details}). The magnetic state of CeAlSi controls the directionality of the NODE enabling a switchable nonlinear optical diode.}
	\label{Fig2}
\end{figure}

In Figs.~\ref{Fig2}a,b we show the polarization dependence of $I_\mathrm{+y}^{\ra}$ and $I^{\la}_\mathrm{+y}$, respectively, at a fundamental photon energy of \SI{715}{meV} for the magnetic $\mathbf{M}_\mathrm{+y}$ state of CeAlSi. The polarization dependence also clearly reveals the NODE in the SHG response. To further corroborate the nonreciprocal character of the observed SHG, we define the integrated SHG intensity $\langle I\rangle$ as the intensity measured without polarization analysis, thus simultaneously detecting s- and p-polarized SHG light, and averaging over all polarizations of the incident fundamental light. Comparing the integrated SHG intensity between the forward and backward direction gives us a measure of the diode effect independent of the light polarization (i.e., the diode effect for unpolarized light). Interestingly, we find for the magnetic $\mathbf{M}_\mathrm{+y}$ state that the integrated SHG intensity  $\langle I^{\ra}_\mathrm{+y}\rangle$ in the forward direction is significantly higher than $\langle I^{\la}_\mathrm{+y}\rangle$ in the reversed direction (Fig.~\ref{Fig2}c). This shows that the NODE is not restricted to particular polarization configurations but a net diode effect can persist even in the case of unpolarized incident light.

\subsection{Magnetic control of the NODE}

Interestingly, the directionality of the NODE can be controlled by the magnetic order. In Figs.~\ref{Fig2}a-c and ~\ref{Fig2}d-e, we compare the results for the $\mathbf{M}_\mathrm{+y}$ and $\mathbf{M}_\mathrm{-y}$ states, from which we see that the directionality of the NODE is flipped as we flip the magnetization. Thus, while the broken \calP\ symmetry in general activates the NODE, the broken \calT\ symmetry enables controlling its directionality. By selecting either the $\mathbf{M}_\mathrm{+y}$ or $\mathbf{M}_\mathrm{-y}$ state, we can deterministically set the propagation direction that generates higher integrated SHG intensity (see also Supplementary Movies 1 and 2 for the domain evolution during magnetic-field induced switching).

We can understand the NODE and its magnetic control by the interference of independent SHG tensor components with different origins. In our case, we can use $\chi_\mathrm{xxx}$ and $\chi_\mathrm{xxz}$. $\chi_\mathrm{xxz}$ arises from the noncentrosymmetric polar lattice of CeAlSi; $\chi_\mathrm{xxx}$ arises from the magnetic order. As shown in \red{Supplementary Fig.~5}, the total SHG intensity is given by $\vert \chi_\mathrm{xxx}E_\mathrm{x}^2+\chi_\mathrm{xxz}E_\mathrm{x}E_\mathrm{z}\vert^2$. Considering p-polarized incident light, the electric field $\mathbf{E}$ of the light wave is parallel to $\mathbf{k}\times\hat{\mathbf{y}}$ leading to a $\mathbf{k}$ dependence of the SHG response. Specifically, as we reverse the light propagation direction, we reverse the relative sign between $E_x$ and $E_z$. Hence, we find $I_\mathrm{+y}^\rightarrow(2\omega) \propto \vert\chi_\mathrm{xxx}+\chi_\mathrm{xxz}\vert^2$ and $I_\mathrm{+y}^\leftarrow(2\omega) \propto \vert\chi_\mathrm{xxx}-\chi_\mathrm{xxz}\vert^2$. Thus, the NODE is based on a directionally dependent mixing between tensor components. It is interesting to note that the origin of the symmetry breaking that enables those tensor components is not crucial. This is in contrast to the established SHG interference imaging technique which strictly relies on interference between order-parameter dependent and order-parameter independent $\chi$ tensor components \cite{Fiebig05}. As a consequence, a NODE can in principle exist even in non-magnetic materials. Here, however, we explicitly utilize the different origin of $\chi_{xxx}$ and $\chi_{xxz}$ to control the directionality of the NODE. As we reverse the magnetization direction, we reverse the sign of the magnetic SHG tensor $\chi_{xxx}$ but the crystalline SHG tensor component $\chi_{xxz}$ remains unchanged. Thus, we find $I_\mathrm{-y}^\ra(2\omega) \propto \vert-\chi_\mathrm{xxx}+\chi_\mathrm{xxz}\vert^2$ and $I_\mathrm{-y}^\la(2\omega) \propto \vert-\chi_\mathrm{xxx}-\mathrm\chi_\mathrm{xxz}\vert^2$, which explains why the NODE directionality flips as we flip the magnetization direction. 

\subsection{Electrical control of the NODE}

\begin{figure}
	\includegraphics[width = 90 mm]{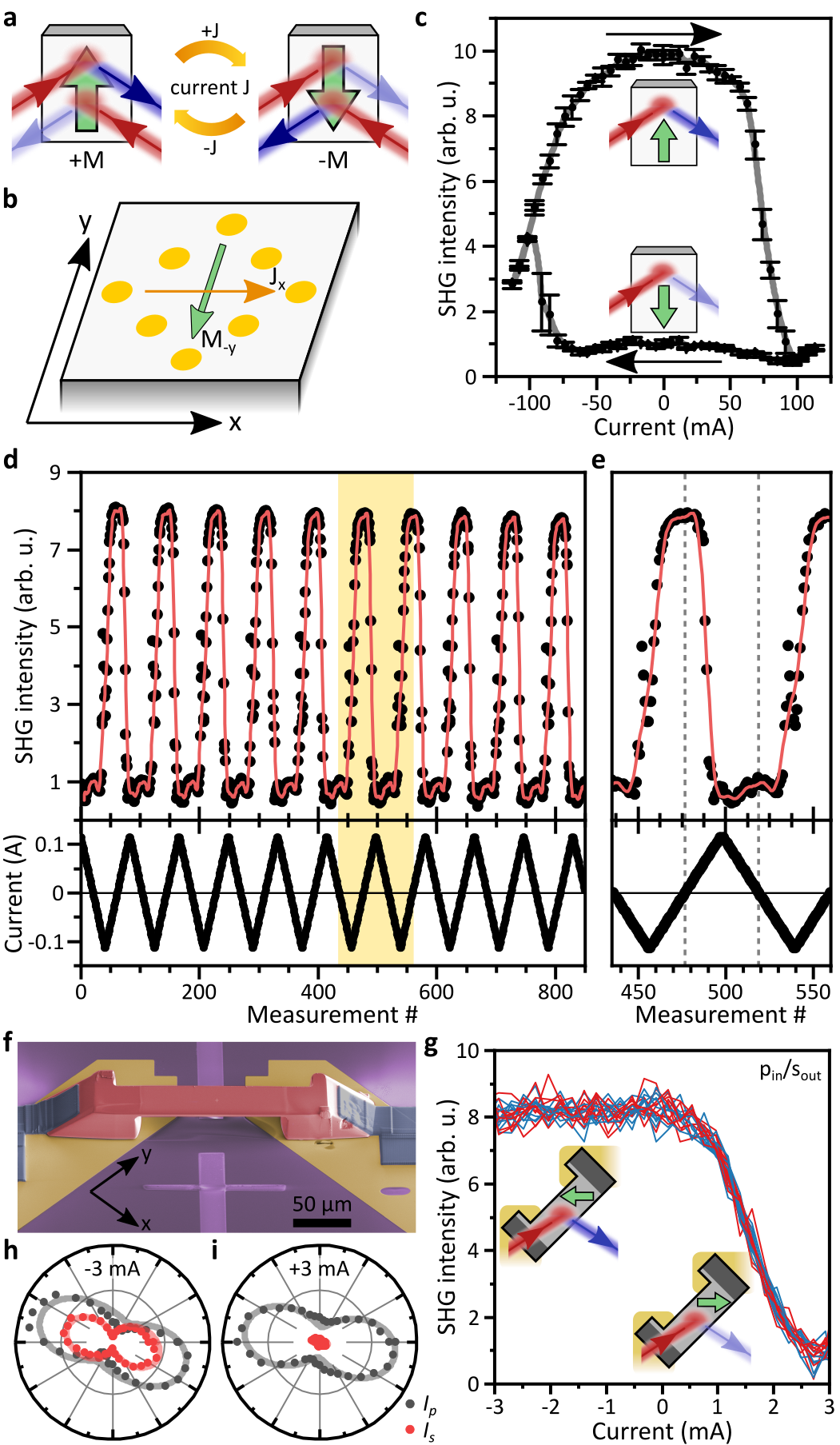}
	\caption{\textbf{Electrical control of the NODE.} \textbf{a} Schematic of a device concept. Current-induced magnetization switching enables electrical control of the NODE. \textbf{b} We deposit eight gold electrodes on the top surface of a bulk mm-sized CeAlSi crystal, which allows us to pass current through the sample. Here, we focus on current along the $\hat{\mathbf{x}}$ direction, which causes magnetization switching along the $\hat{\mathbf{y}}$ direction. \textbf{c} Current hysteresis loop. A current of \SI{100}{mA} fully switches the magnetization. We measure p-polarized SHG in forward direction with s-polarized incident light. Therefore, high and low intensity correspond to $\mathbf{M}_\mathrm{+y}$ and $\mathbf{M}_\mathrm{-y}$ states, respectively. \red{Error bars denote the standard deviation of three consecutive measurements.} \textbf{d} Current-induced magnetization switching is highly reproducible. Here, we show 10 consecutive cycles. \textbf{e} Enlarged view of the highlighted region in d. Dashed lines correspond to the remanent states. \textbf{f} Scanning electron microscopy image of a micro-sized device manufactured by focused ion beam milling of CeAlSi (red) on gold electrodes (gold). Platinum contacts (gray) ensure good electrical connections. \textbf{g} Currents as small as \SI{3}{mA} can control the magnetization state. Red and blue curves correspond to consecutive measurements with increasing and decreasing current, respectively. We detect s-polarized SHG, which can discriminate $\mathbf{M}_\mathrm{\pm x}$ states. A hysteresis is suppressed possibly due to shape anisotropy or residual strain in the sample \cite{Dubowik1996,Wang2013,Xu2021}. \textbf{h,i} SHG polarization dependence under application of \SI{-3}{mA} and +\SI{3}{mA}, respectively. Changes are most prominent for s-polarized SHG (red). \red{Solid lines are fits corresponding to magnetic $\mathbf{M}_\mathrm{-x}$ and $\mathbf{M}_\mathrm{+x}$ states, respectively (Supplementary Section 9).}}
	\label{Fig3}
\end{figure}

The noncentrosymmetric symmetry of CeAlSi supports a magnetoelectric coupling between the magnetization $\mathbf{M}$ and a current $\mathbf{J}$. Microscopically, such a coupling may be either intrinsic \cite{Birss66, Johansson2018}, mediated by strain \cite{Xu2021}, or relying on the spin-Hall effect\cite{Sinova2015} and assisted by the Oersted field of the current \cite{Krause2007,Mendil2019}. In all cases, the magnetoelectric coupling is such that a current in the $xy$-plane favors a magnetization perpendicular to the current in the $xy$-plane with a fixed handedness as depicted in Fig.~\ref{Fig3}b, providing a way to control the magnetic domain configuration through an electrical current.  

We supply a current in the $xy$-plane through a pattern of gold electrodes on a $(001)$ facet of the sample as illustrated in Fig.~\ref{Fig3}b. We can then use SHG to optically detect the magnetization state. Specifically, we choose s-polarized incident light at \SI{715}{meV} and detect p-polarized SHG, where $\mathbf{M}_\mathrm{\pm y}$ states correspond to high and low SHG intensity, respectively (Fig. \ref{Fig2}a,b. See also \red{Supplementary Section 5} for more systematic characterizations).

Remarkably, we find that passing a current in CeAlSi can directly choose its magnetization direction (see also Supplementary Movie 3 for \textit{operando} imaging of the domain pattern). Moreover, the selected magnetization persists when the current is removed (i.e., the control is nonvolatile). Specifically, as shown in Fig.~\ref{Fig3}c, if we pass a negative current of \SI{-100}{mA} along the $\hat{\mathbf{x}}$ direction and then remove this current, we control the magnetization to be along $+\hat{\mathbf{y}}$ ($\mathbf{M}_\mathrm{+y}$); if we pass a positive current of \SI{+100}{mA} and then remove this current, we control the magnetization to be along $-\hat{\mathbf{y}}$ ($\mathbf{M}_\mathrm{-y}$). Therefore, the data in Fig.~\ref{Fig3}c shows that the magnetization of CeAlSi can be controlled by current. Moreover, Fig.~\ref{Fig3}d shows the results over the course of $>10$ consecutive current cycles along opposite directions. We found that the control is highly deterministic, i.e., $-\hat{\mathbf{x}}$ current selects $+\hat{\mathbf{y}}$ magnetization whereas $+\hat{\mathbf{x}}$ current selects $-\hat{\mathbf{y}}$ magnetization. The control is also clearly nonvolatile, i.e., the selected magnetization persists when the current is reduced to zero. By electrically controlling the NODE, we realize here a basic \red{directional} nonlinear opto-spintronic functionality.

We note that the above results are obtained on a large mm-sized single crystal. So although the total current needed for the magnetic control appears large (\SI{100}{mA}), the current density is very small. However, as the contacts are deposited on the top surface (Fig.~\ref{Fig3}b), the spatial distribution of current flow is likely to be inhomogeneous for such a thick bulk crystal (especially along the $z$ direction), so it is hard to reliably estimate the current density. \red{In addition, although the sample remains in the magnetically ordered phase even at the applied highest currents (Supplementary Section 3), we notice significant Joule heating. In order to minimize Joule heating and at the same time reliably estimate the current density, it is highly desirable to fabricate a miniaturized device.} Such a micro-device also helps us to explore the tantalizing potential of electrically controlling the nonlinear optical response of CeAlSi for spintronic applications. 

In Fig.~\ref{Fig3}f, we use focused ion-beam (FIB) milling to fabricate such a micro-device of CeAlSi. The core part of the structure is a free-standing slab of CeAlSi of $\SI{126}{\micro m}\times\SI{29}{\micro m}\times\SI{2}{\micro m}$. In order to counter-act the significant magnetic shape anisotropy due to the large aspect ratio and keep the device switchable, we prepared the slab with the long axis along the magnetic hard axis, i.e. at an angle of $45^\circ$ relative to the $x$ and $y$ directions. The device can be poled into a magnetic single domain state with magnetic fields of about \SI{30}{mT}. Analogously to the previous bulk measurements, the four magnetic states can be distinguished by their SHG polarization dependence (see also Supplementary Section 9).

Most strikingly, in the absence of external magnetic fields, we can electrically control the magnetic state of the micro-device with dramatically reduced currents. As shown in Fig.~\ref{Fig3}g, we find that we can change the magnetic state in the FIB device with currents as low as \SI{3}{mA} corresponding to current densities of around \SI{5}{kA/cm^2} in the slab (in this case, we do not see remanent switching possibly due to shape anisotropy and residual strain in the sample \cite{Dubowik1996,Wang2013,Xu2021}). The magnetic switching is clearly evidenced by a change of the SHG polarization dependence indicative of switching between the $\mathbf{M}_\mathrm{-x}$ state at \SI{-3}{mA} and a $\mathbf{M}_\mathrm{+x}$ state at \SI{+3}{mA} \red{(Supplementary Section 9). Current-induced Joule heating is negligble at such low currents (Supplementary Section 9).} Such a reversible electrical control of a magnetic state may be highly desirable for novel device concepts.

\subsection{Observation of a broadband NODE}

Note that so far we observed a pronounced NODE for a specific SHG component at \SI{715}{meV} without deliberate optimization with respect to polarization or photon energy. However, a comparison of Figs.~\ref{Fig2}a and b reveals that changing the magnetization induces large changes of SHG intensity for certain polarization components and vanishing changes for others. To quantify the NODE, we define the \red{directional} contrast $\eta$ as 

\begin{equation}
    \eta = \frac{ I^{\ra}_\mathrm{+y}-I^{\la}_\mathrm{+y}}{I^{\ra}_\mathrm{+y}+I^{\la}_\mathrm{+y}}.
\end{equation}

The \red{directional} contrast $\eta = 0$ corresponds to absence of the NODE (i.e., the SHG intensity along forward and backward directions is equal); the \red{directional} contrast $\eta = \pm 1$ corresponds to extreme NODE (i.e., SHG can only occur along one direction).

In Fig.~\ref{Fig4}a, we show the evolution of $\left\vert\eta\right\vert$ for p-polarized SHG as a function of both incoming laser polarization and SHG photon energy. Here, we compare $I^{\ra}_\mathrm{+y}$ and $I^{\ra}_\mathrm{-y}$, which is equivalent to comparing $I^{\ra}_\mathrm{+y}$ and $I^{\la}_\mathrm{+y}$ but experimentally more straightforwardly accessible (see also Supplementary Section 4.C for other assessments of $\eta$).  The figure is characterized by a rich behavior that is mirror symmetric along the 0 and 90 deg directions due to mirror symmetries in CeAlSi. At \SI{715}{meV} we find that $\eta$ exhibits maxima at 0 deg/180 deg (p-polarized light) as well as near 90 deg (s-polarized light) but vanishes near 45/135 deg, in agreement with Fig.~\ref{Fig2}a and b. In Fig.~\ref{Fig4}b, we show the maximum achievable contrast for each SHG photon energy. The \red{directional} contrast peaks around \SI{750}{meV} at $97.2\,\%$ corresponding to a more than 70-fold change in SHG intensity. Moreover, we find that a \red{directional} contrast of at least $73\,\%$ (corresponding to a more than 6-fold change) is achievable for all photon energies in the considered spectral range between \SI{650}{meV} and \SI{900}{meV}.

\subsection{Broadband NODE from first principles}

In order to understand the microscopic origin of the NODE, we directly compute the NODE from the electronic structure of CeAlSi. Specifically, the NODE is the difference in SHG intensity between opposite propagation directions, which is determined by the two momentum-space resolved quantities $\kappa^\ra$ and $\kappa^\la$ as (see methods, Eq.~(\ref{eqn:SHG})):

\begin{equation}
	\mathrm{NODE} = I^\ra - I^\la = \vert \int_{BZ}\kappa^\ra\, d^3\mathbf{k}\vert^2 - \vert \int_{BZ}\kappa^\la\, d^3\mathbf{k}\vert^2.
	\label{eqn:SHGDFT}
\end{equation}

Here, $\kappa^{\ra(\la)} = \sum_{i,j,k}\xi_{ijk}E_j^{\ra(\la)}(\omega)E_k^{\ra(\la)}(\omega)A_i^{\ra(\la)}$,
where $\mathbf{E}^{\ra(\la)}$ is the electric field of the incident light for the forward and backward propagation direction, $\mathbf{A}^{\ra(\la)}$ is a projector to select either the $s$ or $p$ polarized SHG response, and $\xi_{ijk}$ is the \textbf{k}-space resolved contribution to the nonlinear optical susceptibility ($\xi_{ijk}$ is directly computed from the band structure of CeAlSi, see methods Eqs.~(\ref{eqn:supp1}) and (\ref{eqn:supp2})). 

\begin{figure}
	\includegraphics[width = 110 mm]{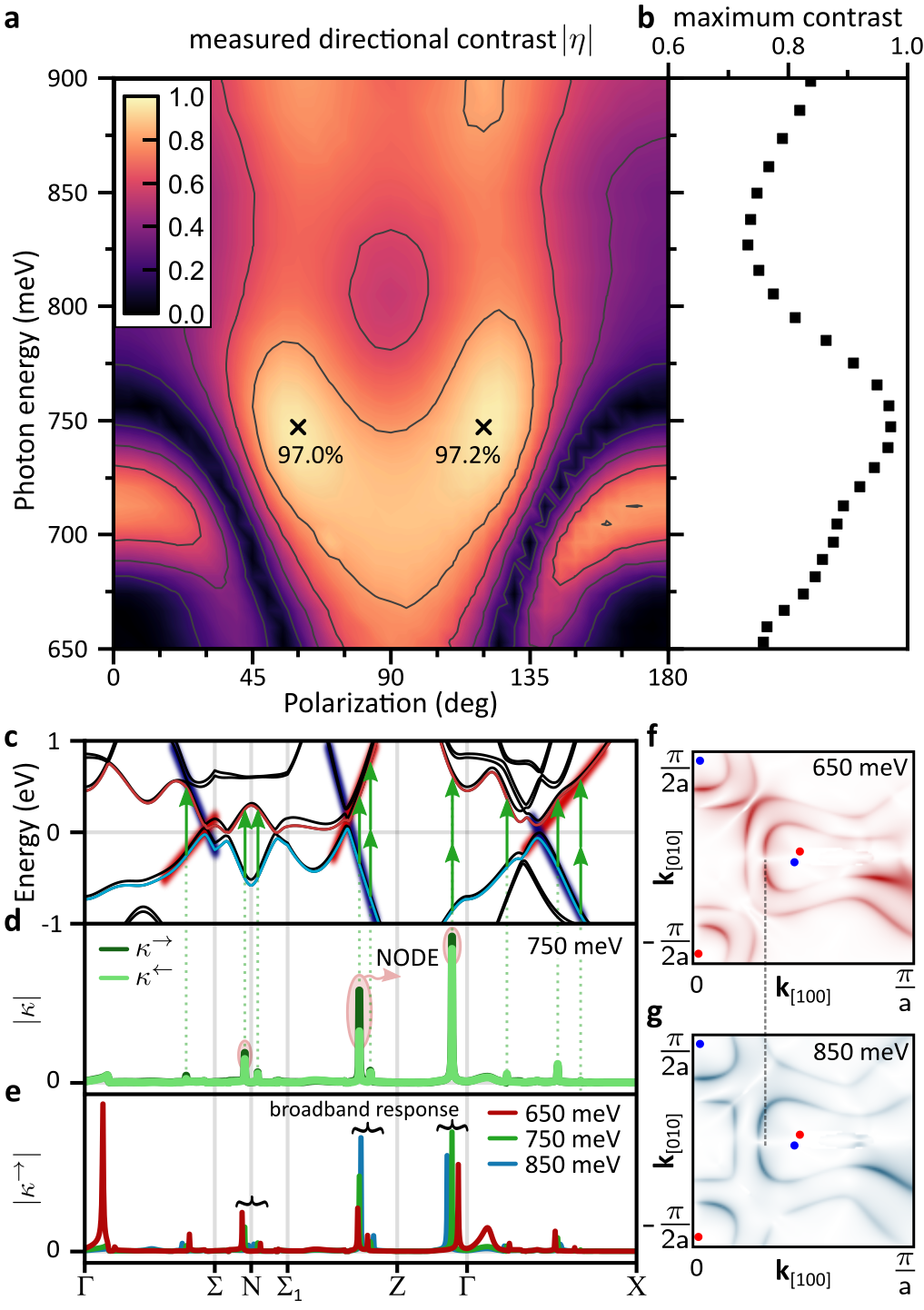}
	\caption{\textbf{NODE spectroscopy and its microscopic origin from the linearly dispersive bands.} \textbf{a} Measured absolute value of the \red{directional} contrast $\vert\eta\vert$ between $\mathbf{M}_\mathrm{\pm y}$ states for p-polarized SHG in forward direction as a function of incident photon energy and polarization. \textbf{b} Maximum achievable \red{directional} contrast by varying incident polarization as function of photon energy. A contrast of at least $73\%$ corresponding to a six-fold change in SHG intensity is achievable over the whole considered spectral range. \textbf{c} Band structure of CeAlSi calculated from first principles in the $\mathbf{M}_\mathrm{+y}$ state. \textbf{d} Momentum-space resolved contributions $\kappa^\ra$ and $\kappa^\la$ at \SI{750}{meV} for forward and backward propagating beams. The difference between $\kappa^\ra$ and $\kappa^\la$ leads to the NODE (see text). For clarity, we only show here contributions arising from the valence band marked in blue and the conduction band marked in red. \textbf{e} Momentum-space resolved contributions to the SHG intensity in forward direction at incident photon energies $\hbar\omega$ of \SI{650}{meV}, \SI{750}{meV}, and \SI{850}{meV}. Most contributions arise at $\mathbf{k}$ points where electronic bands are linearly dispersive thus giving rise to a broadband response. \textbf{f, g} Distribution of $\vert\xi_\mathrm{xxy}\vert$ (in arbitrary units) in a section of the Brillouin zone plane $k_\mathrm{z} = 0.295 \AA^{-1}$ for $\hbar\omega = \SI{650}{meV}$ and \SI{850}{meV}, respectively. Red and blue dots indicate the projection of the $W_2$ Weyl nodes of different chirality onto the considered plane. As the photon energy increases, contributions to $\vert\xi_\mathrm{xxy}\vert$ occur at larger distance from the Weyl nodes -- a direct consequence of linearly dispersive bands. Dashed line is a guide to the eye highlighting the expansion of the contribution line with increased photon energy.}
	\label{Fig4}
\end{figure}

Figure~\ref{Fig4}c shows a section of the band structure of CeAlSi consistent with previous first-principles calculations \cite{Yang2021, Sakhya2023} (see methods for further details). Along with the band structure, we show in Fig.~\ref{Fig4}d the magnitude of $\kappa^\ra$ and $\kappa^\la$ in the band structure of CeAlSi (for $\hbar\omega = \SI{750}{meV}$). We see from Fig.~\ref{Fig4}d that discrete $\mathbf{k}$ points contribute significantly. By comparing Figs.~\ref{Fig4}c and d, it is clear that these are the states which allow for optical transitions at $\omega$ or $2\omega$ across the Fermi energy. Importantly, the magnitude of $\kappa^\ra$ and $\kappa^\la$ differs significantly, giving rise to the observed NODE.

To understand the band structure origin of the broadband nature of the NODE in CeAlSi, we vary the incident photon energy from \SI{650}{meV} to \SI{850}{meV}. We can observe two qualitatively different behaviors (Fig.~\ref{Fig4}e). On the one hand, we notice a few contributions such as near the $\Gamma$ point along the $\Gamma-\Sigma$ line that occur only at specific photon energies (here \SI{650}{meV}). The bands at that $\mathbf{k}$ point are relatively flat. As the photon energy changes, such contributions disappear, which is indicative of a typical electronic resonance. Despite the apparent size of this resonant contribution, the spectrum of all SHG tensor components evolves smoothly towards \SI{650}{meV} indicating an overall negligible effect (Supplementary Section 10). On the other hand, the majority of the SHG response arises at $\mathbf{k}$ points where the electronic bands are linearly dispersive. Accordingly, as we vary the incident photon energy, those contributions only shift slightly to different $\mathbf{k}$ points but remain comparable in magnitude. This observation of SHG originating from linearly dispersive bands is qualitatively different from the conventional resonance picture. 

\red{We further elucidate the $k$-space origin of the broadband SHG response in Fig.~\ref{Fig4}f and g. We consider as an example the distribution of $\vert\xi_\mathrm{xxy}\vert$ in the plane $k_\mathrm{z} = \SI{0.295}{\AA^{-1}}$. In the paramagnetic phase of CeAlSi, this plane contains the $W_2$ Weyl nodes. In the ferromagnetic phase, the Weyl nodes shift out of the plane. Red and blue dots therefore indicate the projection of the Weyl nodes onto the plane  $k_\mathrm{z} = \SI{0.295}{\AA^{-1}}$. The pair of Weyl nodes in Fig.~\ref{Fig4}f are enveloped by two lines of strong contributions to $\vert\xi_\mathrm{xxy}\vert$. In analogy to Fig.~\ref{Fig4}d, we identify the inner and outer lines as transitions that are resonant at $\hbar\omega$ and $2\hbar\omega$, respectively. As we increase the incident photon energy from \SI{650}{meV} (Fig.~\ref{Fig4}f) to \SI{850}{meV} (Fig.~\ref{Fig4}g), we observe that the lines of strong contributions occur at an increased distance from the Weyl nodes (as indicated by the dashed line). This observation is a direct consequence of the linearly dispersive bands that disperse over hundreds of meV in CeAlSi. We thus see that the broadband NODE in CeAlSi ultimately arises from the linearly dispersive bands that naturally occur in the vicinity of the Weyl nodes (see also Supplementary Section 10 for an extended discussion).}

\section{Discussion}

In summary, we explored the nonlinear optical response of the magnetic Weyl semimetal CeAlSi using SHG spectroscopy. In addition to the crystallographic (time-reversal symmetric) SHG, we observe magnetic (time-reversal asymmetric) SHG, which in a topologically nontrivial material. By physically reversing the light path, we revealed a pronounced NODE. 

\red{We introduce the NODE as a novel concept that utilizes a directionally dependent SHG intensity. The effect is rooted in a directionally dependent interference that is conceptually different from the well-known magnetic SHG contrast \cite{Pavlov03,Fiebig05, Toyoda23} or nonreciprocal SHG \cite{Toyoda2021,Mund2021}. In particular, the NODE does not strictly require a broken time-reversal symmetry \calT\ nor the simultaneous presence of electric-dipole SHG and magnetic-dipole/electric-quadrupole SHG (which requires a broken spatial inversion symmetry \calP). Instead, the NODE is microscopically enabled by a broken mirror symmetry (Supplementary Section 1D).}

\red{Remarkably, we found a sizable \red{directional} contrast of at least $73\%$ over a spectral range exceeding \SI{250}{meV} --- two orders of magnitude wider than previous reports based on nonreciprocal SHG. We note that the directional contrast can in principle be further optimized by varying external parameters like the sample temperature or the angle of incidence.}

Microscopically, our DFT calculations show that the broadband NODE is directly related to the electronic band structure. The SHG response emerges from linearly dispersive bands that naturally occur near the Fermi energy in a Weyl semimetal. The linear dispersion allows for strong optical responses over a wide range of frequencies.

Our observations open up a number of intriguing possibilities. First, it suggests the NODE as a powerful method to measure the phase of optical properties. All optical processes (e.g., SHG, Raman) are governed by the corresponding susceptibility tensor, which is determined by the material. Hence, inversely, by measuring the SHG, Raman, or other optical process, we can learn about the symmetry and electronic properties of a material. However, typically we only access the magnitude of the susceptibility, whereas the sign or phase is hard to probe. Here, as we showed above, the NODE arises from the directionally dependent mixing between tensor components which provides insights into their relative phases. It would be interesting to test the optical diode effect for other optical processes such as high-harmonic generation \cite{Heide2022}, Raman scattering \cite{Wang2022a}, and optically generated spin currents \cite{Li2020}, where interesting phenomena hinting nonreciprocity have been observed. \red{Second, in addition to searching for new diode effects based on other nonlinear optical processes, our results open up for a search for other materials exhibiting diode effects. So far, all observation of directional SHG responses were demonstrated in noncentrosymmetric, time-reversal-broken materials, although this should not strictly be necessary. It will be interesting to demonstrate the NODE in a time-reversal symmetric (e.g., antiferromagnetic) or inversion-symmetric systems. Third}, we note the low current density associated with the switching of magnetic order in CeAlSi. It would be of interest to perform future studies to understand its microscopic mechanism and to test it on other magnetic and noncentrosymmetric Weyl fermion systems. The low-current switching in combination with the broadband character of the NODE, may enable fundamentally new device concepts for photonic circuits. From a fundamental point of view, our observation of a broadband NODE is testament to the exotic electromagnetic responses that can be discovered in novel quantum magnets. 

\section{Methods}

\subsection{Crystal growth}
Single crystals of CeAlSi were grown by a self-flux method. The starting materials were Ce and Al ingots, and Si pieces, mixed in ratio Ce:Al:Si = 1:10:1 in an alumina crucible. The crucible was sealed in an evacuated quartz tube, and went through the following heating sequence: the sample was heated from \SI{25}{\degreeCelsius} to \SI{1000}{\degreeCelsius} at \SI{3}{\degreeCelsius \per \minute}, stayed for \SI{12}{h}, cooled to \SI{700}{\degreeCelsius} at \SI{0.1}{\degreeCelsius \per \minute}, stayed for \SI{12}{h}, and finally centrifuged to remove the residual Al flux at \SI{700}{\degreeCelsius}.

\subsection{SHG measurements}

\red{All measurements were performed at \SI{3}{K} unless explicitly stated differently. In particular, all measurements shown in the main text were performed at \SI{3}{K}.}

Second-harmonic generation (SHG) is a nonlinear optical process that describes the interaction of two photons at frequency $\omega$ in a material leading to the re-emission of one photon at frequency $2\omega$. In the lowest-order electric-dipole approximation, the process can be formally expressed as

\begin{equation}
	P_i(2\omega) = \sum_{j,k}\mathcal{X}_{ijk}E_j(\omega)E_k(\omega),
\end{equation}

where $E_{j}(\omega)$ and $E_{k}(\omega)$ denote the electric-field components of the incident light-wave and $P_i$ the components of the induced nonlinear polarization oscillating at $2\omega$ (indices $i,j,k$ can be $x$, $y$, or $z$). The process is mediated by the second-order susceptibility tensor $\mathcal{X}_{ijk}$, which can only have non-vanishing components when \calP\ is broken \cite{Birss66}. 

All SHG measurements were obtained using a laser system by LightConversion consisting of an amplified femtosecond laser (Pharos, photon energy \SI{1.2}{eV}, maximum power \SI{10}{W}, maximum repetition rate \SI{100}{kHz}) in combination with an optical parametric amplifier (Orpheus One HE). In order to avoid heating effects, we reduce the repetition rate of the laser and employ attenuating filters to achieve an average power on the sample of less than \SI{2}{mW}. For domain imaging (e.g., Fig~\ref{Fig2}i), we chose a spot size of approximately \SI{800}{\micro m}. For all other measurements, we focused the laser to a spot size of approximately \SI{150}{\micro m}. All SHG measurements were performed in reflection under an angle of incidence of $45^\circ$. We adjust the laser polarization of the incident fundamental beam using an achromatic half-waveplate, while selectively detecting s- and p-polarized SHG intensity (red and gray data points, respectively) using a Glan-laser polarizer and a thermo-electrically cooled electron-multiplying CCD camera (EMCCD). For spectrally resolved SHG measurements, a grating spectrometer was mounted in front of the EMCCD camera. For all other measurements, the EMCCD camera was mounted behind a narrow band pass filter centered at \SI{1.425}{eV} (bandwidth \SI{30}{meV}). We used additional filters directly before the sample to block parasitic SHG signals from all previous optical components as well as directly after the sample to block the fundamental beam.

The measured SHG intensity is proportional to 

\begin{equation}
	I(2\omega)	= \vert \mathbf{P}(2\omega)\cdot\mathbf{A}\vert^2 = \vert \sum_{i,j,k}\mathcal{X}_{ijk}E_j(\omega)E_k(\omega)A_i\vert^2,
	\label{eqn:SHGSupp}
\end{equation}

where $\mathbf{A}$ is the direction of the transmitted polarization axis of the analyzer (here $s$ or $p$ polarization).

\subsection{Focused-ion-beam microstructuring}

An oriented large single crystal of CeAlSi was milled using a Xe-Plasma FIB (Thermofisher Helios PFIB). At 2.5mA, 30kV, first a rectangular bar was milled from the parent crystal (bar length 300mm, width 50mm, height 30mm). This bar was attached with Pt deposition to an in-situ micromanipulator and rotated by 90°, to access the side face of the bar. At this angle, the trapezoidal outer shape and the central bridge were patterned at 500nA, 30kV for coarse and 60nA, 30kV for fine patterning. Furthermore, the outer feet were polished flat at 60nA, 30kV to ensure a flat mating with the substrate. A sapphire substrate (2x2mm) with two large, lithographically prepared Au leads (10nm Ti / 100nm Au) was introduced into the chamber and the bridge rotated back. Using the in-situ manipulator, the bridge was placed ontop of these Au pads and connected by Pt deposition (60nA, 12kV), on the right foot in main Fig.~\ref{Fig3}f. The tip of the manipulator was cut off (60nA, 30kV), and the remains of the manipulator are well visible on the front section of the right foot. Despite best efforts of alignment, an approximately 500nm gap appeared between the left foot and the Au pad. The micromanipulator was used to gently push the structure flat, yet without forming a solid bond via deposition. Using the same settings as on the right side, the left side was connected with Pt. The sizable depositions resulted in visible overspray, which at these channel lengths does not notably conduct. Out of caution, the overspray was removed in a rectangular channel on both sides of the bridge (60nA, 30kV). The FIB process at 30kV usually results in a ~20nm thick amorphous layer which may cause issues with the SHG experiments. This layer was strongly reduced using low-voltage polishing. At an angle of 52° between the beam and the surface normal, the central top of the bridge was irradiated at 60nA, 5kV in a final cleaning step. As the entire process is performed using Xe beams, no implantation of the primary ions is expected.

\subsection{First-principles calculations}

We performed density-functional theory calculations \cite{Hohenberg1964} using the \textit{Vienna Ab initio Simulation Package} (VASP) \cite{Kresse1996} with the Perdew-Berke-Ernzerhof's (PBE) pseudopotential \cite{Perdew1996} in the generalized-gradient approximation \cite{Langreth1983}. We adopted $U_\mathrm{eff} = \SI{6}{eV}$ for the Hubbard U-term acting on the Ce $f$-orbital electrons \cite{Yang2021}. The Wannier tight-binding Hamiltonians were established using the $Wannier90$ package \cite{Mostofi2014}. The Wannier bands were symmetrized to restore the symmetry restrictions of the CeAlSi space group \cite{Gresch2018}. The magnetic order of CeAlSi is described by two magnetic sublattices of equal magnitude with noncollinear magnetic moments $\mathbf{m}_\mathrm{1}$ and $\mathbf{m}_\mathrm{2}$ on Ce $4f$ orbitals such that $(\mathbf{m}_\mathrm{1}+\mathbf{m}_\mathrm{2})\parallel \left[110\right]$. The angle between $\mathbf{m}_\mathrm{1}$ and $\mathbf{m}_\mathrm{2}$ is $70^\circ$ as indicated in Ref. \cite{Yang2021}.

\subsection{Calculation of $\chi_{ijk}$}
We calculated the nonlinear optical susceptibility $\chi_{ijk}$ for electric-dipole SHG according to the diagrammatic approach to the nonlinear optical response \cite{Parker2019, Takasan2021}. Interband transitions cause two contributions $\xi^{I}$ and $\xi^{II}$ such that $\chi_{ijk} = \int_{BZ}d^3\mathbf{k} \left(\xi^I_{ijk}+\xi^{II}_{ijk}\right)$, where

\begin{eqnarray}
	\xi^I_{ijk} &=& C \sum_{m\neq n} f_{mn}\left(\frac{h^{ij}_{nm}h^k_{mn} + h^{ik}_{nm}h^j_{mn}}{\omega+i\eta - \omega_{mn}} + \frac{h^{jk}_{mn}h^i_{nm}}{2\omega+i\eta - \omega_{mn}}\right) \label{eqn:supp1}, \\
	\xi^{II}_{ijk} &=& C \sum_{m\neq n\neq p} \frac{h^i_{pm}\left(h^j_{mn}h^k_{np}+h^k_{mn}h^j_{np}\right)}{\omega_{mn}+\omega_{np}}\left(\frac{f_{np}}{\omega + i\eta -\omega_{pn}} + \frac{f_{nm}}{\omega + i\eta -\omega_{nm}} + \frac{2f_{mp}}{\omega + i\eta -\omega_{pm}}\right)\label{eqn:supp2}
\end{eqnarray}  

Here, $\left\{i,j,k\right\}$ run over $\left\{x,y,z\right\}$ directions, $C = \frac{e^3}{2i\epsilon_0\hbar^2\omega^3}$, $\left\{m,n,p\right\}$ are the band indices, $h^i_{mn} = \frac{1}{\hbar}\langle m \vert \partial_{k_i} h(\mathbf{k})\vert n\rangle$, $h^{ij}_{mn} = \frac{1}{\hbar}\langle m \vert \partial_{k_i}\partial_{k_j} h(\mathbf{k}) \vert n\rangle$,  $\omega_{mn} = \frac{1}{\hbar}\left(\epsilon_m-\epsilon_n\right)$ represents the transition frequency between bands $m$ and $n$, $f_{mn}$ is the difference in band occupation according to the Fermi-Dirac distribution, and $\epsilon_0$ is the vacuum permittivity. The derivatives $h^i_{mn}$ and $h^{ij}_{mn}$ can be rewritten in terms of the generalized Berry connection $\mathcal{A}^{i}_{mn} = i\langle m\vert \partial_{k_i} \vert n\rangle$ \cite{Morimoto2016}:

\begin{equation}
	h^i_{mn} = \frac{1}{\hbar}\langle m \vert \partial_{k_i} h(\mathbf{k})\vert n\rangle = \partial_{k_i}\langle m \vert h(\mathbf{k})\vert n\rangle - \langle \partial_{k_i} m \vert  h(\mathbf{k})\vert n\rangle - \langle m \vert  h(\mathbf{k})\vert \partial_{k_i} n\rangle = \delta_{mn} \partial_{k_i} \epsilon_m + i\hbar\omega_{mn}\mathcal{A}^{i}_{mn}.
\end{equation}

As the summation in Eqs. (\ref{eqn:supp1}) and (\ref{eqn:supp2}) only considers interband transitions ($m\neq n$), we find $h^i_{mn} =  i\hbar\omega_{mn}\mathcal{A}^{i}_{mn}$. Similarly, we find $h^{ij}_{mn} = i\hbar\mathcal{A}^{i}_{mn}\partial_{k_j}\omega_{mn}$ for the relevant case of $m\neq n$. Therefore, all interband contributions to $\chi_{ijk}$ can be related to the generalized Berry connection $\mathcal{A}^{i}_{mn}$.

Using $\chi_{ijk} = \int_{BZ}d^3\mathbf{k} \xi_{ijk}$ and Eq.~(\ref{eqn:SHGSupp}), we can express the SHG intensity as 

\begin{equation}
	I(2\omega)	= \vert \int_{BZ}\sum_{i,j,k}\xi_{ijk}E_j(\omega)E_k(\omega)A_i d^3\mathbf{k}\vert^2 = \vert \int_{BZ}\kappa d^3\mathbf{k}\vert^2,
	\label{eqn:SHG}
\end{equation}

where we define $\kappa = \sum_{i,j,k}\xi_{ijk}E_j(\omega)E_k(\omega)A_i$.

\section{Data Availability}

Source data are provided in the Source Data file. All data are available upon request from the corresponding authors.

\section{References}
\newpage
\renewcommand{\baselinestretch}{1.2}
\bibliographystyle{Science}

\providecommand{\noopsort}[1]{}\providecommand{\singleletter}[1]{#1}%

\clearpage

\renewcommand{\baselinestretch}{1.8}

\section{Acknowledgment}
Work in the SYX group was partly supported through the Center for the Advancement of Topological Semimetals (CATS), an Energy Frontier Research Center (EFRC) funded by the U.S. Department of Energy (DOE) Office of Science (fabrication and measurements), through the Ames National Laboratory under contract DE-AC0207CH11358, and partly by the AFOSR grant FA9550-23-1-0040 (data analysis), and partly by the NSF Career DMR-2143177 (manuscript writing). SYX also acknowledges the Corning Fund for Faculty Development. The work in the QM group was supported through the CATS, an EFRC funded by the DOE Office of Science (manuscript writing), through the Ames National Laboratory under contract DE-AC0207CH11358. QM also acknowledges the support from NSF through a CAREER award DMR-2143426 (material supplies) and the CIFAR Azrieli Global Scholars Program. C.T. acknowledges support from the Swiss National Science Foundation under project no. P2EZP2\_191801 and from the Harvard University Climate Change Solutions Fund. J.N. acknowledges support from the Swiss National Science Foundation under project no. P2EZP2\_195686. F.H. received funding by the National Natural Science Foundation of China under grant 52103353. P.J.W.M. acknowledges funding by the European Research Council (ERC) under the European Union’s Horizon 2020 research and innovation programme (MiTopMat, grant agreement no. 715730). C.G. received funding by the Swiss National Science Foundation (grant no. PP00P2\_176789). This material is based upon work supported by the Air Force Office of Scientific Research under award number FA2386-21-1-4059. K.T.L. acknowledges the support of HKRGC through Grants RFS2021-6S03, C6025-19G, AoE/P-701/20, 16310520, 16310219 and 16307622.

\section{Author Contributions}

C.T. designed and conducted the SHG experiments with assistance from J.-X.Q., H.-C.L., Y.-F.L., A.G., D.B., T.D., and S.-C.H. C.T. evaluated the data with support from J.N. X.-J.G., C.-P.Z., Y.-M.X., and K.T.L. performed the first-principles calculations. H.-Y.Y., Y.F., F.H., and F.T. synthesized CeAlSi single crystals. C.G. and P.J.W.M. prepared the FIB cut. C.T., Q.M. and S.-Y.X. wrote the manuscript with discussions and contributions from all authors.

\section{Competing Interests}

The authors declare no competing interests.

\clearpage

\end{document}


\begin{center}
		\vspace{6cm}\vskip6cm
		\huge{\textbf{Supplementary Information\\}}
		\vspace{3cm}
		\Large{\textbf{Nonlinear optical diode effect in a magnetic Weyl semimetal}}
		\vspace{2cm}
		\flushleft{C. Tzschaschel \textit{et al.}}
	\end{center}
	\tableofcontents
	\clearpage
	
	\section{Materials and Methods}
	
	\subsection{Symmetry considerations in {C\lowercase{e}A\lowercase{l}S\lowercase{i}}}
	
\begin{figure}
		\centering
		\includegraphics[width = 0.7\columnwidth]{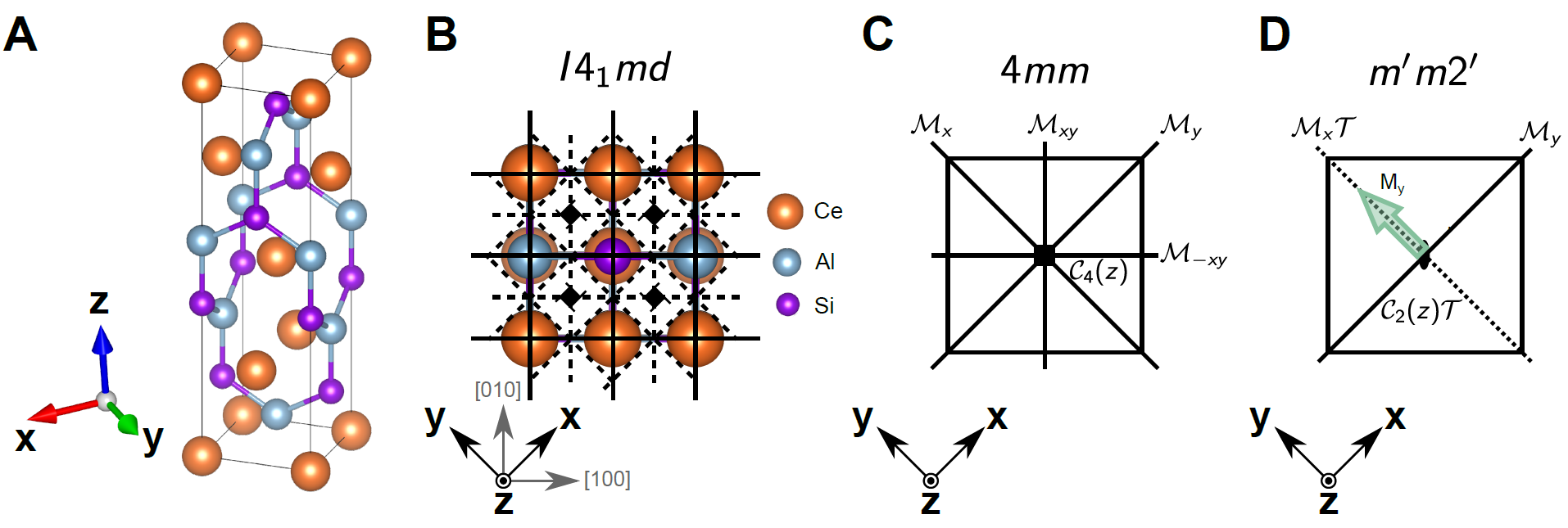}
		\caption{\textbf{Lattice structure and symmetry of CeAlSi.} \textbf{a,} 3D representation of the paramagnetic unit cell of CeAlSi (point group $4mm$, space group \#109: $I4_1md$) \textbf{b,} Projection of the the unit cell along $\hat{\mathbf{z}}$ axis. Solid lines, dashed lines and diamonds indicate the positions of mirror planes, glide-mirror planes and screw rotational axis, respectively. Here, we use an $xyz$ coordinate system that is $45^\circ$ rotated with respect to the conventional tetragonal unit cell (grey). \textbf{c,} Illustration of point group symmetries of the paramagnetic point group $4mm$ projected along $\hat{\mathbf{z}}$ axis.  \textbf{d,} Illustration of point group symmetries of the ferromagnetic point group $m^\prime m 2^\prime$ projected along $\hat{\mathbf{z}}$ axis. The magnetization $M_y$ along the $\hat{\mathbf{y}}$ axis reduces the symmetry from $4mm$ to $m^\prime m 2^\prime$.}
		\label{fig:CeAlSiStructure}
	\end{figure}

	Throughout the Supplementary Material, all Miller indices $[hkl]$ are given with respect to the conventional tetragonal unit cell of the paramagnetic phase (Fig.~\ref{fig:CeAlSiStructure}). The directions $\hat{\mathbf{x}}$, $\hat{\mathbf{y}}$, $\hat{\mathbf{z}}$ are defined as $[110]=+\hat{\mathbf{x}}$, $[\bar{1}10]=+\hat{\mathbf{y}}$, and $[001]=+\hat{\mathbf{z}}$.
	
	In this section, we describe the symmetry of the paramagnetic and ferromagnetic phases of CeAlSi. The symmetry dictates the nonzero components of the SHG tensor $\chi_{ijk}$ as defined by
	
	\begin{equation}
		P_i(2\omega) = \chi_{ijk} E_j(\omega)E_k(\omega).
		\label{eqn:SHGP}
	\end{equation}
	
	The paramagnetic phase of CeAlSi crystallizes in the same tetragonal lattice as TaAs, with the space group \#109 ($I4_1md$) and the point group $4mm$ (Fig.~\ref{fig:CeAlSiStructure}A). Just like in the case of TaAs, all atoms occupy sites belonging to a 4a-Wyckoff orbit. Table~\ref{Paramag_symm} enumerates the symmetry operations of the paramagnetic phase of CeAlSi.
	
	\begin{center}
		\begin{table}[ht]
			\begin{tabular}{p{3.5cm}|p{1.5cm}|p{1.5cm}|p{1.5cm}|p{1.5cm}|p{1.5cm}|p{1.5cm}|p{1.5cm}}
				Paramagnetic CeAlSi & $\mathcal{E}$ & $\mathcal{C}_4(z)$ & $\mathcal{C}_2(z)$ & $\mathcal{M}_x$ & $\mathcal{M}_y$ & $\mathcal{M}_{xy}$ & $\mathcal{M}_{-xy}$\\
				\hline
				& $\mathcal{T}$ & $\mathcal{C}_4(z)\mathcal{T}$ & $\mathcal{C}_2(z)\mathcal{T}$ & $\mathcal{M}_x\mathcal{T}$ & $\mathcal{M}_y\mathcal{T}$ & $\mathcal{M}_{xy}\mathcal{T}$ & $\mathcal{M}_{-xy}\mathcal{T}$\\
			\end{tabular}
			\caption{\label{Paramag_symm} \textbf{Point group symmetry of the paramagnetic phase of CeAlSi ($4mm$).}}
		\end{table}
	\end{center}
	
	$\mathcal{E}$ is identity, $\mathcal{T}$ is time-reversal symmetry, $\mathcal{C}_4(z)$ and $\mathcal{C}_2(z)$ are the $4$-fold rotation and $2$-fold rotation around the $\hat{\mathbf{z}}$ axis, and $\mathcal{M}_i$ represents the mirror reflection from $i$ to $-i$ ($i=x,y,xy,-xy$). Because the system has time-reversal symmetry $\mathcal{T}$, any spatial symmetry (mirror, rotation, etc) combined with $\mathcal{T}$ is also a good symmetry, as reflected in the second line above.
	
	\vspace{0.5cm}
	The ferromagnetic phase of CeAlSi exhibits a spontaneous magnetization along $\hat{\mathbf{x}}$ or $\hat{\mathbf{y}}$. The magnetization breaks not only time-reversal symmetry $\mathcal{T}$ but also a number of other symmetries. As a result, the point group symmetry is reduced to $m^\prime m 2^\prime$. Supplementary Table~\ref{ferromag_symm} shows the remaining magnetic symmetry operations in the ferromagnetic phase for a domain with $\vector{M}^\mathrm{Ce} \parallel \hat{\mathbf{y}}$ \cite{Birss66}.
	\begin{center}
		\begin{table}[h]
			\begin{tabular}{p{3.5cm}|p{1.5cm}|p{1.5cm}}
				Ferromagnetic CeAlSi & $\mathcal{E}$ & $\mathcal{M}_y$\\
				\hline
				& $\mathcal{C}_2(z)\mathcal{T}$ & $\mathcal{M}_x\mathcal{T}$
			\end{tabular}
			\caption{\label{ferromag_symm} \textbf{Point group symmetry of the ferromagnetic phase of CeAlSi ($m^\prime m 2^\prime$) with $\vector{M} \parallel \hat{\mathbf{y}}$ \cite{Birss66}.}}
		\end{table}
	\end{center}
	
	\subsection{Electrode patterning}\label{Sec:patterning}
\begin{figure}[ht]
		\centering
		\includegraphics[width = 119 mm]{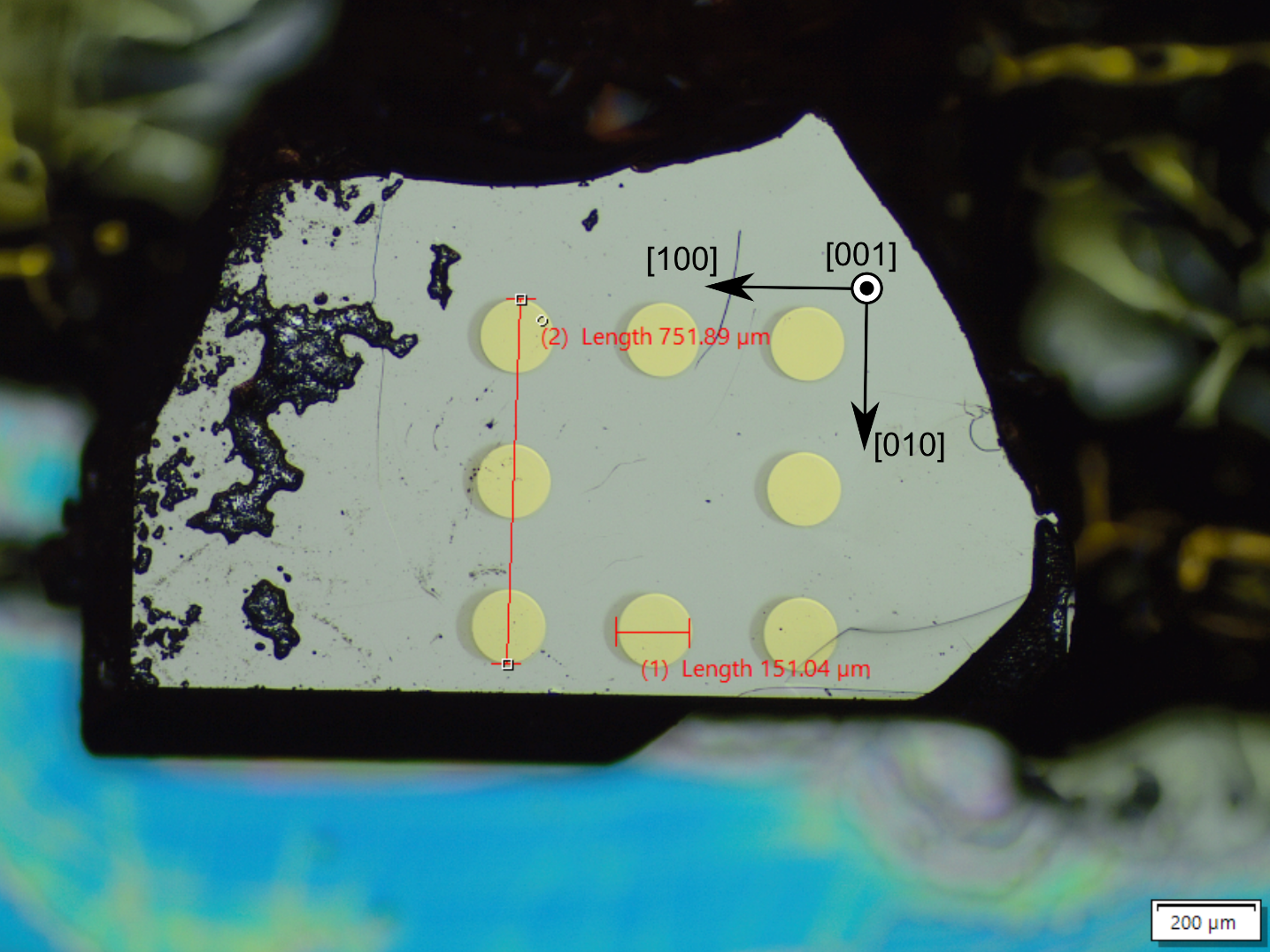}
		\caption{\textbf{Optical microscope image of CeAlSi device with electrode pattern.}}
		\label{fig:SampleImage}
	\end{figure}

	In order to apply current in the crystallographic $xy$ plane, we deposited electrical contacts on a naturally cleaved $(001)$ facet (see Fig.~\ref{fig:SampleImage}). A square pattern of eight gold electrodes was deposited at room temperature by e-beam evaporation through a shadow mask. Aligning the shadow mask with a naturally cleaved edge allows us to apply current along the crystallographic $\langle 110 \rangle$ or $\langle 100 \rangle$ direction, which corresponds to perpendicular/parallel or at an angle of $45^\circ$ relative to the magnetization direction in CeAlSi, respectively. Subsequently, the crystal was glued and wire bonded onto a chip carrier. Current was applied using a Keithley Model 2400 SourceMeter.
	
	\subsection{SHG tensor components of CeAlSi}
	
	The symmetry of the paramagnetic phase of CeAlSi has the point group $4mm$. A symmetry analysis similar to the previous section shows that the allowed components of the SHG tensor introduced in Eq.~(\ref{eqn:SHGP}) are \cite{Birss66}
	
	\begin{equation}
		\chi_{zzz} \qquad \chi_{zxx} = \chi_{zyy} \qquad \chi_{xxz} = \chi_{xzx} = \chi_{yyz} = \chi_{yzy}.
	\end{equation}
	
	All other tensor components vanish. Specifically, no electric-dipole SHG is allowed for light propagating along the $z$ direction.
	
	Below $T_\mathrm{C}$, the magnetic order reduces the symmetry to $Fd^\prime d 2^\prime$ (\#43.3.322, point group $m^\prime m 2^\prime$) \cite{Yang2021}, which gives rise to additional contributions to the second-order susceptibility. In particular, the allowed components below $T_\mathrm{C}$ are (for a magnetization along the $y$ direction):
	
	\begin{eqnarray}
		&\chi_{zzz}& \qquad \chi_{zxx} \qquad \chi_{zyy} \qquad \chi_{xxz} = \chi_{xzx} \qquad \chi_{yyz} = \chi_{yzy}\\
		&\chi_{xxx}& \qquad \chi_{xyy} \qquad \chi_{xzz} \qquad \chi_{zzx} = \chi_{zxz} \qquad \chi_{yyx} = \chi_{yxy}.
	\end{eqnarray}
	
	Note that the allowed $\chi$ tensor components for the four magnetic states of CeAlSi are related by symmetry as shown in Supplementary Table~\ref{Tab:SHGcomp}.
	\begin{center}
		\begin{table}
			\begin{tabular}{c|c c c c}
				\hline
				\hline
				& $\mathbf{M}_\mathrm{+y}$ & $\mathbf{M}_\mathrm{-y}$ & $\mathbf{M}_\mathrm{+x}$ & $\mathbf{M}_\mathrm{-x}$ \\
				\hline
				$\chi_{xxx}$ & $\chi_{xxx}$ & $-\chi_{xxx}$ & 0 & 0 \\
				$\chi_{xyy}$ & $\chi_{xyy}$ & $-\chi_{xyy}$ & 0 & 0 \\
				$\chi_{xzz}$ & $\chi_{xzz}$ & $-\chi_{xzz}$ & 0 & 0 \\
				$\chi_{xyz}$ & 0 & 0 & 0 & 0 \\
				$\chi_{xxz}$ & $\chi_{xxz}$ & $\chi_{xxz}$ & $\chi_{yyz}$ & $\chi_{yyz}$ \\
				$\chi_{xxy}$ & 0 & 0 & $-\chi_{yxy}$ & $\chi_{yxy}$ \\
				\hline
				$\chi_{yxx}$ & 0 & 0 & $-\chi_{xyy}$ & $\chi_{xyy}$ \\
				$\chi_{yyy}$ & 0 & 0 & $-\chi_{xxx}$ & $\chi_{xxx}$ \\
				$\chi_{yzz}$ & 0 & 0 & $-\chi_{xzz}$ & $\chi_{xzz}$ \\
				$\chi_{yyz}$ & $\chi_{yyz}$ & $\chi_{yyz}$ & $\chi_{xxz}$ & $\chi_{xxz}$ \\
				$\chi_{yxz}$ & 0 & 0 & 0 & 0 \\
				$\chi_{yxy}$ & $\chi_{yxy}$ & $-\chi_{yxy}$ & 0 & 0 \\
				\hline
				$\chi_{zxx}$ & $\chi_{zxx}$ & $\chi_{zxx}$ & $\chi_{zyy}$ & $\chi_{zyy}$ \\
				$\chi_{zyy}$ & $\chi_{zyy}$ & $\chi_{zyy}$ & $\chi_{zxx}$ & $\chi_{zxx}$ \\
				$\chi_{zzz}$ & $\chi_{zzz}$ & $\chi_{zzz}$ & $\chi_{zzz}$ & $\chi_{zzz}$ \\
				$\chi_{zyz}$ & 0 & 0 & $-\chi_{zxz}$ & $\chi_{zxz}$ \\
				$\chi_{zxz}$ & $\chi_{zxz}$ & $-\chi_{zxz}$ & 0 & 0 \\
				$\chi_{zxy}$ & 0 & 0 & 0 & 0\\
				\hline
			\end{tabular}
			\caption{Allowed SHG tensor components in the four magnetic states of CeAlSi and their symmetry relationship.}
			\label{Tab:SHGcomp}
		\end{table}
	\end{center}
	
	The polarization dependence of the measured SHG intensity can be simulated as
	
	\begin{equation}
		I_{2\omega} \propto \vert\vector{P}(2\omega)\cdot\vector{A}\vert^2,
	\end{equation}
	
	\noindent where $\vector{P}(2\omega)$ is the second-order polarization as described by Eq.~(\ref{eqn:SHGP}) and $\vector{A}$ is a unit-vector defining the transmission axis of the analyzer crystal for the SHG light. All tensor components are in general complex quantities, i.e. $\chi_{ijk} = \vert\chi_{ijk}\vert \exp\left(i\Phi_{ijk}\right)$. 
	
	\subsection{SHG interference phenomena}
	
	Interference between the various contributions to the nonlinear optical polarization $\vector{P}(2\omega)$ can give rise to several intriguing phenomena including the nonlinear optical diode effect described in the main text. 
	
\begin{figure}
		\includegraphics[width = 80 mm]{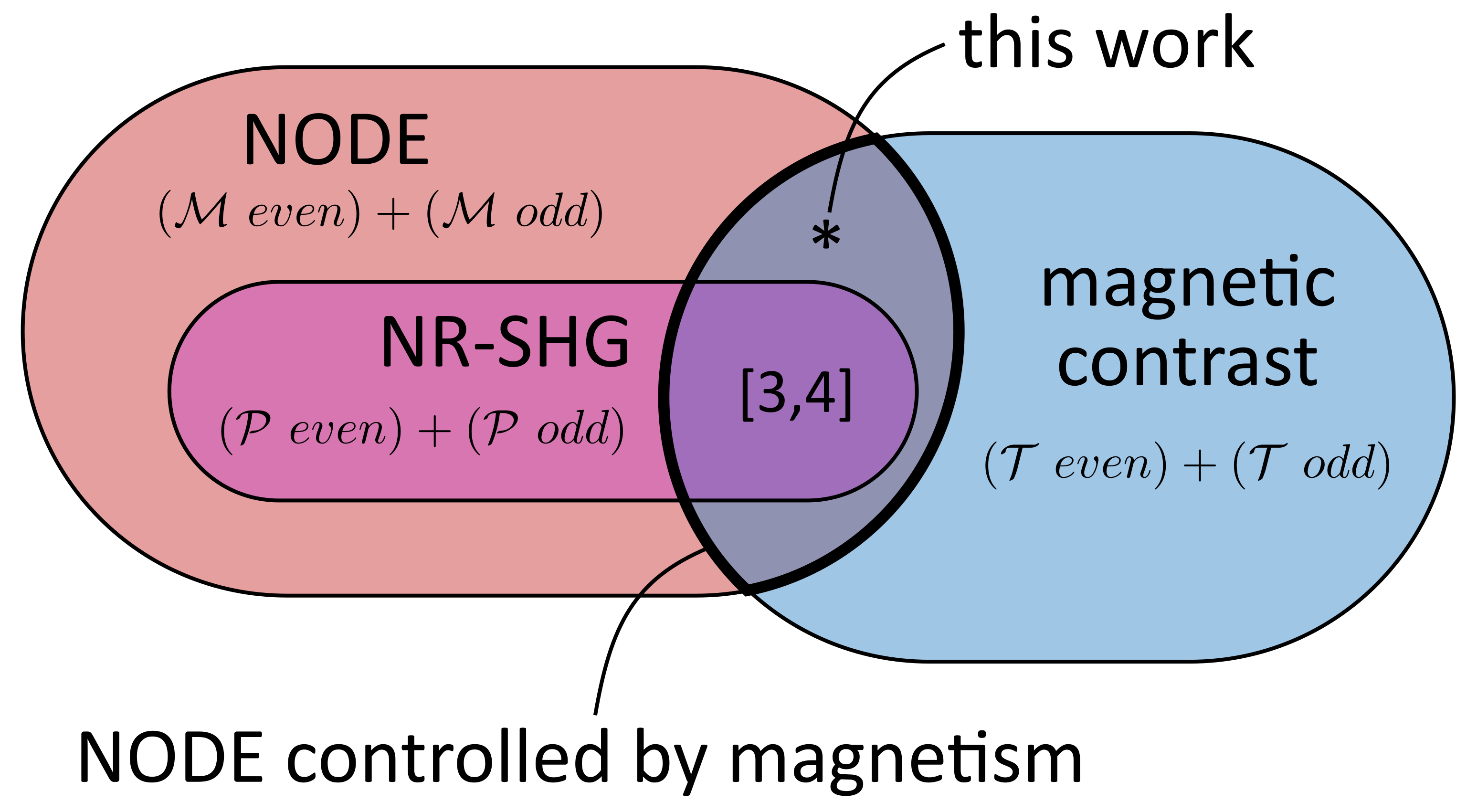}
		\caption{\textbf{SHG interference phenomena} Interference between different contributions to the total SHG response can give rise to a variety of optical phenomena. The phenomena can be distinguished based on the microscopic origin of the contributions. Magnetic contrast arises from a mixing of contributions that are even/odd under time reversal \calT; nonreciprocal SHG arises from mixing SHG contributions that are even/odd under spatial inversion \calP; the NODE emerges from interference between SHG contributions that are even/odd under a mirror operation $\mathcal{M}$. Previous demonstrations of NR-SHG are in the cross section of all three phenomena \cite{Toyoda2021, Mund2021}. As a broken inversion symmetry (required for \calP\ odd SHG) implies a broken mirror symmetry, any material that exhibits NR-SHG also allows for a NODE, i.e. NODE is a generalization of NR-SHG.}
		\label{fig:Venn}
	\end{figure}

	In the following we distinguish in particular between three separate, but sometimes overlapping, effects: magnetic domain contrast, nonreciprocal SHG (NR-SHG), and the nonlinear optical diode effect (NODE). We illustrate their relative relationship in a Venn diagram in Fig.~\ref{fig:Venn}
	
\begin{figure}
		\includegraphics[width = 110 mm]{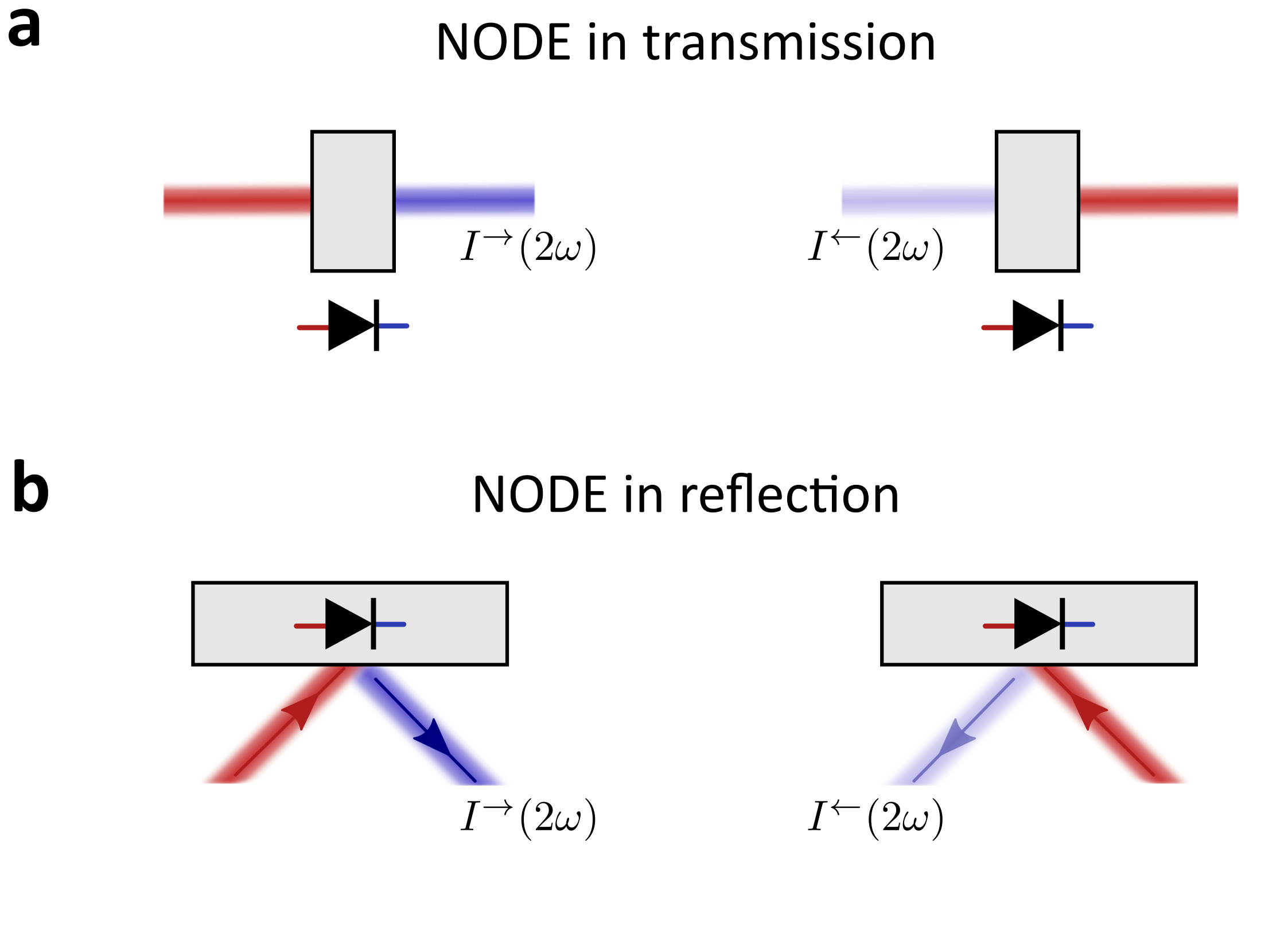}
		\caption{\textbf{Definition of the NODE.} We define the \textbf{nonlinear optical diode effect (NODE)} purely phenomenologically as an effect whereby a nonlinear optical response (such as SHG), is stronger for one light propagation direction than for the reversed light propagation direction. Such an effect can be defined both in \textbf{a,} transmission and in \textbf{b,} reflection geometry. In both cases, we can define a directional contrast $\eta$.}
		\label{ExtDefinition}
	\end{figure}

	\begin{enumerate}
		\item \textbf{Magnetic domain contrast} describes a difference in SHG intensity between different magnetic domains. A magnetic domain contrast can be observed within the electric-dipole approximation. This is due to interference between i-type (time reversal \calT\ even) and c-type (\calT\ odd) SHG contributions. Below the magnetic ordering temperature $T_\mathrm{C}$, the nonlinear optical susceptibility can be expressed as 
		\begin{equation}
			\chi_{ijk}(T<T_\mathrm{C}) = \underbrace{\chi_{ijk}(T>T_\mathrm{C})}_\mathrm{crystallographic\ (i-type)\ SHG} + \underbrace{\chi_{ijkl}(T>T_\mathrm{C})M_l,}_\mathrm{magnetic\ (c-type)\ SHG}
		\end{equation}
		
		\noindent where the indices $i$,$j$,$k$,$l$ refer to the crystallographic directions $x$, $y$, or $z$. Since the measured SHG intensity is proportional to $\vert\chi_{ijk}(T<T_\mathrm{C})\vert^2$, we find that the SHG intensity depends on the interference between crystallographic and magnetic SHG contributions and therefore on the direction of $\vector{M}$. This gives rise to the observed domain contrast. The domain contrast, in turn, depends on the relative amplitude between the crystallographic and magnetic SHG tensor contributions. The contrast is maximized if crystallographic and magnetic SHG have the same amplitude.  
		
		\item \textbf{Nonreciprocal SHG} can occur if the source term $\vector{S}$ in the electromagnetic wave equation contains contributions that are both even and odd functions of the wave vector \vector{k}, such that 
		\begin{equation}
			\vector{S} = \underbrace{\vector{S}_{\mathrm{even}}}_\mathrm{k-even\ (electric\ dipole)\ SHG} + \underbrace{\vector{S}_{\mathrm{odd}}}_\mathrm{k-odd\ (magnetic\ dipole)\ SHG}.
		\end{equation}
		
		Nonreciprocal SHG is not possible to observe in the electric-dipole approximation as all electric-dipole contributions are $\vector{k}$-even. Often, k-even and k-odd contributions correspond to electric dipole and magnetic dipole SHG, respectively \cite{Toyoda2021, Mund2021}, but also higher order multipole contributions can exist \cite{Shen66}. The precise structure of \vector{S}$_{\mathrm{even}}$ and \vector{S}$_{\mathrm{odd}}$ depends on the involved multipole contribution. In terms of point group symmetry operations,
		k-odd contributions like magnetic-dipole SHG are allowed in centrosymmetric materials and are hence \calP\ even. Moreover, the presence of k-even SHG contributions requires a broken inversion symmetry. They are thus \calP\ odd. Whereas the magnetic domain contrast arises from interference between \calT\ even and \calT\ odd SHG contributions, the observation of NR-SHG requires interference between \calP\ even and \calP\ odd SHG contributions, which highlights the different microscopic origin.
		
		\item We introduce the \textbf{nonlinear optical diode effect (NODE)} as a third SHG interference phenomenon. We will provide here both a phenomenological as well as a microscopic definition. Phenomenologically, we define the NODE as an effect whereby a nonlinear optical response (such as SHG), is stronger for one light propagation direction than for the reversed light propagation direction. Such an effect can be defined both in reflection and in transmission geometry (Supp. Fig.~\ref{ExtDefinition}). In both geometries, we can define a directional contrast $\eta$ as 
			\begin{equation}
				\eta = \left(I^\ra - I^\la \right)/\left(I^\ra + I^\la \right).
			\end{equation} 
		\noindent Microscopically, the reversal of the propagation direction is related to a mirror operation $\mathcal{M}$ (the relevant mirror plane runs vertically in Supp. Fig.~\ref{ExtDefinition}a and b). The NODE arises if the total SHG response is comprised of contributions that are even and odd under that mirror operation $\mathcal{M}$.
			
			\begin{equation}
				\chi_{ijk} = \chi_{ijk}(\mathcal{M}\ even) + \chi_{ijk}(\mathcal{M}\ odd)
				\label{eqn:NODEDef}
			\end{equation}
			
		\noindent While nonreciprocal SHG is crucial for the occurrence of a NODE in transmission, the minimal example in Supplementary Fig.~\ref{ExtMinimal} illustrates that a NODE can be observed in reflection even in the electric dipole approximation. This shows that the NODE and nonreciprocal SHG are independent effects that rely on the transformation behavior of the SHG response under different operations. In this example, the relevant mirror operation that enables the NODE is $\mathcal{M}_x$. The component $\chi_{xxx}$ is odd under $\mathcal{M}_x$, whereas $\chi_{xxz}$ is even. Note that the NODE is present in Supplementary Fig.~\ref{ExtMinimal}a,b even if $\chi_{xxx}$ and $\chi_{xxz}$ are i-type. Thus, the NODE is also distinct from the magnetic domain contrast as a NODE does not strictly require breaking of time-reversal symmetry or c-type SHG contributions \cite{Tokura2018}. The presence of a magnetic domain contrast also does not imply the presence of a NODE \cite{Pavlov03,Toyoda23} (both c-type and i-type SHG contributions may have the same transformation behavior with respect to $\mathcal{M}$). Hence, the magnetic contrast in Supp. Fig.~\ref{fig:Venn} is overlapping, but not contained in the NODE. In the cross section, the magnetic orders allows to control the directionality of the NODE, as demonstrated in the present work.
	\end{enumerate}
	
	\begin{figure}
		\includegraphics[width = 110 mm]{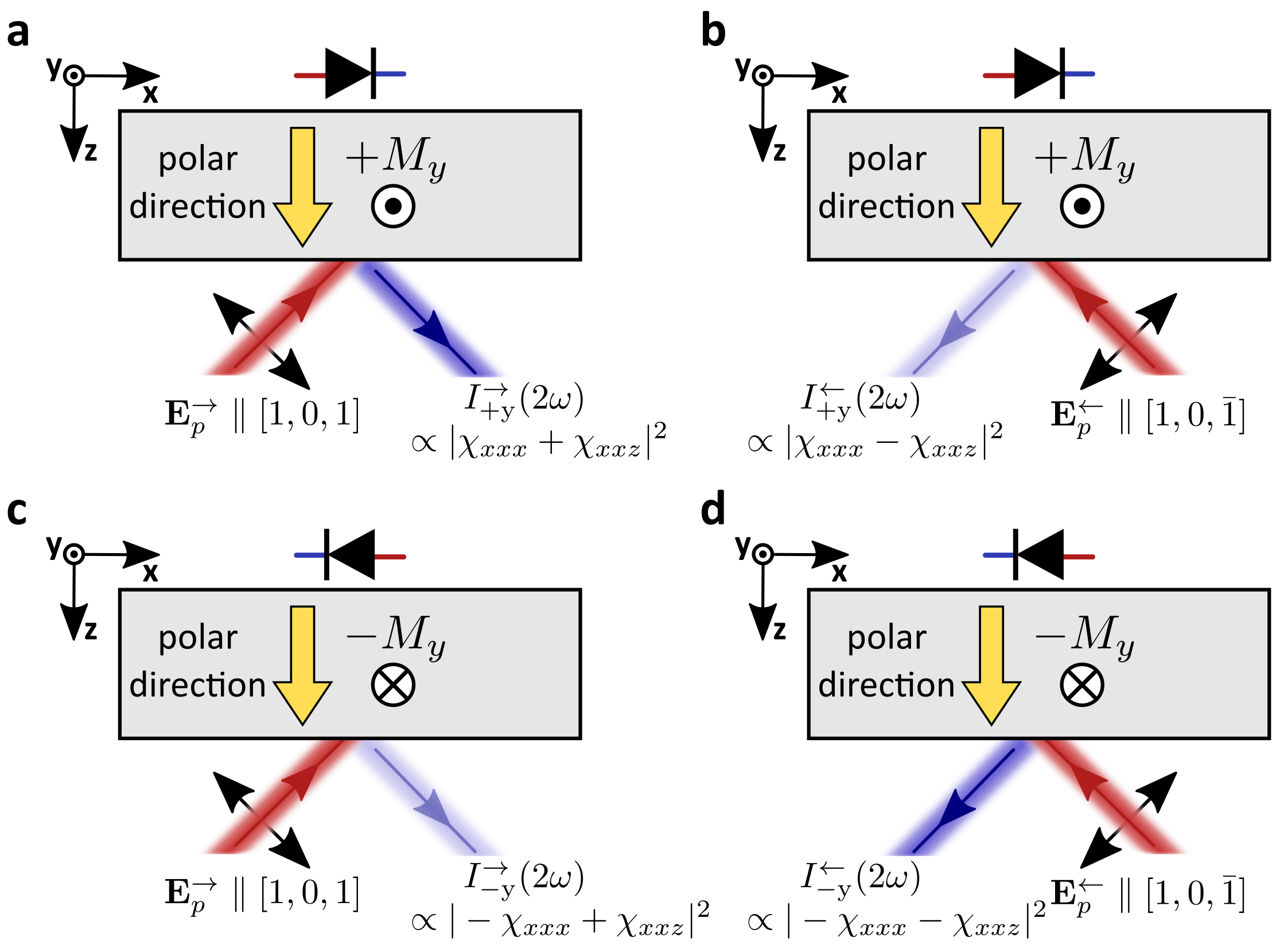}
		\caption{\textbf{Minimal example for the observation and manipulation of the NODE.} We consider a fictitious noncentrosymmetric material for which $\chi_{xxx}$ and $\chi_{xxz}$ are the only nonvanishing SHG tensor components in an orientation analogous to Fig.~1d. For p-polarized light, the electric field $\mathbf{E}_p(\omega)$ is in general parallel to $\mathbf{k}\times\hat{\mathbf{y}}$. The detected SHG intensity is proportional to the induced nonlinear polarization $I(2\omega) \propto \vert\sum_{i}P_i(2\omega)\vert^2\propto\vert\sum_{i,j,k}\chi_{ijk}E_j(\omega)E_k(\omega)\vert^2 = \vert \chi_{xxx}E_{p,x}^2+\chi_{xxz}E_{p,x}E_{p,z}\vert^2$. Specifically for an angle of incidence of $45^\circ$, the electric field is parallel to \textbf{a,} $\left[1,0,1\right]$ in forward direction and \textbf{b,} parallel to $\left[1,0,\bar{1}\right]$ in backward direction. Thus, $I^\rightarrow(2\omega) \propto \vert\chi_{xxx}+\chi_{xxz}\vert^2$ and $I^\leftarrow(2\omega) \propto \vert\chi_{xxx}-\chi_{xxz}\vert^2 \neq I^\rightarrow(2\omega)\vert^2$. We therefore observe a NODE in this minimal example. \textbf{c,d,} In a material where $\chi_{xxz}$ arises from a polar crystal structure whereas $\chi_{xxx}$ is due to magnetic order (as in the case of CeAlSi), switching the magnetization reverses the directionality of the NODE and thus allows its manipulation.}
		\label{ExtMinimal}
	\end{figure}
	
	To summarize, magnetic contrast, NR-SHG, and NODE are distinct interference phenomena that rely on different microscopic symmetries. Magnetic contrast strictly requires a broken time-reversal symmetry \calT\ and NR-SHG can only be observed in materials with broken spatial inversion \calP. The NODE requires neither symmetry to be broken, but is related to mirror symmetries. Note that $\mathcal{P} = \mathcal{M}_x \circ \mathcal{M}_y \circ \mathcal{M}_z$. Thus, a broken spatial inversion symmetry (as required for NR-SHG) implies a broken mirror symmetry, which enables a NODE. Therefore, NR-SHG is fully contained in the NODE in Supp. Fig.~\ref{fig:Venn}.
	\clearpage
	
	\newpage
	\section{Polarization-resolved SHG measurements}
	
	\begin{figure}[ht]
		\centering
		\includegraphics[width = \textwidth]{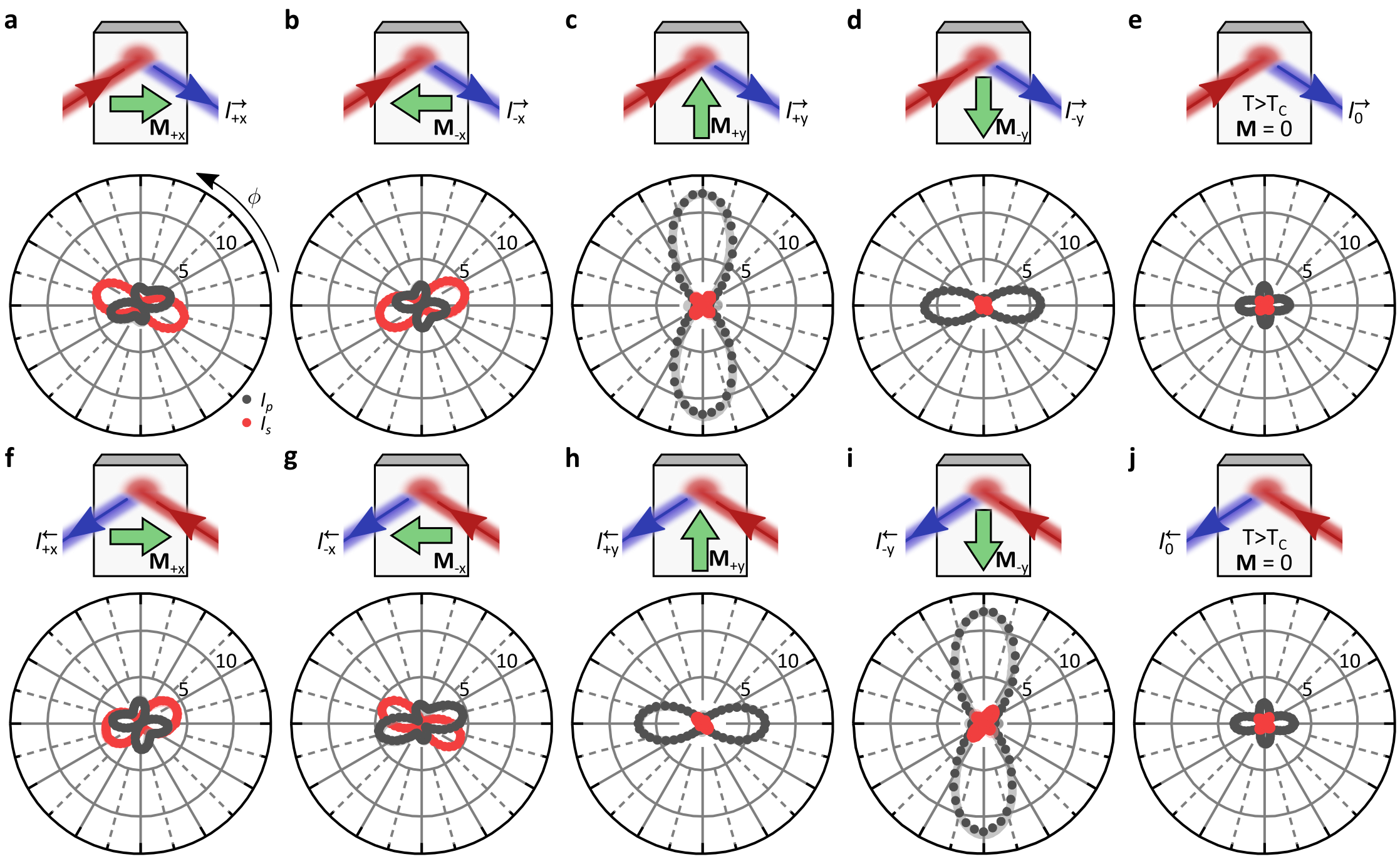}
		\caption{\textbf{SHG Polarization dependencies.} \textbf{a-e,} in forward direction and \textbf{f-j,} in backward direction. $I^{\mathbf{k}}_{\mathbf{M}}(\phi) = I^{-\mathbf{k}}_{-\mathbf{M}} (\phi)$ always holds (see Supplementary Fig.~\ref{ExtSymm} for details). Moreover, reversing the light propagation direction is equivalent to reversing the magnetization. All measurements were taken at $\hbar\omega = \SI{715}{meV}$.}
		\label{fig:PolScans}
	\end{figure}
	
	We show in Supplementary Fig.~\ref{fig:PolScans} the polarization dependence of the SHG intensity for the different magnetic states at $\SI{3}{K} < T_\mathrm{C}$ and in the paramagnetic phase at $\SI{12}{K} > T_\mathrm{C}$ ($T_\mathrm{C} = \SI{8.2}{K}$, see below). We show the polarization dependence for both forward (Fig.~\ref{fig:PolScans}a-e) and backward (Fig.~\ref{fig:PolScans}f-j) propagating light. 
	
	The polarization dependencies were measured such that the incident polarization was rotated, while the outgoing $p$ (grey) and $s$ (red) polarized SHG was detected. All measurements were taken at $\hbar\omega = \SI{715}{meV}$.
	
	In all magnetic states, we observe a NODE  for some light polarization. In the paramagnetic state, we observe an absence of a NODE for all polarizations. Moreover, we note certain symmetries between the polarization dependencies. In particular, $I^{\mathbf{k}}_{\mathbf{M}}(\phi) = I^{-\mathbf{k}}_{-\mathbf{M}} (\phi)$ always holds (see Supplementary Fig.~\ref{ExtSymm} for details).
	
	All 16 measurements in the magnetic phase ('4 magnetic states' x '2 SHG polarizations' x '2 propagation directions') were fit simultaneously with one consistent set of SHG tensor components for ED-SHG according to point group $2^\prime m m^\prime$ (Supplementary Table~\ref{Tab:SHGcomp}). To accommodate experimental uncertainties, we allow for a small rotation of the sample about the $x$, $y$, and $z$ axis corresponding to a pitch of the sample, a deviation of the angle of incidence from $45^\circ$, and an in-plane rotation of the sample, respectively. All deviations from the ideal case are below $2^\circ$. The fit results are shown as solid lines in Supplementary Fig.~\ref{fig:PolScans}. Note that the fit curves are often overlapping with the data points.
	
	The excellent quality of the fit indicates that the detected SHG is dominated by ED-SHG. We provide in table \ref{table:SHGcomponentsbulk} the magnitudes and phases for the fitted tensor $\chi_{ijk} = \vert\chi_{ijk}\vert * e^{i\Phi_{ijk}}$ components relative to $\chi_{zzz}$.
	
	\begin{table}
		\begin{center}
			\begin{tabular}{|c|c|c|}
				\hline
				$\chi_{ijk}$ & $\vert \chi_{ijk}\vert/\vert \chi_{zzz}\vert$ & $(\Phi_{ijk} - \Phi_{zzz})/\pi$ \\
				\hline
				$zzz$ & 1 & 0 \\
				\hline
				$xxz$ & 0.2 & -0.58 \\
				\hline
				$zxx$ & 0.29 & 0.28 \\
				\hline
				$yyz$ & 0.76 & -0.56 \\
				\hline
				$zyy$ & 1.87 & 0.76 \\
				\hline
				$xxx$ & 0.06 & 0.64 \\
				\hline
				$yyx$ & 0.62 & -0.55 \\
				\hline
				$xyy$ & 1.99 & 0.77 \\
				\hline
				$zzx$ & 0.78 & 0.55 \\
				\hline
				$xzz$ & 2.36 & -0.17 \\
				\hline
			\end{tabular}
			\caption{Magnitudes and phases of $\chi$ tensor components relative to $\chi_{zzz}$. Values were found by fitting all 168 measurements below $T_C$ in Supplementary Fig.~\ref{fig:PolScans} simultaneously with one consistent set of $\chi$ tensor components according to point group $2^\prime m m^\prime$.}
			\label{table:SHGcomponentsbulk}
		\end{center}
	\end{table}
	
	Based on the obtained relative magnitudes and phases of all tensor components, we can calculate the SHG intensity and the directional contrast even in configurations that are not easily experimentally accessible. For example, we show in Supplementary Fig.~\ref{fig:AoI} the incident angle dependence of the SHG intensity for s-polarized incident light and p-polarized outgoing SHG (experimental configuration of Fig.~1e in the main text). Although we observe a strong directional contrast at an incident angle of $45^\circ$, this angle is in general not optimal. As the directional contrast must be a continuous function of the incident angle, it is interesting to consider two edge case:
	
	\begin{enumerate}
		\item \underline{Normal incidence (AoI = $0^\circ$)}: In this scenario, incident and outgoing beams overlap. A reversal of the light path (AoI $\rightarrow$ -AoI) does not change the experiment. Hence, a NODE cannot be observed in normal incidence reflection.
		\item \underline{Grazing incidence (AoI = $90^\circ$))}: Also here, a NODE cannot be observed in the electric-dipole approximation. In the coordinate system of Supplementary Fig.~\ref{ExtMinimal}, grazing incidence corresponds to light propagation parallel to $\hat{\mathbf{x}}$. The transversal electric fields thus only couple to tensor components $\chi_{ijk}$ with $i,j,k \in \left\{y,z\right\}$, which are all even with respect to the mirror operation $\mathcal{M}_x$. Hence, the basic symmetry requirement for the observation of a NODE (Eq.~(\ref{eqn:NODEDef})) cannot be realized.
	\end{enumerate}
	
	\begin{figure}
		\includegraphics[width = 119 mm]{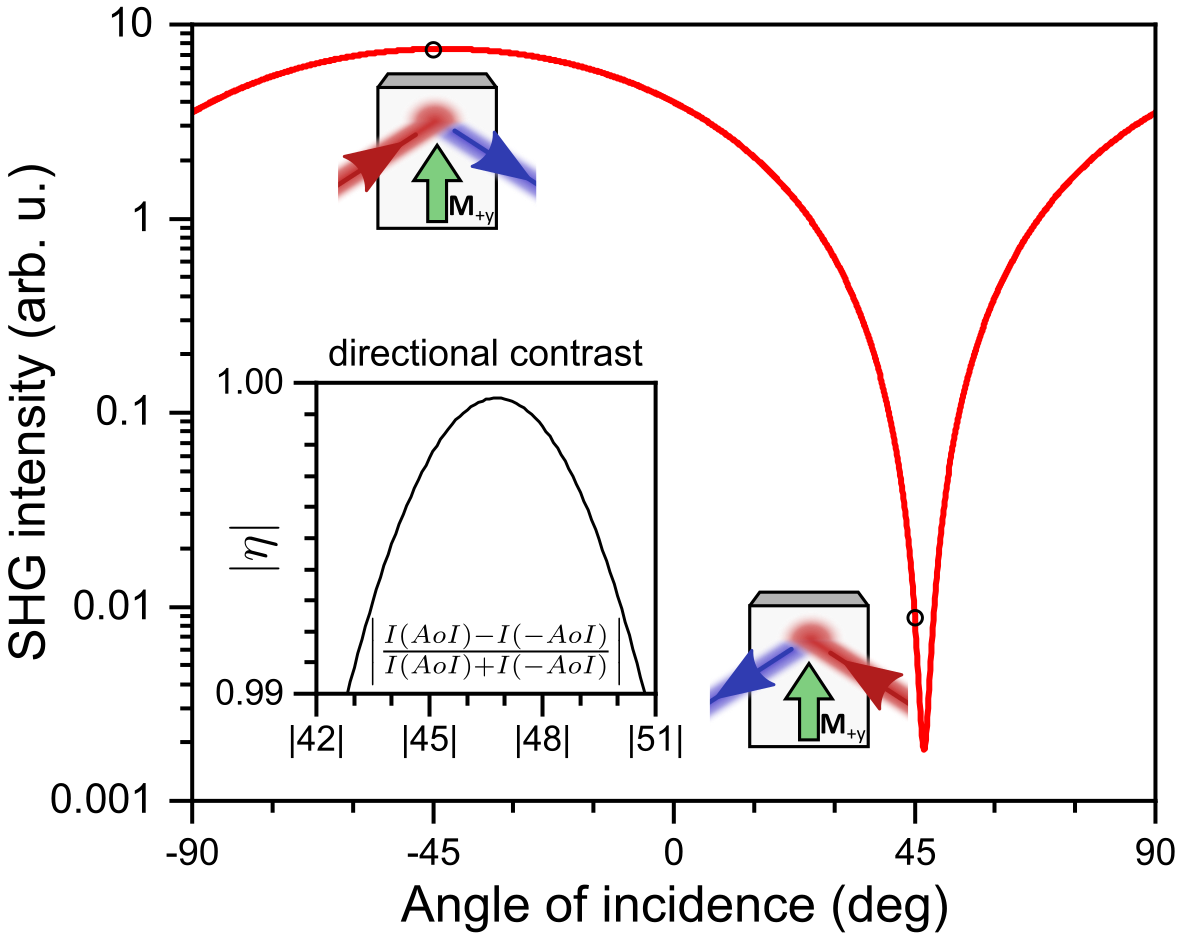}
		\caption{\textbf{Incident angle dependence.} Calculated incident angle dependence based on the relative magnitudes and phases of the SHG tensor component extracted from experimental polarization dependencies. An angle of incidence of $45^\circ$ is not optimal. By adjusting the angle of incidence, a numerically determined directional contrast $>99.9\%$ could be possible (inset). Black circles mark the experimental configuration of $\pm 45^\circ$.}
		\label{fig:AoI}
	\end{figure}
	\newpage
	\section{Temperature dependent characterization}\label{sec:Temp}
	
	Unless explicitly noted differently, all measurements were performed at a sample temperature of \SI{3}{K}. We provide here additional details on the temperature-dependent characterization of CeAlSi.
	
	\begin{figure}[ht]
		\centering
		\includegraphics[width = 119 mm]{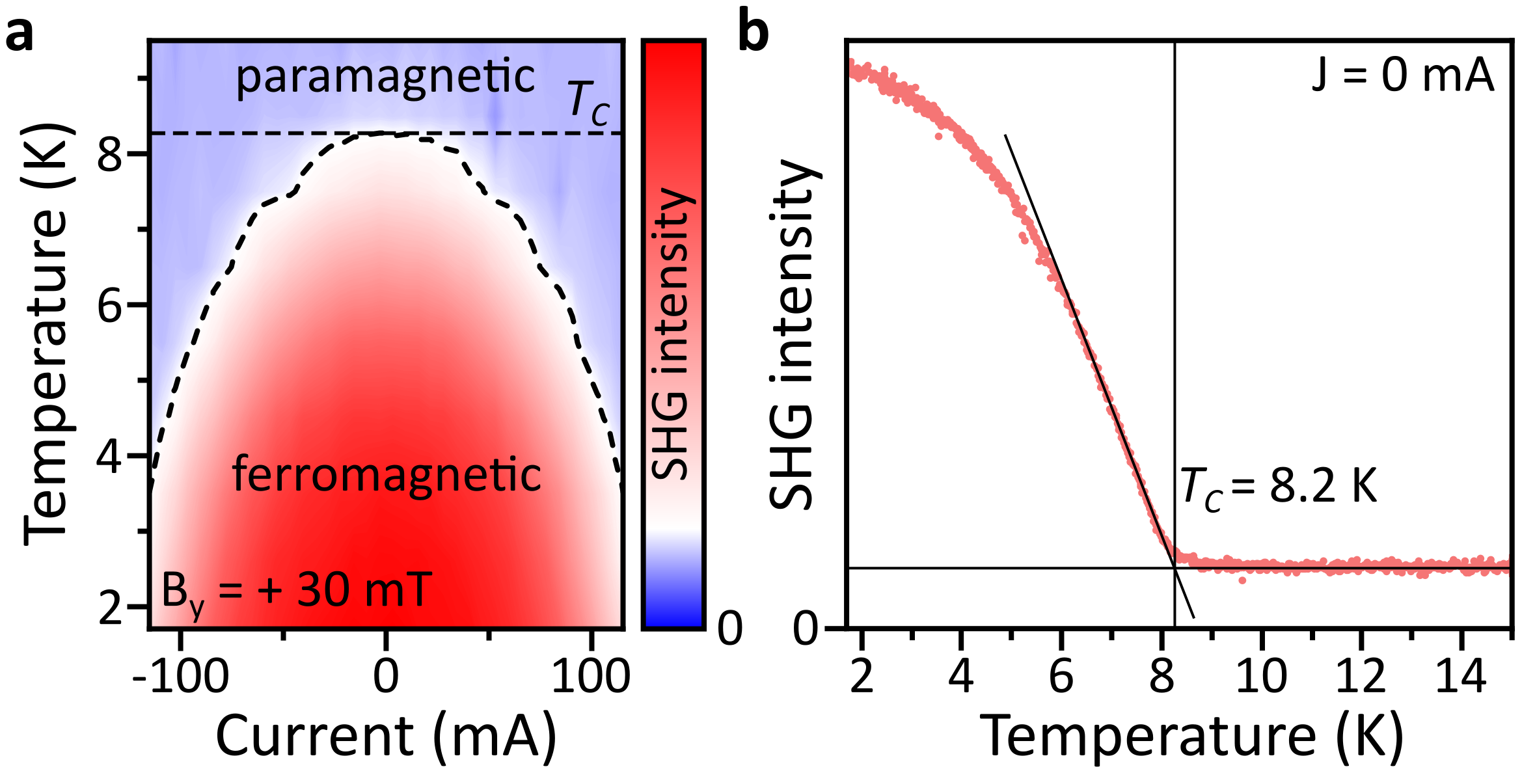}
		\caption{\textbf{Temperature dependent characterization.} \textbf{a,} SHG intensity as a function of temperature and applied electric DC current. For temperatures below \SI{4}{K}, the sample remains in the ferromagnetically ordered phase for all applied currents. \textbf{b,} Temperature dependent SHG intensity in the absence of electric current. We find a magnetic transition temperature of $T_C = \SI{8.2}{K}$ in agreement with literature \cite{Yang2021}.}
		\label{fig:Temperature}
	\end{figure}
	
	We show in the main text that we can switch the magnetization of CeAlSi and control the NODE by applying current (Fig.~3 in the main text). In Supplementary Fig.~\ref{fig:Temperature}a, we show the measured SHG intensity at various temperatures during the application of a DC current. Here, an external magnetic field of \SI{30}{mT} stabilizes a ferromagnetic $\mathbf{M}_\mathrm{+y}$ state such that no current-induced switching can occur. We can clearly distinguish the paramagnetic phase (homogeneously light blue) from the ferromagnetically ordered phase (white to red). 
	
	Note that "Temperature" here refers to the temperature set point instead of the actual sample temperature, which is elevated due to current-induced ohmic heating. The effect of ohmic heating can be seen from the approximately parabolic current-induced temperature increase. This increase is further independent of the applied current direction (consistent with ohmic heating). Under maximum current, the sample enters the paramagnetic phase at a temperature of about \SI{4}{K}. Thus. at \SI{3}{K}, the sample remains in the magnetically ordered phase for all applied DC currents.
	
	Note also that the SHG intensity in the paramagnetic phase is independent of the applied current. In contrast, the SHG intensity at maximum current in Fig.~3c in the main text is different for the opposite polarities. Thus, the sample remained in the magnetically ordered state during the measurement of Fig.~3c.
	
	Supplementary Fig.~\ref{fig:Temperature}b shows the temperature dependence of the SHG intensity in the absence of applied current. We find a transition temperature of \SI{8.2}{K} in agreement with previous characterizations \cite{Yang2021, Sun2021Mapping}. As CeAlSi is noncentrosymmetric above $T_C$, we continue to observe a finite SHG intensity above \SI{8.2}{K}.
	
	\newpage
	\section{Additional NODE spectroscopy}
	
	We report in the main text the observation of a nonlinear optical diode effect (NODE) over a wide spectral range. We demonstrate this observation in the main text on the basis of selected measurements in certain magnetic configurations. In this section, we provide additional systematic experimental evidence in complementary settings. The striking agreement between symmetry-related measurements is evidence for a high data quality and shows that the observations are intrinsic properties of CeAlSi. 
	
	\subsection{Observation of a broadband NODE in the $\mathbf{M}_\mathrm{-y}$ state}
	
	We show in Supplementary Fig.~\ref{fig:NODE} SHG spectra for counter-propagating light beams. The sample is in a magnetic $\mathbf{M}_\mathrm{-y}$ state. Complementary to Fig.~1e of the main text, we again observe a broadband NODE. However, the spectra are inverted relative to Fig.~1e of the main text, which were recorded in the $\mathbf{M}_\mathrm{+y}$ state. This observation is immediate evidence that the magnetization in CeAlSi controls the directionality of the broadband NODE.
	
	Other experimental parameters are identical: spectra were recorded at \SI{3}{K}, incident fundamental light is $s$ polarized, and we detect outgoing $p$ polarized SHG light. 
	
	\begin{figure}[ht]
		\centering
		\includegraphics[width = 0.4\textwidth]{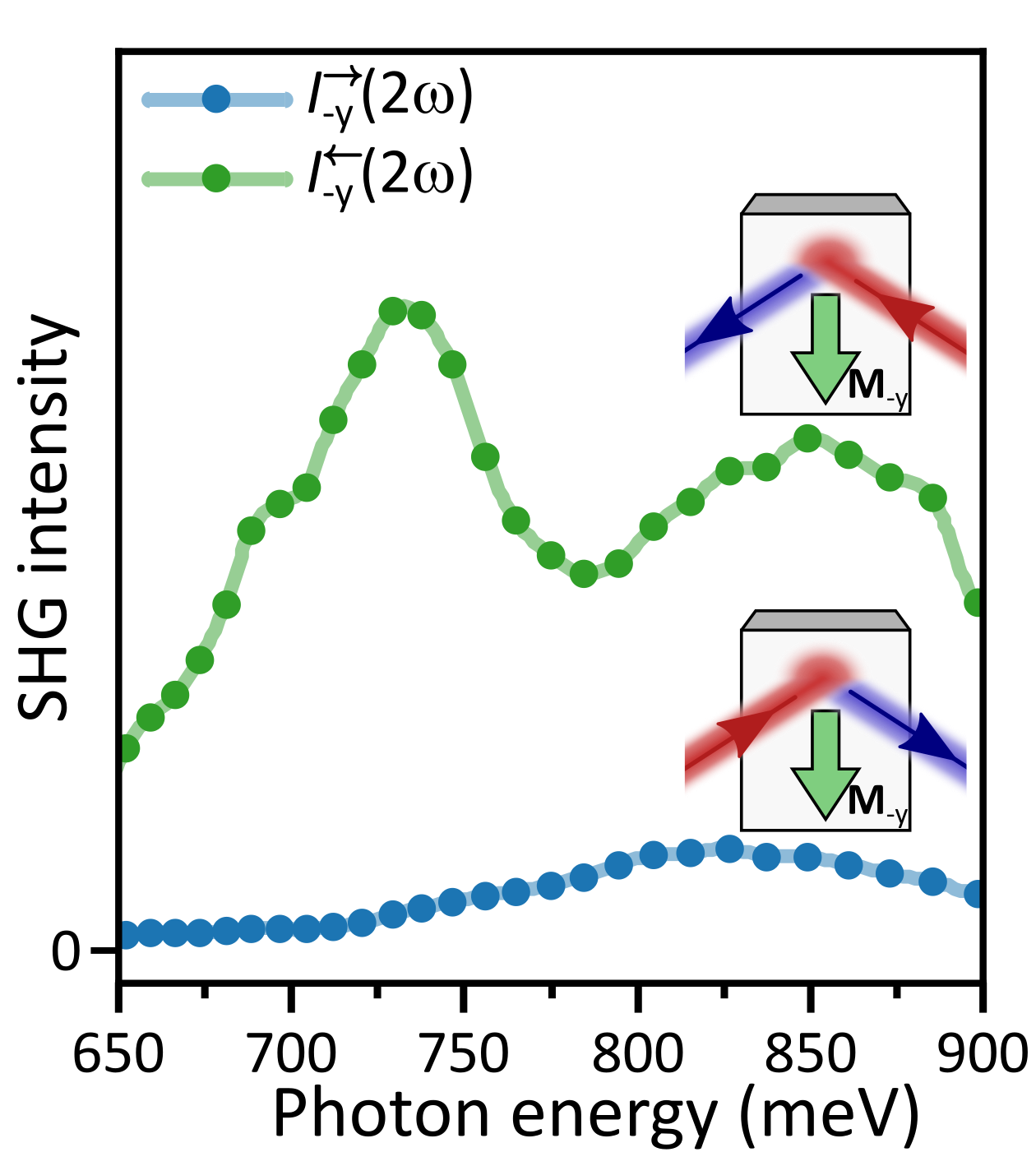}
		\caption{\textbf{Observation of a broadband NODE in the $\mathbf{M}_\mathrm{-y}$ state.} $I^\rightarrow_\mathrm{-y}(2\omega)\ll I^\leftarrow_\mathrm{-y}(2\omega)$ over a broad range $>\SI{250}{meV}$.}
		\label{fig:NODE}
	\end{figure}
	
	\subsection{Determination and robustness of directional contrast}
	
	We introduce in the main text the directional contrast 
	
	\begin{equation}
		\eta = \frac{ I^{\ra}_{+y}-I^{\la}_{+y}}{I^{\ra}_{+y}+I^{\la}_{+y}}. \label{eq:eta1}
	\end{equation}
	
	By symmetry, this expression is equivalent to 
	
	\begin{eqnarray}
		\eta &=& \frac{ I^{\ra}_{+y}-I^{\ra}_{-y}}{I^{\ra}_{+y}+I^{\ra}_{-y}} \label{eq:eta2}\\
		&=&  -\frac{ I^{\la}_{+y}-I^{\la}_{+y}}{I^{\la}_{+y}+I^{\la}_{+y}}\label{eq:eta3}\\
		&=&  -\frac{ I^{\ra}_{-y}-I^{\la}_{-y}}{I^{\ra}_{-y}+I^{\la}_{-y}}\label{eq:eta4}\\
	\end{eqnarray}
	
	Here, we experimentally verify this equivalence. We show in Supplementary Fig.~\ref{fig:Mycontrast}a and b 2D maps of the detected $p$ polarized SHG intensity as a function of incident photon energy and polarization angle ($0^\circ/180^\circ$ - $p$ polarized; $90^\circ$ - $s$ polarized fundamental light). Both maps were obtained with light propagating in forward direction, however, the sample is in a magnetic $\mathbf{M}_\mathrm{+y}$ state in Supplementary Fig.~\ref{fig:Mycontrast}a and  in a $\mathbf{M}_\mathrm{-y}$ state in Supplementary Fig.~\ref{fig:Mycontrast}b. We notice clear differences between the two panels.
	
	Supplementary Fig.~\ref{fig:Mycontrast}c shows the directional contrast according to Eq.~\ref{eq:eta2} showing a strong contrast (for certain polarizations) at all photon energies. To verify the robustness of the directional contrast, i.e., the equivalence of Eqs.~\ref{eq:eta1}-\ref{eq:eta4}, we evaluated the SHG intensity also in the reversed direction for both magnetic states (Supplementary Fig.~\ref{fig:Mycontrast}d and e). 
	
	We notice the (expected) similarity between Supplementary Fig.~\ref{fig:Mycontrast}a and e, as well as Supplementary Fig.~\ref{fig:Mycontrast}b and d. Based on these four measurements, we can determine the directional contrast $\eta$ according to all four definitions in Eqs.~\ref{eq:eta1}-\ref{eq:eta4}. Indeed, as we can see from Supplementary Figs.~\ref{fig:Mycontrast}c, f, g, and h, the obtained values for $\eta$ agree well (up to a sign). The absolute value of Supplementary Fig.~\ref{fig:Mycontrast}c is shown in Fig.~4a in the main text.
	
	In Supplementary Fig.~\ref{fig:Mycontrast}i, we display for all photon energies within the considered spectral range the maximum contrast that is achievable by selecting a specific incident polarization. We note that the maximum of $\vert\eta\vert$ determined according to Eqs.~\ref{eq:eta2} and \ref{eq:eta3} show a striking agreement, whereas the values determined according to Eqs.~\ref{eq:eta1} and \ref{eq:eta4} differ slightly. This is due to experimental uncertainties in reversing the beam path, which we can circumvent by determining $\eta$ from measurements in the same propagation direction but opposite magnetic states. Note that this options exists in CeAlSi as reversing the propagation direction is equivalent to reversing the beam path (see Sec.~\ref{sec:Symmetry}).
	
	\begin{figure}[ht]
		\centering
		\includegraphics[width = \textwidth]{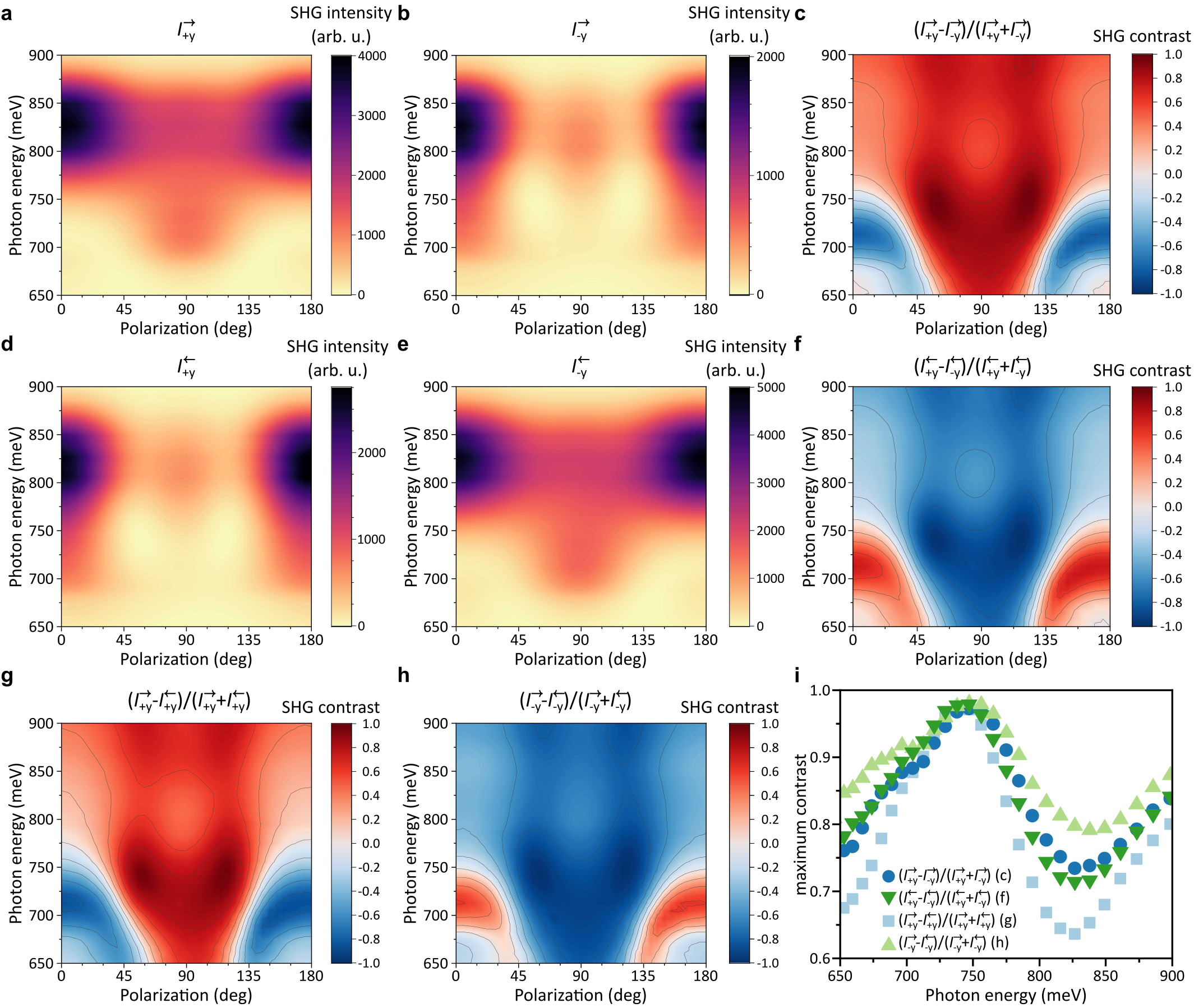}
		\caption{\textbf{Determination and robustness of directional contrast (Fig.~4a).} \textbf{a,} 2D map of measured $p$ polarized SHG intensity in forward direction as a function of incident polarization and photon energy for the $\mathbf{M}_\mathrm{+y}$ state and \textbf{b,} $\mathbf{M}_\mathrm{-y}$ state. \textbf{c,} directional contrast $\eta$ as defined by Eq.~\ref{eq:eta2}. \textbf{d-f,} analogous to \textbf{a-c} in backward direction. \textbf{g,h,} directional contrast defined as contrast between opposite directions in the same magnetic state. Panels c, f, g, and h are all consistent, demonstrating the robustness of determining the directional contrast. \textbf{i,} Maximum contrast analogous to Fig.~4b for panels \textbf{c}, \textbf{f}, \textbf{g}, and \textbf{h}. Maximum contrast extracted from \textbf{c}, \textbf{f} agree well, whereas experimental uncertainties in reversing the beam path cause slight deviations between \textbf{g} and \textbf{h}. The absolute value of panel \textbf{c} is shown in Fig.~4a.}
		\label{fig:Mycontrast}
	\end{figure}
	
	\subsection{Directional contrast in the $\mathbf{M}_\mathrm{\pm x}$ states}
	
	The discussion of the directional contrast in the main text and in the previous section was centered on the magnetic $\mathbf{M}_\mathrm{\pm y}$ states. In this section, we present an analogous analysis for the $\mathbf{M}_\mathrm{\pm x}$ states.
	
	In Supplementary Fig.~\ref{fig:Mxcontrast}a-d, we show 2D maps of the SHG intensity as a function of incident fundamental photon energy and polarization. The panels show data recorded in forward direction in the $\mathbf{M}_\mathrm{+x}$ state (panel a), in forward direction in the $\mathbf{M}_\mathrm{-x}$ state (panel b), in reversed direction in the $\mathbf{M}_\mathrm{+x}$ state (panel c), and in reversed direction in the $\mathbf{M}_\mathrm{-x}$ state (panel d). We recognize the clear (expected) pairwise similarity between panels a and d, as well as panels b and c.
	
	As we showed in the previous section that a determination of $\eta$ according to Eqs.~\ref{eq:eta1}-\ref{eq:eta4} yields analogous results, we restrict ourselves here to Eq.~\ref{eq:eta2}. The obtained 2D map for the directional contrast is shown in Supplementary Fig.~\ref{fig:Mxcontrast}e. By selecting specific incident light polarizations, we can ensure a directional contrast of at least 72\% for all photon energies and up to 96.6\%, very similar to the achievable contrast in the magnetic $\mathbf{M}_\mathrm{\pm y}$ states (73\% to 97.2\%).
	
	\begin{figure}[ht]
		\centering
		\includegraphics[width = 118 mm]{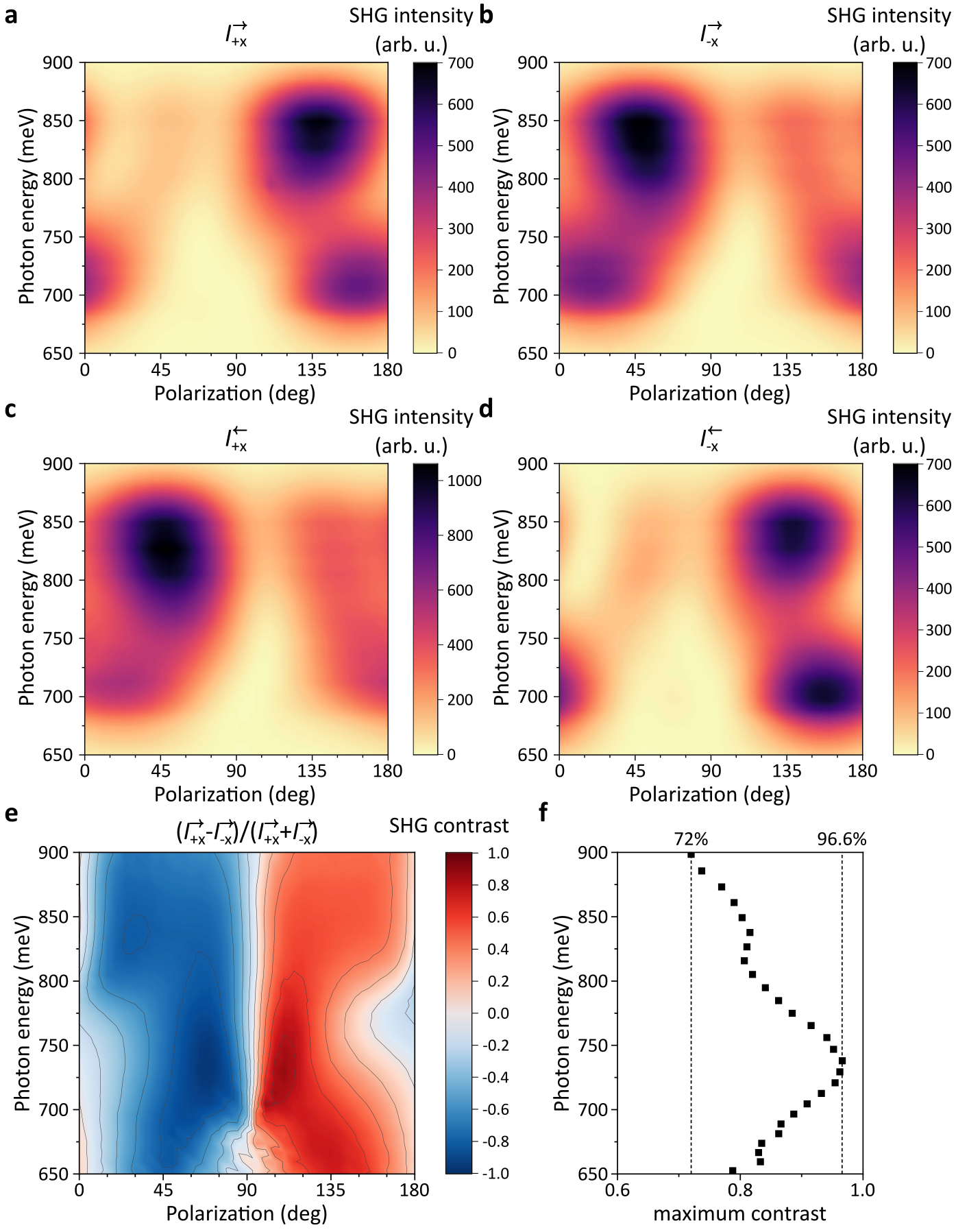}
		\caption{\textbf{Directional contrast in the $\mathbf{M}_\mathrm{\pm x}$ states.} \textbf{a,} 2D map of measured $s$ polarized SHG intensity as a function of incident polarization and photon energy in forward direction for the $\mathbf{M}_\mathrm{+x}$ state and \textbf{b,} $\mathbf{M}_\mathrm{-x}$ state and in reversed direction for \textbf{c,} the $\mathbf{M}_\mathrm{+x}$ state and \textbf{d,}  the $\mathbf{M}_\mathrm{-x}$ state. \textbf{e,} directional contrast $\eta$ as defined by Eq.~\ref{eq:eta2}. \textbf{f,} Maximum achievable contrast analogous to Fig.~4b obtain from panel e.}
		\label{fig:Mxcontrast}
	\end{figure}

	\subsection{A note on symmetry}
	
	By comparing the 2D maps of the directional contrast in the $\mathbf{M}_\mathrm{\pm x}$ states (Supplementary Fig.~\ref{fig:Mxcontrast}e) and  $\mathbf{M}_\mathrm{\pm y}$ states (e.g, Supplementary Fig.~\ref{fig:Mycontrast}c), we recognize differences in the symmetry of these magnetic states. In the $\mathbf{M}_\mathrm{\pm y}$ states, the 2D map is mirror symmetric about the $0^\circ$ and $90^\circ$ polarization directions. This is due to the mirror symmetry of the polarization dependence in the $\mathbf{M}_\mathrm{\pm y}$ states (Supplementary Fig.~\ref{fig:PolScans}), which is microscopically a consequence of the $m_y$ mirror symmetry in these states. In contrast, in the $\mathbf{M}_\mathrm{\pm y}$ states, the 2D map is mirror symmetric with a sign change, which is an experimental manifestation of the  $m_y^\prime$ symmetry in these states ($m_y^\prime = m_y \circ \mathcal{T}$).
	
	\subsection{Directional contrast in the paramagnetic phase}
	
	Here, we verify that the NODE vanishes in the paramagnetic phase of CeAlSi as required by symmetry (Sec.~\ref{sec:Symmetry}) and indicated by the polarization-dependent measurements at \SI{715}{meV} (Supplementary Fig.~\ref{fig:PolScans}e,f).
	
	In Supplementary Fig.~\ref{fig:12KPolSpectra}, we show 2D maps showing the detected SHG intensity as a function of incident polarization and photon energy. All measurements were taken at \SI{12}{K}, which is well above $T_C = \SI{8.2}{K}$ (Sec.~\ref{sec:Temp}). We consider here both $p$ and $s$ polarized SHG light for both forward and backward propagating light. For both polarizations, measurements with forward and backward propagating light are strikingly similar. We note that for the $s$ polarized SHG response (Supplementary Fig.~\ref{fig:12KPolSpectra}b,d), the overall SHG intensity is relatively low. Therefore, experimental uncertainties in reversing the light become apparent in the measurement. Moreover, in particular for the $0^\circ$ and $90^\circ$ polarization, the SHG intensity approaches zero (as required by $4mm$ symmetry), which impedes a meaningful determination of the directional contrast. We therefore refrain from showing 2D maps for $\eta$, but note the both qualitative and quantitative similarity between forward and backward propagating SHG responses for both $s$ and $p$ polarized SHG light.
	
	\begin{figure}[ht]
		\centering
		\includegraphics[width = 118 mm]{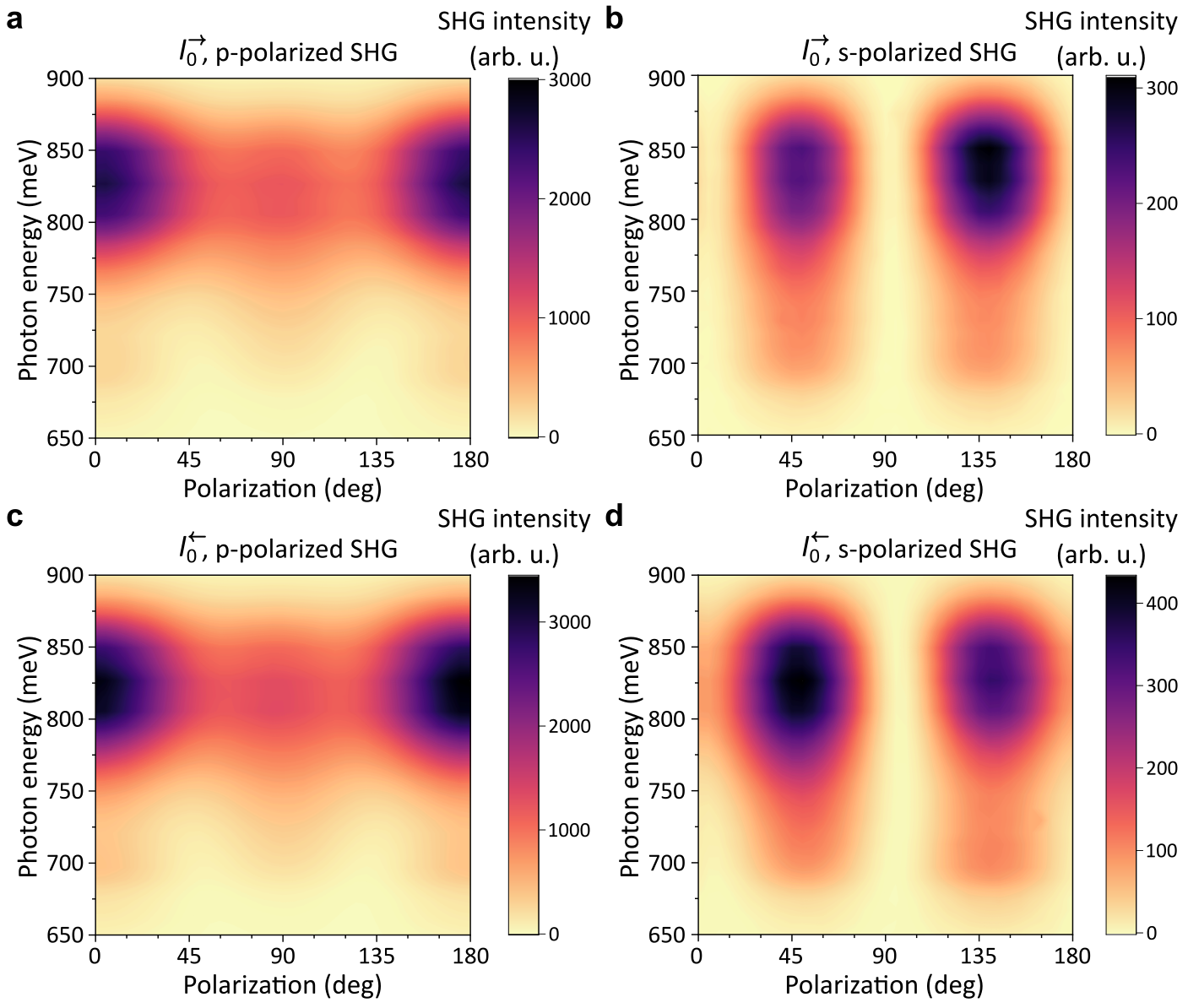}
		\caption{\textbf{Vanishing NODE in the paramagnetic phase.} SHG intensity as a function of incident light polarization and photon energy for \textbf{a,} $p$ polarized SHG in forward direction \textbf{b,} $s$ polarized SHG in forward direction \textbf{c,} $p$ polarized SHG in reversed direction and \textbf{d,} $s$ polarized SHG in reversed direction. Comparing opposite propagation directions for each SHG polarization, we not the absence of a discernible NODE.}
		\label{fig:12KPolSpectra}
	\end{figure}
	\clearpage
	
	\newpage
	\section{Domain assignment and mapping}\label{sec:mapping}
	
\begin{figure}[ht]
		\centering
		\includegraphics[width = \textwidth]{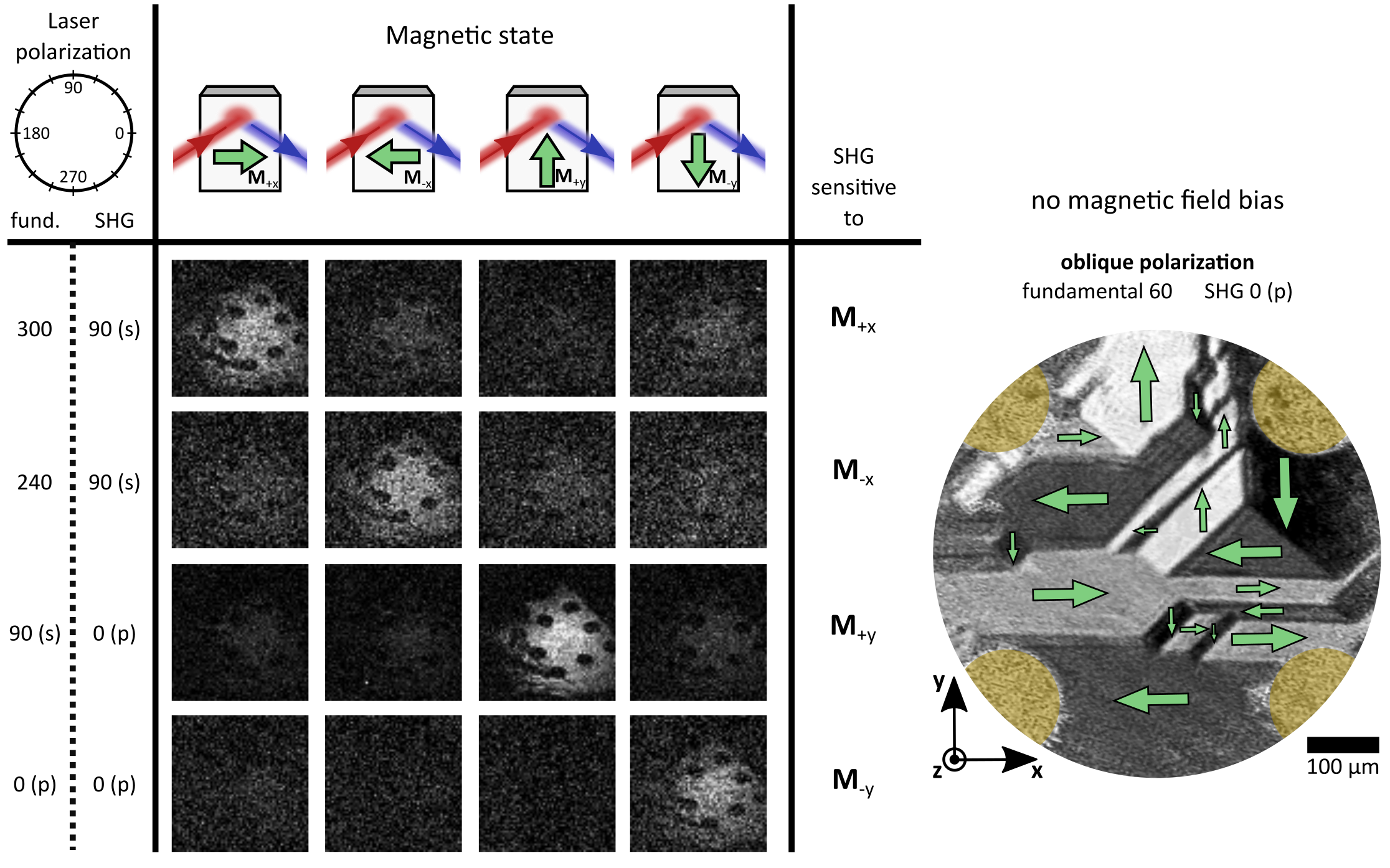}
		\caption{\textbf{Domain assignment.} The unique polarization dependence of the SHG in all four magnetic states (Supplementary Fig.~\ref{fig:PolScans}) allows us to identify four distinct combinations of incident fundamental and outgoing SHG polarizations that are uniquely sensitive to each of the magnetic states. Here, an external magnetic field of approximately \SI{30}{mT} ensures a well-defined magnetic single-domain state. We were thus able to verify that the current-induced switching in Fig.~3c is between $\mathbf{M}_\mathrm{\pm y}$ states. In the absence of a magnetic field bias, we can identify an oblique polarization angle, where the intensity of all four magnetic states is different and assign the domains (right hand side). Yellow circles show positions of gold electrodes.}
		\label{fig:DomainAssignment}
	\end{figure}

	For a fixed propagation direction, we can utilize the SHG domain contrast in CeAlSi to visualize the four different magnetic states. In particular, the unique polarization dependence of the SHG in all four magnetic states (Supplementary Fig.~\ref{fig:PolScans}) allows us to identify four distinct combinations of incident fundamental and outgoing SHG polarizations that are uniquely sensitive to each of the magnetic states (Supplementary Fig.~\ref{fig:DomainAssignment}). This enables the capability to image the magnetic domain distribution of CeAlSi and assign the magnetic states. 
	
	We frequently exploit the capability to map and assign domain states in CeAlSi. For example:
	
	\begin{enumerate}
		\item It enables the \textit{operando} visualization of the domain distribution during the application of current (Supplementary Movie 3).
		\item It allows us to verify that the current-induced switching in Fig.~3c of the main text is between $\mathbf{M}_\mathrm{\pm y}$ states.
		\item It allows us to verify that the current-induced switching in Fig.~3g of the main text is between $\mathbf{M}_\mathrm{\pm x}$ states.
		\item It enables the visualization of magnetic-field induced switching (Section~\ref{sec:Bfield})
	\end{enumerate}
	
	To obtain the domain image on the right-hand side of Supplementary Fig.~\ref{fig:DomainAssignment}, we normalized the SHG image recorded at \SI{3}{K} to the SHG image recorded at \SI{12}{K}. Thus, spatial inhomogeneities in the SHG intensity due to the approximately Gaussian beam profile are removed. Since such a normalization is only possible within the illuminated area, the image is cropped to area approximately corresponding to the beam spot.
	\clearpage
	
	\newpage
	\section{Magnetic field dependent SHG characterization}\label{sec:Bfield}
	
	Utilizing the capability to selectively visualize magnetic domains by SHG imaging (Sec.~\ref{sec:mapping}), we determined the domain state for magnetic fields in the range of $\pm\SI{20}{mT}$. The measurement was performed at \SI{1.6}{K}. We show the area fraction of the different states in Supplementary Fig.~\ref{fig:Bfield}a.
	
	We chose here two setting for the laser polarization: (1) the 90/0 configuration that yields a high SHG intensity only for $\mathbf{M}_\mathrm{+y}$ domains and (2) the 90/90 configuration yields a high SHG intensity for both $\mathbf{M}_\mathrm{+x}$ and $\mathbf{M}_\mathrm{-x}$ domains. The area fraction for $\mathbf{M}_\mathrm{-y}$ domains was obtained numerically by completing the determined area fractions to 1.
	
	Consistent with literature, we find a narrow hysteresis from which we deduce a low coercive field of 1-\SI{2}{mT} (estimated from the hysteresis at an area fraction of 50\%). Note that the $\mathbf{M}_\mathrm{+y}$ hysteresis loop is not symmetric around \SI{0}{mT} due to CeAlSi exhibiting four magnetic states. Thus, we expect an area fraction of 25\% for all domains in a fully randomized state at \SI{0}{mT} (i.e., 25\% $\mathbf{M}_\mathrm{+y}$, 25\% $\mathbf{M}_\mathrm{-y}$, and 50\% $\mathbf{M}_\mathrm{+x}$ + $\mathbf{M}_\mathrm{-x}$). Thus, the $\mathbf{M}_\mathrm{+y}$ ($\mathbf{M}_\mathrm{-y}$) hysteresis loop appears shifted towards positive (negative) magnetic fields. Moreover, we find that $\mathbf{M}_\mathrm{\pm x}$ domains are only present at low magnetic fields and fully suppressed for $\vert B_y\vert > \SI{10}{mT}$.
	
	In panels b-e of Supplementary Fig.~\ref{fig:Bfield}, we show SHG domain images at selected magnetic fields for laser polarization setting (1) sensitive only to $\mathbf{M}_\mathrm{+y}$ domains. At \SI{-20}{mT}, the sample appears homogenously dark (panel b). In contrast, at \SI{+20}{mT}, the sample appears homogenously bright (panel d). These states correspond to single domain $\mathbf{M}_\mathrm{+y}$ and $\mathbf{M}_\mathrm{-y}$ states, respectively. Panels c and e show the domain image at \SI{2.7}{mT} taken while increasing and decreasing the magnetic field, respectively. As expected, the area fraction of $\mathbf{M}_\mathrm{+y}$ domains is larger after decreasing the field from \SI{+20}{mT} (panel e). The series of all images for both laser polarization settings are combined into Supplementary Movies 1 (polarization setting 1) and 2 (polarization setting 2).
	
	\begin{figure}
		\includegraphics[width = 0.8\textwidth]{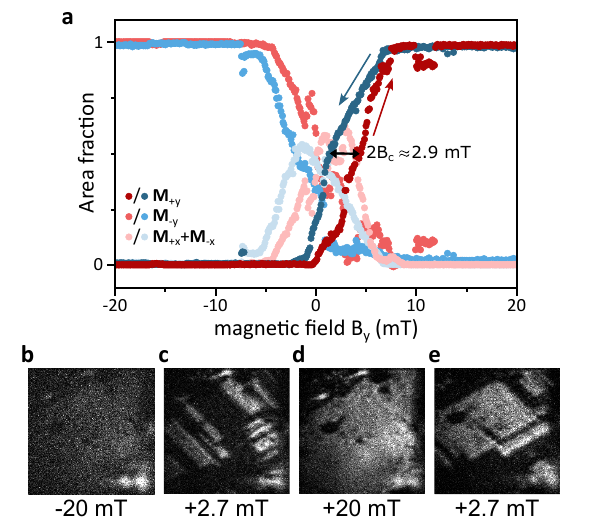}
		\caption{\textbf{Magnetic field dependent characterization.} \textbf{a,} magnetic hysteresis. Red shaded data points are taken with increasing field; blue shaded data points with decreasing field. Consistent with literature \cite{Yang2021}, we find that CeAlSi is a soft magnet with a coercive field of just a few mT. \textbf{b-e,} magnetic domain imaging by SHG. The laser polarization is chosen such that bright areas correspond $\mathbf{M}_\mathrm{+y}$ domains. The area fractions in panel a were extracted from such images.}
		\label{fig:Bfield}
	\end{figure}
	\clearpage
	
	\newpage
	\section{Symmetry of the NODE}\label{sec:Symmetry}
	
\begin{figure}[h]
		\includegraphics[width = 110 mm]{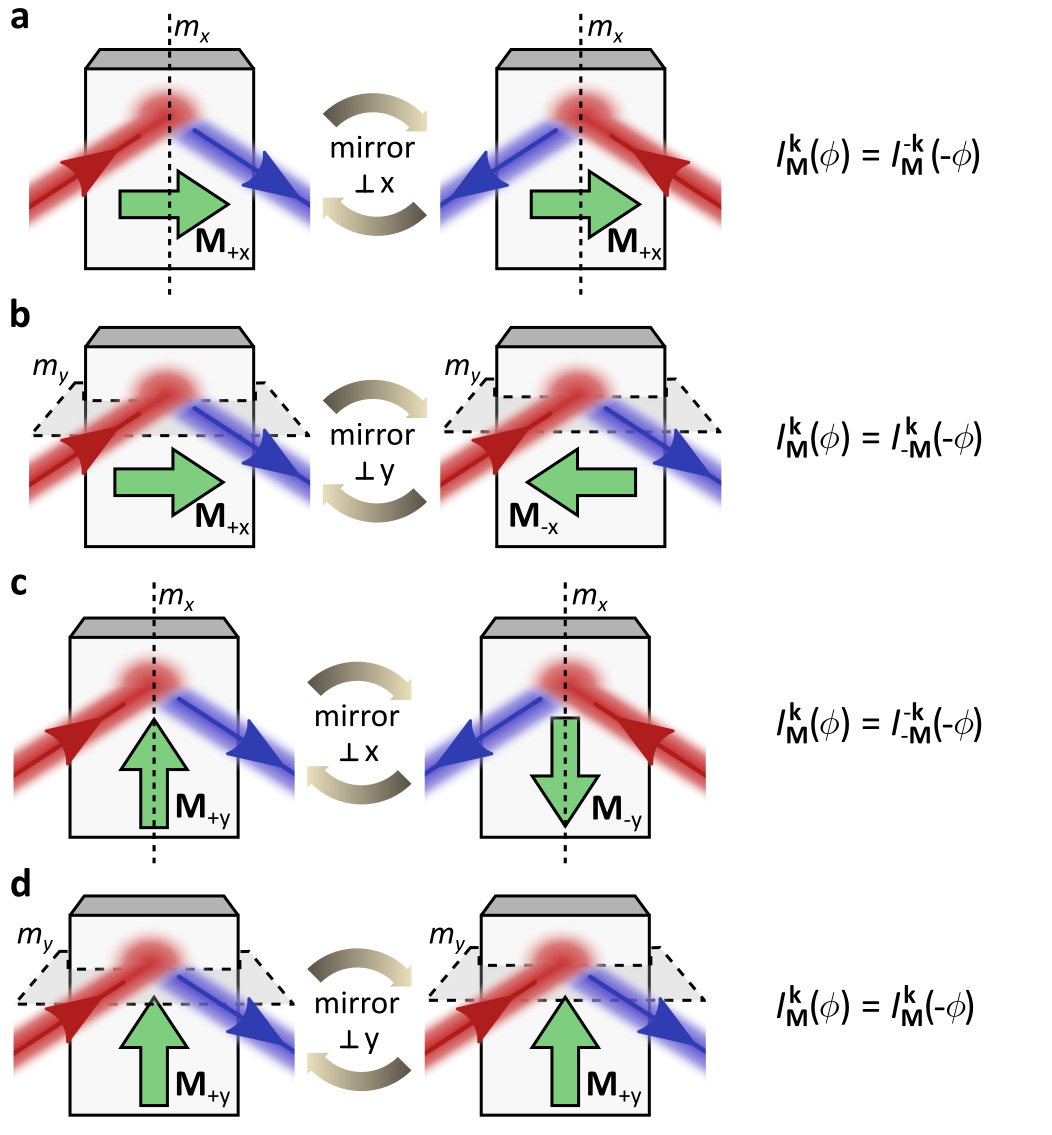}
		\caption{\textbf{Consequences of mirror symmetries in CeAlSi.} \textbf{a,} In the $\mathbf{M}_{\pm x}$ states, $m_x$ preserves the magnetization along $\hat{\mathbf{x}}$ ($\mathbf{M}_{\pm x}\rightarrow \mathbf{M}_{\pm x}$). Therefore, the SHG intensity $I_\mathbf{M}^\mathbf{k}(\phi)$ has the symmetry $I_\mathbf{M}^\mathbf{k}(\phi) = I_\mathbf{M}^\mathbf{-k}(-\phi)$ (Supplementary Figs.~\ref{fig:PolScans}(a,b) are mirror symmetric to Supplementary.~\ref{fig:PolScans}(f,g)). \textbf{b,} $m_y$ reverses the magnetization along $\hat{\mathbf{x}}$ ($\mathbf{M}_{\pm x}\rightarrow \mathbf{M}_{\mp x}$). Thus, $I_\mathbf{M}^\mathbf{k}(\phi) = I_\mathbf{-M}^\mathbf{k}(-\phi)$ (Supplementary Figs.~\ref{fig:PolScans}(a,f) are mirror symmetric to Supplementary Figs.~\ref{fig:PolScans}(b,g)). \textbf{c,} In the $\mathbf{M}_{\pm y}$ states, $m_x$ reverses the magnetization along $\hat{\mathbf{y}}$ ($\mathbf{M}_{\pm y}\rightarrow \mathbf{M}_{\mp y}$).Thus, $I_\mathbf{M}^\mathbf{k}(\phi) = I_\mathbf{-M}^\mathbf{-k}(-\phi)$ (Supplementary Figs.~\ref{fig:PolScans}(c,d) are symmetric to of Supplementary Figs.~\ref{fig:PolScans}(i,h)). \textbf{d,} $m_y$ preserves the magnetization along $\hat{\mathbf{y}}$ ($\mathbf{M}_{\pm y}\rightarrow \mathbf{M}_{\pm y}$). Thus, $I_\mathbf{M}^\mathbf{k}(\phi) = I_\mathbf{M}^\mathbf{k}(-\phi)$ (Supplementary Figs.~\ref{fig:PolScans}(c,d,h,i) are individually mirror symmetric). The application of both mirror operations in any magnetic state enforces $I_\mathbf{M}^\mathbf{k}(\phi) = I_\mathbf{-M}^\mathbf{-k}(\phi)$. Thus, reversing both the magnetization and the propagation direction always preserves the SHG polarization dependence.}
		\label{ExtSymm}
	\end{figure}

	In addition to the discussions in the main text, we discuss here  the symmetry of the NODE in more detail. We find that mirror symmetries of the paramagnetic crystal lattice restore certain symmetries in the SHG polarization dependencies (Fig.~2 in the main text and Supplementary Fig.~\ref{fig:PolScans}). To see this, we consider the effect of the mirror operations $m_x$ and $m_y$ on the magnetization, propagation direction, and light polarization (Supplementary Fig.~\ref{ExtSymm}). 
	
	In general, the mirror operation $m_x$ reverses the propagation direction ($\mathbf{k}\rightarrow -\mathbf{k}$) (Supplementary Fig.~\ref{ExtSymm}a and c) while the mirror operation $m_y$ preserves $\mathbf{k}$ (Supplementary Fig.~\ref{ExtSymm}a and c). Both mirror operations individually transform the linear light polarization from an angle $\phi$ to $-\phi$ relative to p-polarized light. We will consider in the following discussion the SHG intensity $I_\mathbf{M}^\mathbf{k}(\phi)$. The arguments below are valid both for $s$ and $p$-polarized SHG light.
	
	In the $\mathbf{M}_{\pm x}$ states (Supplementary Fig.~\ref{ExtSymm}a), the mirror operation $m_x$ preserves the magnetization along $\hat{\mathbf{x}}$ ($\mathbf{M}_{\pm x}\rightarrow \mathbf{M}_{\pm x}$). Therefore, the SHG intensity $I_\mathbf{M}^\mathbf{k}(\phi)$ has the symmetry $I_\mathbf{M}^\mathbf{k}(\phi) = I_\mathbf{M}^\mathbf{-k}(-\phi)$. As a consequence in the experiment, the polarization dependencies in Supplementary Figs.~\ref{fig:PolScans}a and b are mirror symmetric to Supplementary Figs.~\ref{fig:PolScans}f and g.
	
	In contrast to $m_x$, $m_y$ reverses the magnetization along $\hat{\mathbf{x}}$ as shown in Supplementary Fig.~\ref{ExtSymm}b ($\mathbf{M}_{\pm x}\rightarrow \mathbf{M}_{\mp x}$). Thus, $I_\mathbf{M}^\mathbf{k}(\phi) = I_\mathbf{-M}^\mathbf{k}(-\phi)$. Therefore, Supplementary Figs.~\ref{fig:PolScans}a and f are mirror symmetric to Supplementary Figs.~\ref{fig:PolScans}b and g. 
	
	The situation in the $\mathbf{M}_{\pm y}$ states is slightly different. As shown in Supplementary Fig.~\ref{ExtSymm}c, $m_x$ reverses the magnetization along $\hat{\mathbf{y}}$ ($\mathbf{M}_{\pm y}\rightarrow \mathbf{M}_{\mp y}$). Thus, $I_\mathbf{M}^\mathbf{k}(\phi) = I_\mathbf{-M}^\mathbf{-k}(-\phi)$ (Supplementary Figs.~\ref{fig:PolScans}c and d are symmetric to of Supplementary Figs.~\ref{fig:PolScans}i and h. 
	
	Finally, $m_y$ preserves the magnetization along $\hat{\mathbf{y}}$ as illustrated in Supplementary Fig.~\ref{ExtSymm}d ($\mathbf{M}_{\pm y}\rightarrow \mathbf{M}_{\pm y}$). Thus, $I_\mathbf{M}^\mathbf{k}(\phi) = I_\mathbf{M}^\mathbf{k}(-\phi)$. Therefore, Supplementary Figs.~\ref{fig:PolScans}(c,d,h,i) are individually mirror symmetric. 
	
	Irrespective of the magnetic state, the application of both mirror operations (equivalent to a two-fold rotational axis $2_z$ along the $\hat{\mathbf{z}}$ axis) in any magnetic state always enforces $I_\mathbf{M}^\mathbf{k}(\phi) = I_\mathbf{-M}^\mathbf{-k}(\phi)$. Thus, reversing both the magnetization and the propagation direction always preserves the SHG intensity for any polarization dependence. Note that this property hinges on the presence of a two-fold rotational axis $2_z$ along the $\hat{\mathbf{z}}$ axis in the paramagnetic phase. For a general material, it is therefore not automatically guaranteed that reversing the magnetization is equivalent to reversing the light propagation. As a further consequence of the $2_z$ axis in the paramagnetic phase, the NODE is absent in the paramagnetic phase.
	
	\clearpage
	
	\newpage
	\section{Integrated SHG spectrum}
	
	In analogy to the NDD as a polarization-independent, directional contrast in the absorption coefficient, we introduced in the main text the integrated SHG intensity as a polarization-independent nonlinear optical observable. In this section, we present additional data illustrating properties of the integrated SHG intensity. In particular, we show that the NODE is present even for the integrated SHG intensity and it is thus not restricted to specific incident and outgoing light polarizations.
	
	We define the integrated SHG intensity as the intensity measured without polarization analysis (thus detecting simultaneously $s$ and $p$ polarized SHG light) and averaged over all incident polarization angles $\phi$. The such defined observable is independent of both incoming and outgoing light polarizations.
	
	As we showed in Sec.~\ref{sec:Symmetry} and Supplementary Fig.~\ref{fig:PolScans}, the detected SHG intensity in the $\mathbf{M}_\mathrm{\pm x}$ states exhibits the symmetry $I_\mathrm{\pm x}^\mathbf{k}(\phi) = I_\mathrm{\pm x}^\mathbf{-k}(-\phi) = I_\mathrm{\mp x}^\mathbf{k}(-\phi)$. Averaging over all polarization angles $\phi$, we find $\langle I_\mathrm{\pm x}^\mathbf{k}\rangle = \langle I_\mathrm{\pm x}^\mathbf{-k}\rangle = \langle I_\mathrm{\mp x}^\mathbf{k}\rangle$, i.e., the integrated SHG intensity in the $\mathbf{M}_\mathrm{\pm x}$ states is independent of the propagation direction and magnetization direction. Accordingly, we find in Supplementary Fig.~\ref{fig:IntSpectrum}a equal integrated SHG intensities in the $\mathbf{M}_\mathrm{\pm x}$ states. 
	
	In contrast, the integrated SHG intensity in the $\mathbf{M}_\mathrm{\pm y}$ states is different. This is due to the absence of a symmetry operation that relates the two magnetic states while preserving the propagation direction. We have either $I_\mathrm{\pm y}^\mathbf{k}(\phi) = I_\mathrm{\mp y}^\mathbf{-k}(-\phi)$ or $I_\mathrm{\pm y}^\mathbf{k}(\phi) = I_\mathrm{\pm y}^\mathbf{k}(-\phi)$ (Sec.~\ref{sec:Symmetry}). Thus, for a fixed propagation direction, the integrated SHG intensity can differ for the two magnetic states $\mathbf{M}_\mathrm{+y}$ and $\mathbf{M}_\mathrm{-y}$. 
	
	Reversing the light propagation (Supplementary Fig.~\ref{fig:IntSpectrum}b), we find in agreement with the symmetry considerations equal integrated SHG intensities for the $\mathbf{M}_\mathrm{\pm x}$ states, but the relation between $\langle I_\mathrm{+y}\rangle$ and $\langle I_\mathrm{-y}\rangle$ is inverted relative to panel a. This property is a direct consequence of the relation $I_\mathrm{\pm y}^\mathbf{k}(\phi) = I_\mathrm{\mp y}^\mathbf{-k}(-\phi)$.
	
	In Supplementary Fig.~\ref{fig:IntSpectrum}c, we show the spectral evolution of the integrated SHG intensity for one propagation direction. Very strikingly, we find $\langle I^{\ra}_{+x}\rangle = \langle I^{\ra}_{-x}\rangle$ for all photon energies.
	
	Overall, we note the following properties of the integrated SHG intensity:
	
	\begin{itemize}
		\item if the SHG intensity does not exhibit a NODE (here: SHG response above $T_C$), the integrated SHG intensity does not exhibit a NODE.
		\item if the SHG intensity shows a NODE, mirror symmetries may still enforce a vanishing NODE in the integrated SHG intensity.
	\end{itemize}
	
	\begin{figure}[ht]
		\centering
		\includegraphics[width = 0.7\textwidth]{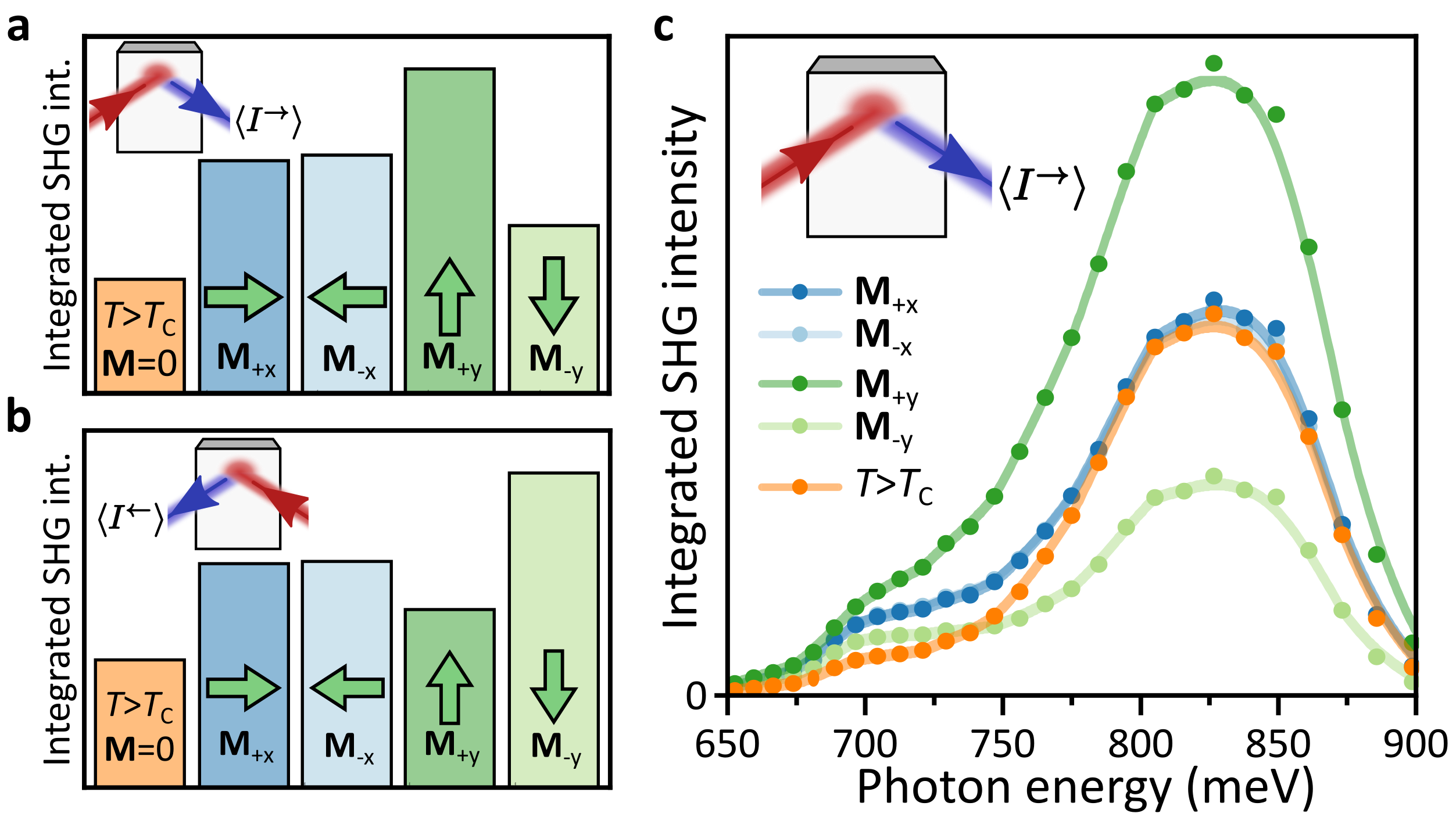}
		\caption{\textbf{NODE in the integrated SHG intensity.} \textbf{a,} Integrated SHG in forward direction at \SI{715}{meV}. \textbf{b,} Integrated SHG in backward direction at \SI{715}{meV}. In contrast to the forward direction, $\langle I^{\la}_{+y}\rangle < \langle I^{\la}_{-y}\rangle$. The integrated SHG intensity in the $\mathbf{M}_\mathrm{\pm x}$ states is always identical for the two propagation directions. \textbf{c,} Spectra of the integrated SHG intensity $\langle I^{\ra}\rangle$ in forward direction for all magnetic states and in the paramagnetic phase. $\langle I^{\ra}_{+x}\rangle = \langle I^{\ra}_{-x}\rangle$ due to $m_y$ mirror operation (Sec.~\ref{sec:Symmetry}).}
		\label{fig:IntSpectrum}
	\end{figure}
	\clearpage
	
	\newpage
	\section{FIB sample}
	
	In an effort to demonstrate the NODE in a device-like structure and minimize the effect of Joule heating observed in the current switching of the bulk single crystal, we microstructured a CeAlSi crystal into a free-standing bar of approximately \SI{126}{\micro\metre} length, \SI{29}{\micro\metre} width, and \SI{2}{\micro\metre} thickness by focused-ion-beam (FIB) milling (see methods for details). A false-colored SEM picture is shown in Fig.~3f in the main text. In order to identify the magnetic state during the application of electrical current (Fig.~3g in the main text), we pre-characterized the SHG response of the FIB device by applying a static magnetic field of approximately \SI{30}{mT} along the four magnetic easy axes. The incident polarization dependence for the $s$ and $p$ polarized SHG light is shown in Supplementary Fig.~\ref{ExtFIB}. 
	
	Although the polarization dependencies differ from the scans on the bulk single crystal (Supplementary Fig.~\ref{fig:PolScans}), we can identify four significantly different polarization patterns for the four magnetic states. Possible reasons for the differences between Supplementary Figs.~\ref{fig:PolScans} and \ref{ExtFIB} include changes to the surface morphology of the material caused by FIB milling (see methods) and deviations of the angle of incidence from $45^\circ$ (a few degree tilt between the surface of the substrate and the surface of the FIB device was unavoidable due to the extreme aspect ratio of the FIB sample).
	
	A comparison of the magnetic-field poled polarization dependencies in Supplementary Fig.~\ref{ExtFIB} and the electric current induced states (Fig.~3h,i in the main text) reveals a current-induced switching between $\mathbf{M}_\mathrm{+x}$ and $\mathbf{M}_\mathrm{-x}$. Solid lines in Supplementary Fig.~\ref{ExtFIB} are fits. In analogy to Supplementary Fig.~\ref{fig:PolScans}, we fit all measurements simultaneously with one consistent set of $\chi$ tensor components. We list the fit parameters relative to $\chi_{zzz}$ in Supplementary Table~\ref{tab:FIBcomponents} below. A surface tilt of approximately $8^\circ$ was included in the fitting procedure. The excellent fit quality is consistent with the $2^\prime m m^\prime$ symmetry for the magnetic state of the FIB sample.
	
	Moreover, we performed a temperature-dependent characterization of the FIB sample. A magnetic field of approximately \SI{30}{mT} stabilizes a magnetic  $\mathbf{M}_\mathrm{+y}$ state. In Supplementary Fig.~\ref{ExtFIBtemp}a, we show the SHG intensity as a function of temperature in the absence of any electrical current. We find a transition temperature in agreement with the bulk sample. 
	
	In Supplementary Fig.~\ref{ExtFIBtemp}b, we show the SHG intensity in the $\mathbf{M}_\mathrm{+y}$ state at \SI{3}{K} under the application of an electric DC current. We extrapolate a critical current of \SI{8.5}{mA} above which Joule heating raises the sample temperature above the transition temperature. Joule heating is negligible for the currents relevant for Fig.~3g in the main text ( $< \SI{3}{mA}$). 
	
	\begin{table}
		\begin{center}
			\begin{tabular}{|c|c|c|}
				\hline
				$\chi_{ijk}$ & $\vert \chi_{ijk}\vert/\vert \chi_{zzz}\vert$ & $(\Phi_{ijk} - \Phi_{zzz})/\pi$ \\
				\hline
				$zzz$ & 1 & 0 \\
				\hline
				$xxz$ & 0.069 & 0.774 \\
				\hline
				$zxx$ & 0.313 & 0.182 \\
				\hline
				$yyz$ & 0.106 & -0.268 \\
				\hline
				$zyy$ & 0.284 & 0.377 \\
				\hline
				$xxx$ & 0.081 & -0.26 \\
				\hline
				$yyx$ & 0.05 & -0.11 \\
				\hline
				$xyy$ & 0.204 & 0.366 \\
				\hline
				$zzx$ & 0.125 & 0.605 \\
				\hline
				$xzz$ & 0.127 & -0.08 \\
				\hline
			\end{tabular}
			\caption{Magnitudes and phases of $\chi$ tensor components relative to $\chi_{zzz}$. Values were found by fitting all 8 measurements below $T_C$ in Supplementary Fig.~\ref{ExtFIB} simultaneously with one consistent set of $\chi$ tensor components according to point group $2^\prime m m^\prime$.}
			\label{tab:FIBcomponents}
		\end{center}
	\end{table}
	
	\begin{figure}[h]
		\includegraphics[width = \textwidth]{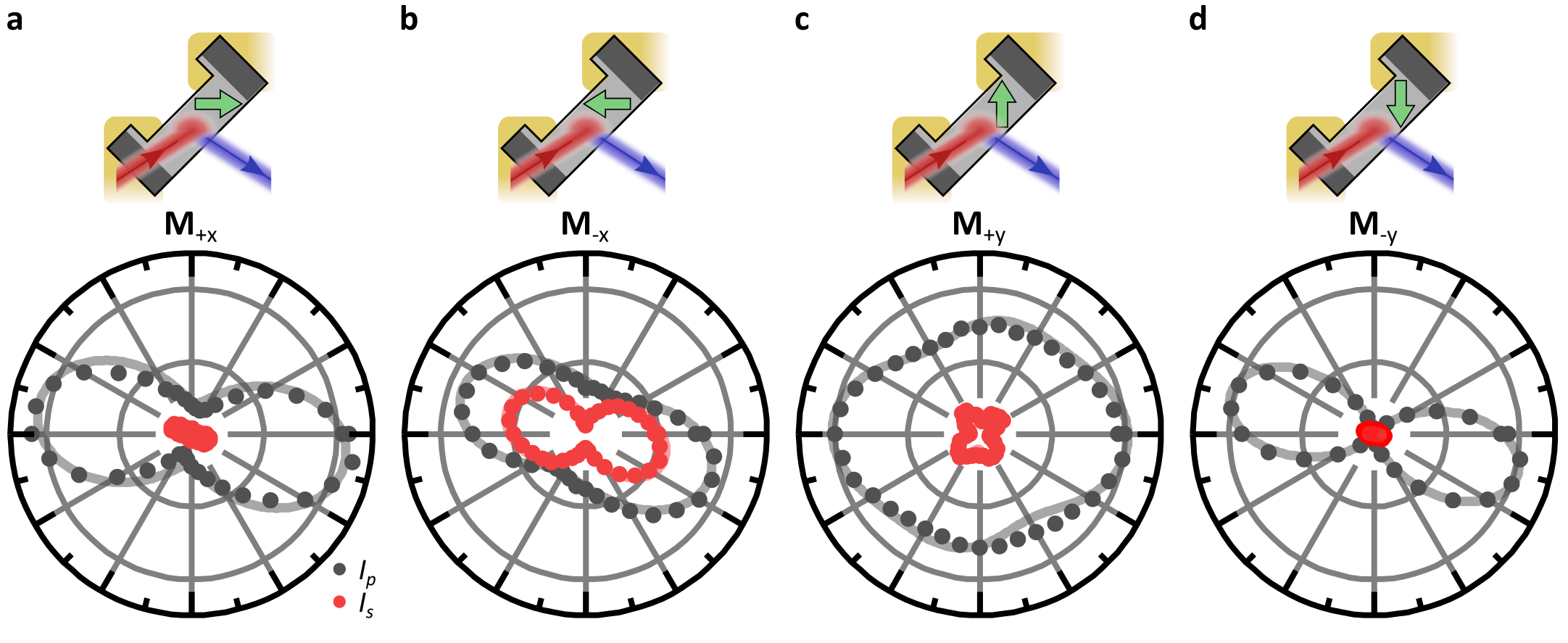}
		\caption{\textbf{SHG polarization dependencies for FIB sample.} \textbf{a}-\textbf{d} SHG polarization dependence in the magnetic $\mathbf{M}_\mathrm{+x}$, $\mathbf{M}_\mathrm{-x}$, $\mathbf{M}_\mathrm{+y}$, and $\mathbf{M}_\mathrm{-y}$ states, respectively. Analogously to the SHG signal from CeAlSi bulk single crystals (Supplementary Fig.~\ref{fig:PolScans}), the SHG polarization dependence allows us to distinguish the four magnetization directions in micro-machined CeAlSi samples. Here, the magnetic state was stabilized by an external magnetic field of \SI{30}{mT}. Solid lines are fits. A comparison of field-polarized measurements to the field-free measurements shown in Figs.~3h,i reveals electrical switching between $\mathbf{M}_\mathrm{-x}$ and $\mathbf{M}_\mathrm{+x}$ states, respectively.}
		\label{ExtFIB}
	\end{figure}
	
	\begin{figure}[h]
		\includegraphics[width = 0.7\textwidth]{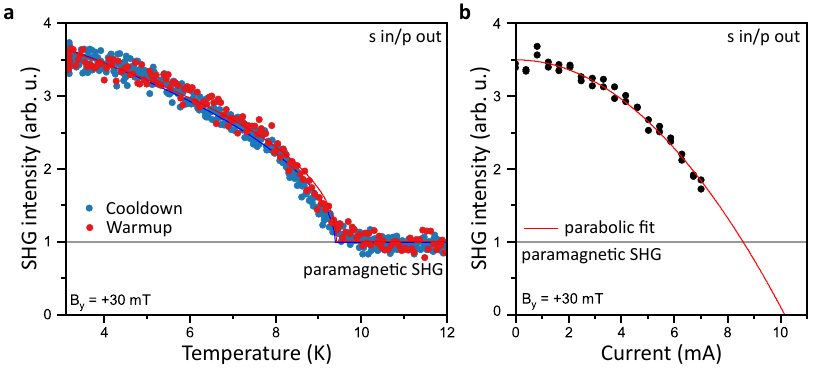}
		\caption{\textbf{Temperature-dependent characterization of the FIB sample.} In analogy to the SHG response of CeAlSi bulk single crystals (Supplementary Fig.~\ref{fig:Temperature}), we performed \textbf{a,} a temperature-dependent characterization of the FIB sample and \textbf{b,} characterized the effect of Joule heating on the FIB sample. We find a transition temperature in agreement with the bulk. From panel b, we find that the FIB sample remains in the magnetically ordered phase below \SI{8.5}{mA}. The effect of Joule heating is negligible below \SI{3}{mA}. The magnetic state was stabilized by an external magnetic field of \SI{30}{mT}.}
		\label{ExtFIBtemp}
	\end{figure}
	\clearpage
	
	\section{First-principles calculations}
	
	\subsection{Electronic structure of CeAlSi}
	
	We provide here further details about the first-principles calculations described in the main text. In particular, in this section we describe the electronic structure of CeAlSi in more detail based on numerical first-principles calculations. We performed calculations both in the paramagnetic and in the ferromagnetically ordered phase of CeAlSi. A comparison of the band structure is shown in Supplementary Fig.~\ref{fig:DFT_band_structure}. Our calculations agree well with other recent calculations and experimental observations \cite{Sakhya2023}.
	
	Due to the noncentrosymmetric lattice structure, CeAlSi is already a Weyl semimetal in the paramagnetic phase. Going to the ferromagnetic phase, CeAlSi remains to be a Weyl semimetal, with relatively small modifications to the details of the band structure. Figure~\ref{fig:DFT_band_structure} shows that the overall band structures of the paramagnetic and ferromagnetic phases are similar: CeAlSi is a semimetal in both states. 
	
	We found a total of 40 Weyl nodes in the band structure of CeAlSi in the paramagnetic phase. We list their positions and energies in Table~\ref{table:PM_WeylL} and illustrate their positions in Supplementary Fig.~\ref{fig:weyl_distribute}. The magnetic order does not change the number of Weyl nodes, but the position of the 40 Weyl nodes shifts both in energy and in momentum upon the inclusion of the magnetic order (Table~\ref{table:FM_WeylL}). 
	
	Based on the electronic structure in the magnetically ordered phase, we calculate the components of the nonlinear optical susceptibility $\chi_{ijk}$ following the diagrammatic approach to nonlinear optical responses \cite{Parker2019,Takasan2021} (see also methods). We show the real and imaginary parts of the obtained susceptibilities in supplementary Fig.~\ref{Ext5} in the energy range of \SI{650}{meV} to \SI{900}{meV} (incident photon energy). We find that only the components that are allowed by symmetry exhibit finite susceptibilities \cite{Birss66}.
	
	We note here also the absence of any sharp resonances in any tensor component. This is in contrast to wide bandgap insulators, where often sharp resonances occur in the energy range close to the band gap \cite{Xiao2023}. Our numerical calculations thus reproduce a broadband SHG response. In Supplementary Fig.~\ref{fig:DFTNODE}, we show the calculated spectrum of the SHG intensity in a setting that reproduces the experiment in Fig.~1e of the main text ($s$ polarized incident light, $p$ polarized SHG response, $45^\circ$ angle of incidence, $\mathbf{M}_\mathrm{+y}$ state). We observe a clear difference in the SHG intensity between the two propagation directions. As a consequence of the absence of resonances in $\chi_{ijk}$, the SHG intensity evolves smoothly as a function of photon energy.
	
	\begin{figure}[h]
		\includegraphics[width = 0.7\columnwidth]{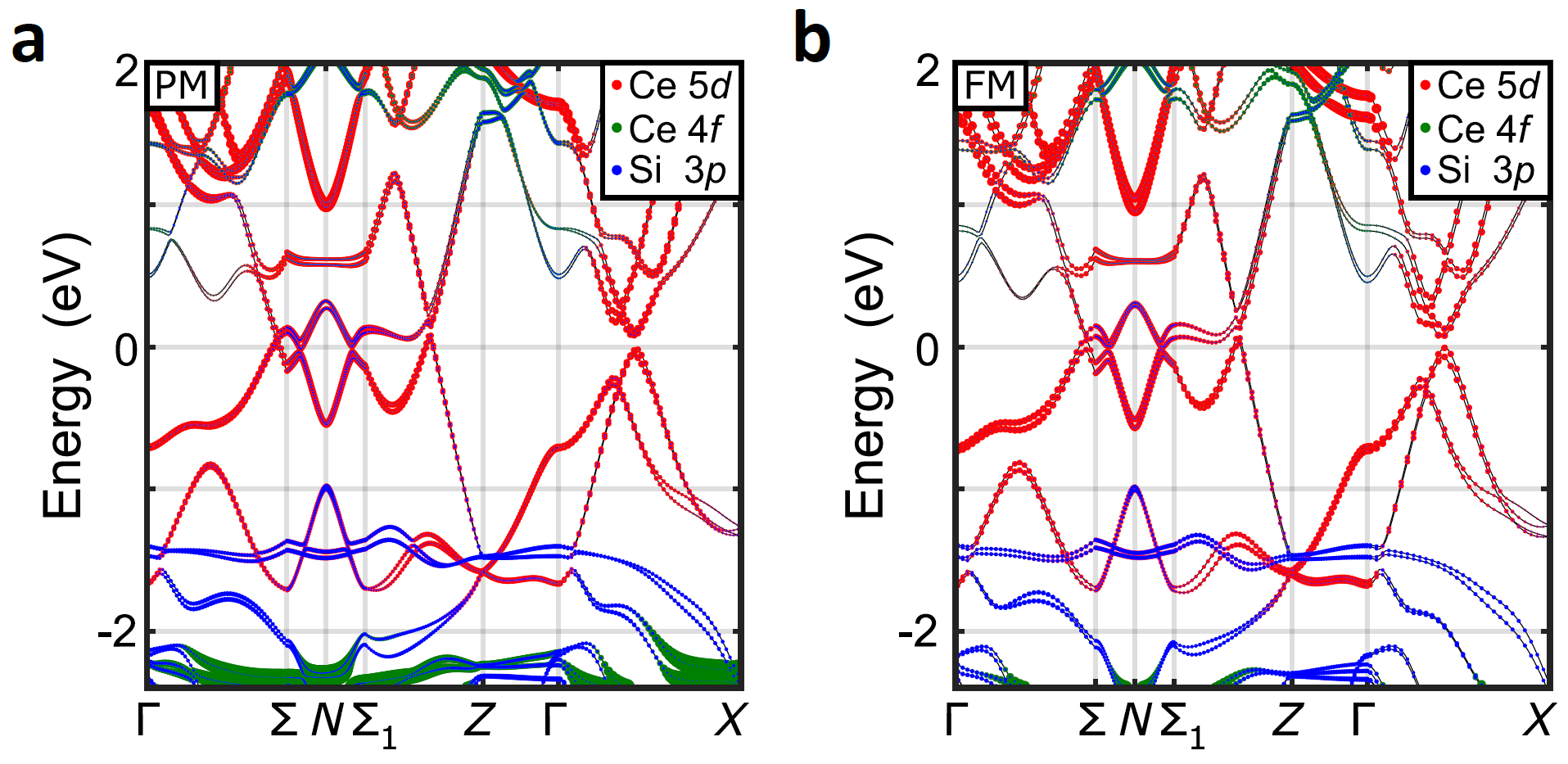}
		\caption{Band structure of CeAlSi in the \textbf{a,} paramagnetic phase and \textbf{b,} ferromagnetic phase. Color coded is the orbital composition of the electronic bands.}
		\label{fig:DFT_band_structure}
	\end{figure}
	
	\begin{figure}[!htb]
		\begin{center}
			\includegraphics[width = 0.7\columnwidth]{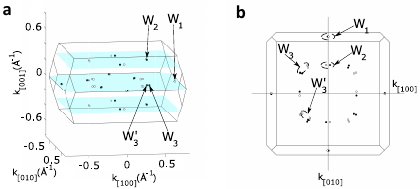}
			\caption{\textbf{Distribution of Weyl points in the Brillouin zone.}  \textbf{a,} The distribution of the 40 Weyl nodes in the Brillouin zone. Black and white colors represent opposite Weyl node chirality. \textbf{b,} 2D projection of all Weyl nodes. }
			\label{fig:weyl_distribute}
		\end{center}
	\end{figure}
	
	\begin{table}
		\begin{tabular}{cccc}
			\hline\hline
			{Weyl Nodes} & \multicolumn{2}{c}{Position: $(k_x, k_y, k_z) (1/\textrm{\AA})$} & {Energy (meV)} \\
			& Chirality: -1 & Chirality: +1 & \\
			\hline
			$\textrm{W}_1$ & $(0.007,0.744,0.000)$ & $(-0.744,-0.007,0.000)$ & 58 \\
			& $(0.744,-0.007,0.000)$ & $(0.007,-0.744,0.000)$ & 58 \\
			& $ (-0.007,-0.744,0.000)$ & $(0.744,0.007,0.000)$ & 58 \\
			&  $ (-0.744,0.007,0.000)$ & $(-0.007,0.744,0.000)$ & 58 \\
			\hline
			$\textrm{W}_2$  & $(-0.026,0.372,\pm0.295)$ & $(-0.372,0.026,\pm0.295)$ & 31 \\
			& $(0.372,0.026,\pm0.295)$ & $(-0.026,-0.372,\pm0.295)$ &  31 \\
			& $(0.026,-0.372,\pm0.295)$ & $(0.372,-0.026,\pm0.295)$ & 31 \\
			& $(-0.372,-0.026,\pm0.295)$ & $(0.026,0.372,\pm0.295)$ & 31 \\
			\hline
			$\textrm{W}_3$  & $(0.274,-0.367,0.000)$ & $(0.367,-0.274,0.000)$ & 46 \\
			& $(-0.367,-0.274,0.000)$ & $(0.274,0.367,0.000)$ & 46 \\
			& $(0.367,0.274,0.000)$ & $(-0.274,-0.367,0.000)$ & 46 \\
			& $(-0.274,0.367,0.000)$ & $(-0.367, 0.274,0.000)$ & 46 \\
			\hline
			$\textrm{W}_3^{'}$  & $(0.339,0.263,0.000)$ & $(-0.263,-0.339,0.000)$ & 31 \\
			& $(-0.339,-0.263,0.000)$ & $(0.263,0.339,0.000)$ & 31 \\
			& $(0.263,-0.339,0.000)$ & $(0.339,-0.263,0.000)$ & 31 \\
			& $(-0.263,0.339,0.000)$ & $(-0.339,0.263,0.000)$ & 31 \\
			\hline
			\hline
		\end{tabular}
		\caption{\textbf{Weyl node positions and energies in the paramagnetic phase of CeAlSi.} Weyl node energies are given relative to the Fermi energy.}
		\label{table:PM_WeylL}
	\end{table}
	
	\begin{table}
		\begin{tabular}{cccc}
			\hline\hline
			{Weyl Nodes} & \multicolumn{2}{c}{Position: $(k_x, k_y, k_z)(1/\textrm{\AA})$} & {Energy(meV)} \\
			& Chirality: -1 & Chirality: +1 & \\
			\hline
			$\textrm{W}_1$ & $(0.005,0.760,0.000)$ & $(-0.760,-0.005,0.000)$ & 39 \\
			& $(0.757,-0.004,0.000)$ & $(0.004,-0.757,0.000)$ & 65 \\
			& $ (-0.007,-0.738,0.000)$ & $(0.738,0.007,0.000)$ & 70 \\
			&  $ (-0.743,0.004,0.000)$ & $(-0.004,0.743,0.000)$ & 81 \\
			\hline
			$\textrm{W}_2$  & $(-0.023,0.395,\pm0.284)$ & $(-0.395,0.023,\pm0.284)$ & -24 \\
			& $(0.388,0.023,\pm0.282)$ & $(-0.023,-0.388,\pm0.282)$ & -23 \\
			& $(0.018,-0.369,\pm0.305)$ & $(0.369,-0.018,\pm0.305)$ & 23 \\
			& $(-0.372,-0.024,\pm0.294)$ & $(0.024,0.372,\pm0.294)$ & 26 \\
			\hline
			$\textrm{W}_3$  & $(0.276,-0.366,0.000)$ & $(0.366,-0.276,0.000)$ & 8 \\
			& $(-0.355,-0.282,0.000)$ & $(0.282,0.355,0.000)$ & 37 \\
			& $(0.387,0.267,0.000)$ & $(-0.267,-0.387,0.000)$ & 48 \\
			& $(-0.270,0.372,0.000)$ & $(-0.372,0.270,0.000)$ & 69 \\
			\hline
			$\textrm{W}_3^{'}$  & $(0.333,0.257,0.000)$ & $(-0.257,-0.333,0.000)$ & 10 \\
			& $(-0.340,-0.257,0.000)$ & $(0.257,0.340,0.000)$ & 26 \\
			& $(0.270,-0.339,0.000)$ & $(0.339,-0.270,0.000)$ & 33 \\
			& $(-0.278,0.353,0.000)$ & $(-0.353,0.278,0.000)$ & 52 \\
			\hline
			\hline
		\end{tabular}
		\caption{\textbf{Weyl node positions and energies in the ferromagnetic phase of CeAlSi.} Weyl node energies are given relative to the Fermi energy.}
		\label{table:FM_WeylL}
	\end{table}
	
	\begin{figure}
		\includegraphics[width = \textwidth]{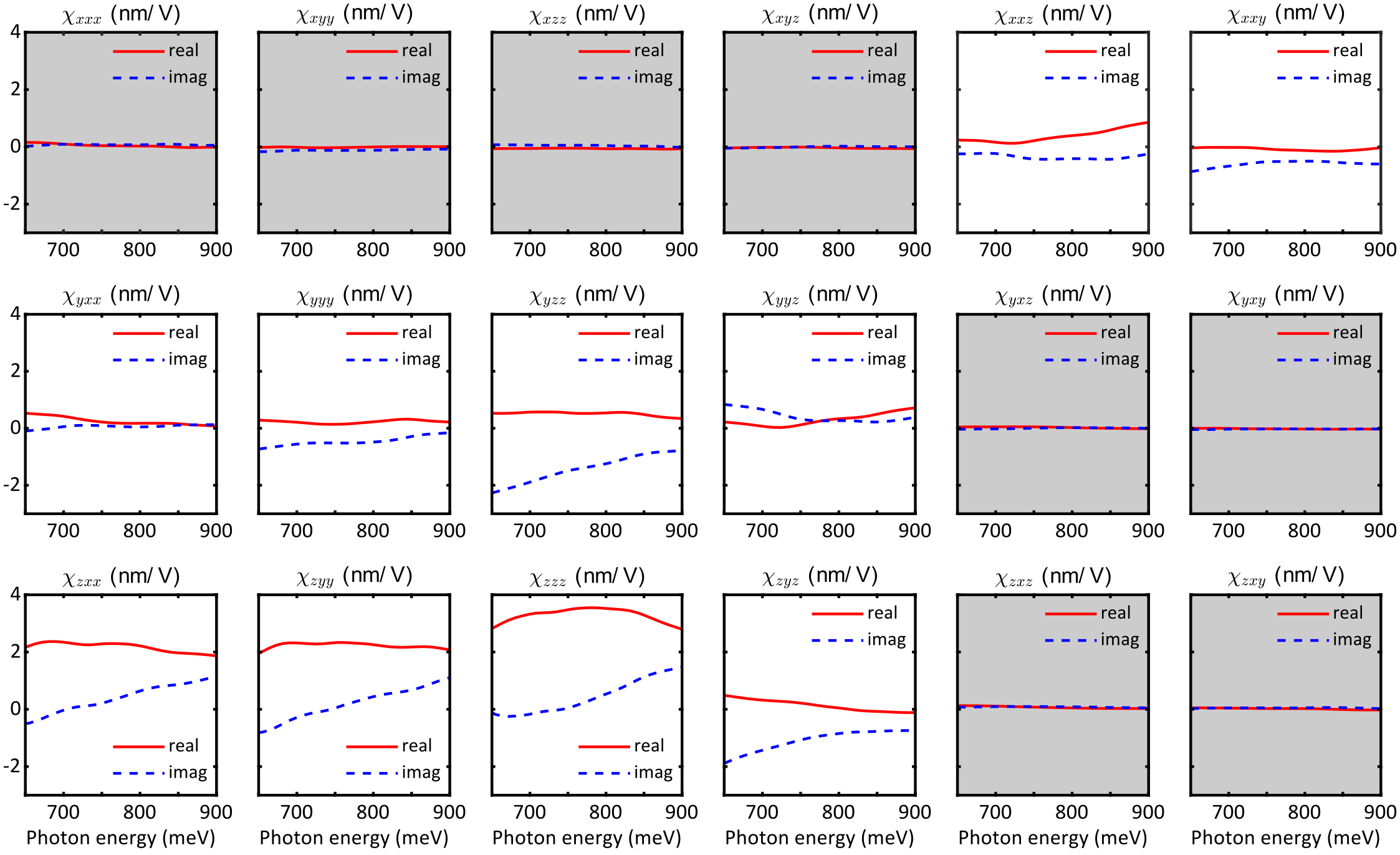}
		\caption{\textbf{First-principles calculations of SHG tensor component spectra in the magnetic phase of CeAlSi.} All spectra are shown on the same scale. In agreement with the $2^\prime m m^\prime$ point group symmetry, the grayed out components vanish. The allowed components evolve continuously without any pronounced resonance in agreement with our understanding of broadband SHG due to linearly dispersive bands in CeAlSi.}
		\label{Ext5}
	\end{figure}
	
	\begin{figure}
		\includegraphics[width = 0.3\textwidth]{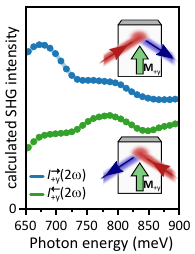}
		\caption{\textbf{Observation of the NODE from first principles.} Spectral dependence of SHG intensity for counter-propagating light paths. The calculations are done based on the numerically obtained tensor components (Supplementary Fig.~\ref{Ext5}) and reproducing the experimental setting of Fig. 1e in the main text. We observe a clear NODE.}
		\label{fig:DFTNODE}
	\end{figure}
	\clearpage
	
	\subsection{$k$-space resolved $\chi$ contributions}
	
	As described in the methods section, we calculate $\chi$ as $\chi_{ijk} = \int_{BZ}d^3\mathbf{k} \xi_{ijk}$, where $\xi_{ijk} = \xi^I_{ijk}+\xi^{II}_{ijk}$ with
	
	\begin{eqnarray}
		\xi^I_{ijk} &=& C \sum_{m\neq n} f_{mn}\left(\frac{h^{ij}_{nm}h^k_{mn} + h^{ik}_{nm}h^j_{mn}}{\omega+i\eta - \omega_{mn}} + \frac{h^{jk}_{mn}h^i_{nm}}{2\omega+i\eta - \omega_{mn}}\right) \label{eqn:supp1}, \\
		\xi^{II}_{ijk} &=& C \sum_{m\neq n\neq p} \frac{h^i_{pm}\left(h^j_{mn}h^k_{np}+h^k_{mn}h^j_{np}\right)}{\omega_{mn}+\omega_{np}}\left(\frac{f_{np}}{\omega + i\eta -\omega_{pn}} + \frac{f_{nm}}{\omega + i\eta -\omega_{nm}} + \frac{2f_{mp}}{\omega + i\eta -\omega_{pm}}\right)\label{eqn:supp2}
	\end{eqnarray}  
	
	To gain further insights into the relation of the SHG response and the electronic structure of CeAlSi, we investigate here the $k$ space distribution of $\xi_{ijk}$. As an example, we focus here in particular on $\xi_{xxy}$ in the ferromagnetic state with magnetization $\mathbf{M}_{+x}$ pointing along $\hat{\mathbf{x}} = [110]$. We show in Supplementary Fig.~\ref{SuppFig:DFTkspace}a the magnitude of $\xi_{xxy}$ at an incident photon energy of \SI{650}{meV}. We chose a plane in the Brillouin zone with $k_z = 0.295 \AA^{-1}$. In the paramagnetic phase of CeAlSi, this plane contains the $W_2$ Weyl nodes. In the ferromagnetic phase, the Weyl nodes shift out of the plane (Supplementary Table~\ref{table:FM_WeylL}). Red and blue dots therefore indicate the projection of the Weyl nodes onto the plane $k_z = 0.295 \AA^{-1}$ in the ferromagnetic phase.
	
	We notice a rather complicated distribution of $\vert\xi_{xxy}\vert$ in the plane, but major contributions appear to be related to the Weyl nodes. Note that $\xi_{xxy}$ contains a sum over all bands (Eqs.~\ref{eqn:supp1} and \ref{eqn:supp2}). The main features of Supplementary Fig.~\ref{SuppFig:DFTkspace}a, however, can already be recognized in Supplementary Fig.~\ref{SuppFig:DFTkspace}b, where we only consider contributions from electronic transitions between bands -1 and +1 (marked red and blue in Fig.~4c of the main text) \footnote{According to the numerically determined band structure (Fig.~4c in the main text), bands -1 and 0 follow each other closely throughout the Brillouin zone (as do bands +1 and +2). Thus, any of the transitions $-1 \rightarrow +1$, $-1 \rightarrow +2$, $0 \rightarrow +1$, or $0 \rightarrow +2$ yields a figure similar to Supplementary Fig.~\ref{SuppFig:DFTkspace}b. The choice of bands -1 and +1 for Supplementary Fig.~\ref{SuppFig:DFTkspace}b is arbitrary and serves as an example.}.
	
	Note that each pair of Weyl nodes is surrounded by two lines of strong contributions to $\xi_{xxy}$. The inner line corresponds to electronic transitions that are resonant at the incident photon energy (here: \SI{650}{meV}); the outer line corresponds to transitions resonant at the second-harmonic photon energy (here: $2\times\SI{650}{meV}$).
	
	As we vary the incident photon energy between \SI{550}{meV} (Supplementary Fig.~\ref{SuppFig:DFTkspace}c) and  \SI{950}{meV} (Supplementary Fig.~\ref{SuppFig:DFTkspace}g), we do not observe major changes, but the distance of the inner and outer lines to the Weyl node pair changes. In fact, the distance of the lines of $\xi_{xxy}$ contributions from the Weyl nodes increases approximately linearly with the incident photon energy (dashed black lines in panels c-g), which is due to the linearly dispersive bands in the vicinity of the Weyl nodes.
	
	In Supplementary Fig.~\ref{SuppFig:DFTkspace}h, we show a schematic illustration of a possible band structure near the line with $k_\mathrm{[010]} = 0$, which would qualitatively reproduce our observations from the DFT calculations.
	
	Interestingly, the magnitude of $\xi_{xxy}$ does not change significantly within the considered energy range (in line with the smooth changes of $\chi_{ijk}$, Supplementary Fig. 17). The comparable magnitude of $\xi$ over a wide energy range in combination with smooth changes of the $\xi$ contributions throughout the Brillouin zone rationalize the broadband SHG response that we observed in the experiment.
	
	\begin{figure}
		\includegraphics[width = 133.5 mm]{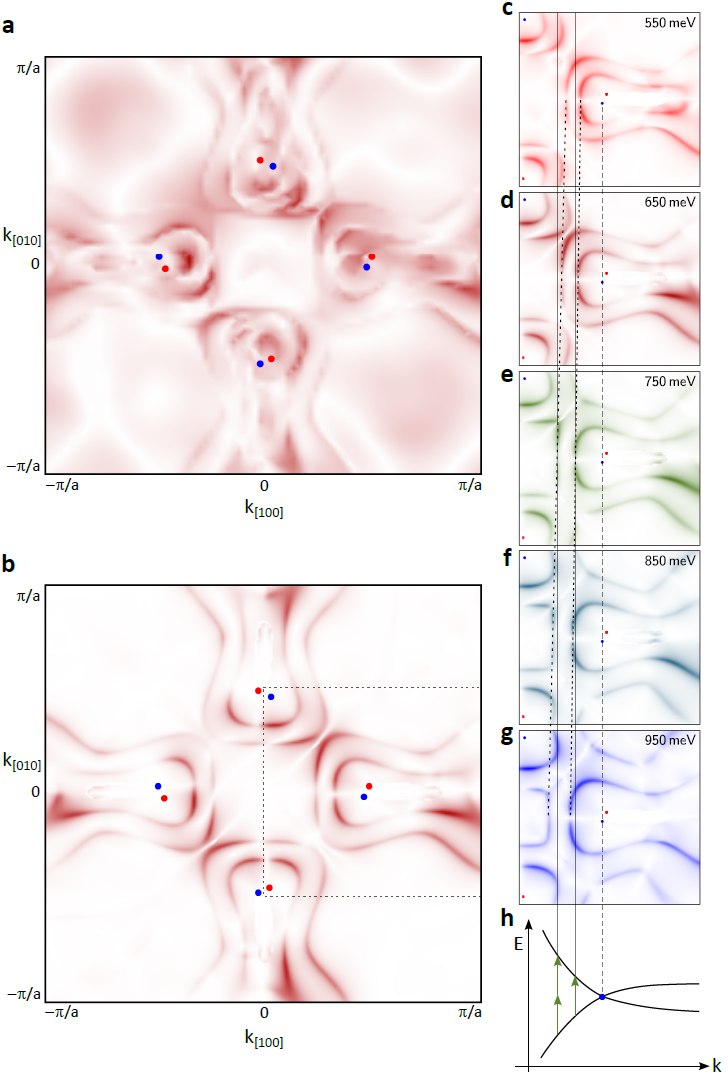}
		\caption{\textbf{$k$ space distribution of $\xi_{ijk}$ from first principles.} \textbf{a,} $k$ space distribution of $\xi_{xxy}$ in the plane $k_z = 0.295 \AA^{-1}$ for an incident photon energy of \SI{650}{meV}. Red and blue dots indicate the projection of the $W_2$ Weyl nodes of different chirality onto the considered plane. \textbf{b,} contributions to $\xi_{xxy}$ from electronic transitions between electronic bands -1 and +1. \textbf{c-g,} Contributions to $\xi_{xxy}$ at various incident photon energies between \SI{550}{meV} and \SI{950}{meV}. Only the section within the dashed box of panel b is shown. \textbf{h,} schematic band structure illustrating the origin of the two bands surrounding the Weyl node pair in panels c-g: the closer band corresponds to an electronic transition at the incident photon energy $\hbar\omega$, whereas the outer band corresponds to an electronic transition at $\hbar 2\omega$.}
		\label{SuppFig:DFTkspace}
	\end{figure}
	
	
	
	\clearpage
	\renewcommand{\baselinestretch}{1.2}
	\bibliographystyle{Science}	

	\providecommand{\noopsort}[1]{}\providecommand{\singleletter}[1]{#1}%